\documentclass[a4paper,11pt]{article}
\usepackage{jheppub}

\usepackage{mathtools}  
\usepackage{xcolor}
\usepackage[bb=stix, cal=boondoxupr]{mathalpha}    
\usepackage{array, longtable, booktabs, multirow}
\usepackage[smalltableaux]{ytableau}
\usepackage{tikz}
\usetikzlibrary{calc}

\newenvironment{eqnalign}{
\equation\aligned
}{
\endaligned\endequation
}

\newcommand{\defeq}{\vcentcolon=}

\definecolor{typeT}{HTML}{E06836}
\definecolor{typeS}{HTML}{4B9423}
\definecolor{typeN}{HTML}{5159B8}
\newcommand{\tT}[1]{{\color{typeT}{\ifmmode{\mathrm{#1}}\else{#1}\fi}}}
\newcommand{\tS}[1]{{\color{typeS}{\ifmmode{\mathrm{#1}}\else{#1}\fi}}}
\newcommand{\tN}[1]{{\color{typeN}{\ifmmode{\mathrm{#1}}\else{#1}\fi}}}

\definecolor{myRed}{HTML}{D1495B}
\definecolor{myYellow}{HTML}{EDAE49}
\definecolor{myTeal}{HTML}{00798C}

\newcommand{\R}{\mathbb{R}}
\newcommand{\Z}{\mathbb{Z}}
\newcommand{\calR}{\mathcal{R}}
\newcommand{\calF}{\mathcal{F}}
\renewcommand{\H}{\mathcal{H}}  
\newcommand{\ch}{\mathrm{ch}}   

\DeclareMathOperator{\diag}{diag}
\DeclareMathOperator{\tr}{tr}
\DeclareMathOperator{\rank}{rank}

\renewcommand{\O}{\operatorname{O}}
\DeclareMathOperator{\U}{U}
\DeclareMathOperator{\SU}{SU}
\DeclareMathOperator{\SO}{SO}
\DeclareMathOperator{\Sp}{Sp}

\newcommand{\triv}{\rep{1}}
\newcommand{\adj}{\mathrm{Adj}}
\newcommand{\rep}[2][]{\mathbf{\underline{#2}^{#1}}}    
\newcommand{\repss}[2]{\mathbf{\underline{#1}_{#2}}}    
\newcommand{\repbar}[2][]{                              
    \mathbf{\underline{\overline{#2}}^{#1}}
} 

\newcommand{\G}{\mathcal{G}}    
\renewcommand{\v}{\mathcal{v}}  
\newcommand{\V}{\mathcal{V}}    
\newcommand{\e}{\mathcal{e}}    
\newcommand{\E}{\mathcal{E}}    
\newcommand{\K}{\mathcal{K}}    

\title{\boldmath Enumerating 6D supergravities with $T\leq 1$}

\author[\clubsuit\spadesuit]{Yuta Hamada}
\author[\clubsuit]{and Gregory J.\ Loges}
\affiliation[\clubsuit]{
    Theory Center, IPNS, High Energy Accelerator Research Organization (KEK), \\
    1-1 Oho, Tsukuba, Ibaraki 305-0801, Japan
}
\affiliation[\spadesuit]{
    Graduate Institute for Advanced Studies, SOKENDAI, 1-1 Oho, Tsukuba, Ibaraki 305-0801, Japan
}

\emailAdd{yhamada@post.kek.jp}
\emailAdd{gloges@post.kek.jp}

\preprint{KEK-TH-2615}

\abstract{
    The space of 6D supergravities with minimal supersymmetry is greatly constrained by anomaly cancellation. Nevertheless, a large number of models satisfy all low-energy consistency conditions and in this work we make progress towards exhaustively enumerating all anomaly-free models with at most one tensor multiplet.
    Generalizing previous techniques, we describe a general algorithm using multi-hypergraphs and simplicial complexes to systematically enumerate anomaly-free models with gauge groups of any number of simple factors and with hypermultiplets falling into any representations.
    Using these new ideas, we obtain a \emph{complete} list of anomaly-free models for $T\leq 1$, the only simplifying assumption being that the gauge group contains no $\operatorname{U}(1)$, $\operatorname{SU}(2)$, $\operatorname{SU}(3)$ or $\operatorname{Sp}(2)$ factors.
    We also study how many/which models in this ensemble satisfy several UV and swampland bounds which have been proposed and previously utilized to great effect, finding that none are ruled out for $T=0$ and $\approx\!50\%$ are inconsistent with quantum gravity for $T=1$.
}

\begin{document}

\maketitle
\flushbottom

\section{Introduction}
\label{sec:introduction}

In the past few years there has been a renewed interest in the use of anomalies to understand the string (and more generally, quantum gravity) landscape. As a general consistency requirement, anomalies have proven to be very useful in constraining the landscape of consistent theories, especially in dimension $D\geq 7$ and in conjunction with supersymmetry~\cite{Adams:2010zy, Kim:2019vuc, Montero:2020icj, Cvetic:2020kuw, Hamada:2021bbz, Bedroya:2021fbu}. For example, it has been shown that the gauge group rank is restricted to be one of $1$, $9$ or $17$ for $D=9$ and one of $2$, $10$ or $18$ for $D=8$~\cite{Montero:2020icj}, matching exactly those values arising in string theory.
With fewer supercharges the strength of anomaly-cancellation is somewhat reduced. The largest $D$ which allows for only $8$ supercharges is $D=6$, and there one finds infinite families containing models with unbounded rank. There is, therefore, a quite striking difference between the supergravity landscape and the quantum gravity landscape, which is generally expected to be finite (e.g.\ see~\cite{Vafa:2005ui,Acharya:2006zw,Douglas:2010ic,Grimm:2021vpn,Hamada:2021yxy,Grimm:2023lrf}). However, by bringing in additional UV conditions many of these infinite families have been shown to be inconsistent with quantum gravity~\cite{Kim:2019vuc,Tarazi:2021duw}, although recently some infinite families have been constructed in~\cite{Loges:2024vpz} which evade these constraints.

Even putting these infinite families aside, the landscape of 6D, $\mathcal{N}=1$ supergravity models is quite large. There have been several studies which tabulate anomaly-free models in various settings where the non-abelian gauge group, hypermultiplet representations or number of tensor multiplets are restricted~\cite{Avramis:2005hc,Kumar:2009ac,Kumar:2010am,Becker:2023zyb}. Recently in~\cite{Hamada:2023zol}, we showed how the task of identifying anomaly-free models with hypermultiplets charged under at most two gauge factors can be reinterpreted as finding cliques in a certain multigraph. With this perspective we provided a specialized so-called branch-and-prune algorithm to systematically construct anomaly-free models in a $T$-agnostic way.

In order to allow for hypermultiplets charged under any number of gauge factors and extend our previous methods, we are lead quite naturally to consider multi-hypergraphs and simplicial complexes in place of multigraphs and cliques. Although the structures involved are more general, the algorithms previously described carry over with very little change. We then bring these techniques to bear on the special cases of $T=0$ and $T=1$, presenting a complete enumeration of anomaly-free models compatible with all low-energy consistency conditions, the \emph{only} simplifying assumption being that the gauge group contains no $\U(1)$, $\SU(2)$, $\SU(3)$ or $\Sp(2)$ factors. In appendix~\ref{app:graph-construction} we argue that comprehensively incorporating any of the three omitted non-abelian simple groups would require a huge increase in computational power. We emphasize that our analysis encompasses models with \emph{any} number of gauge factors and hypermultiplets in \emph{any} representation of the gauge group. With a complete list of anomaly-free models (modulo only the omitted simple factors) at our disposal, we then turn to understanding which models survive the bounds from several UV and swampland bounds.
Following tradition, we often refer to semi-simple Lie groups when we really mean the corresponding Lie algebra; the global structure of the gauge group could be studied on a case-by-case basis, but this is not our focus.

\smallskip

The remainder of this paper is organized as follows. In section~\ref{sec:consistency-conditions} we briefly review the low-energy consistency conditions. In section~\ref{sec:multihypergraphs-and-simplicial-complexes} we discuss their reinterpretation in terms of a multi-hypergraphs and simplicial complexes, naturally generalizing our previous work~\cite{Hamada:2023zol}. Some special features for $T\leq 1$ are leveraged in section~\ref{sec:classification} to obtain a tailored pruning condition which allows for a comprehensive enumeration of anomaly-free models. Section~\ref{sec:results} is devoted to exploring various gross features and highlighting some notable examples from the ensembles of anomaly-free models which result from applying the discussed methods to $T=0$ and $T=1$. There, we also explore additional bounds which can rule out a significant fraction of models as inconsistent with quantum gravity. Finally, we conclude in section~\ref{sec:discussion}. Some details, including technical information regarding the construction of the multi-hypergraphs and enumeration of anomaly-free models, as well as some tables of models, has been relegated to the appendices.

\section{Consistency conditions}
\label{sec:consistency-conditions}

\begin{table}[t]
    \centering
    \begin{tabular}{>$c<$|*{9}{>$c<$}}
        \toprule
        G         & A_n & B_n  & C_n & D_n  & E_6 & E_7 & E_8 & F_4 & G_2 \\ \midrule
        \lambda_G & 1   & 2    & 1   & 2    & 6   & 12  & 60  & 6   & 2   \\
        h_G^\lor  & n+1 & 2n-1 & n+1 & 2n-2 & 12  & 18  & 30  & 9   & 4   \\
        \bottomrule
    \end{tabular}
    \caption{Normalization constants and dual Coxeter numbers for simple gauge factors.}
    \label{tab:lambda-constants}
\end{table}

In this section we briefly review the low-energy consistency conditions required of 6D supergravities with eight supercharges and non-abelian gauge group. Gauge and gravitational anomalies arising at 1-loop are captured by an $8$-form anomaly polynomial containing contributions from fermions in the single gravity multiplet, $T$ tensor multiplets, $V$ vector multiplets in the adjoint of $G$ and $H$ hypermultiplets in the representations $\H$ of $G$~\cite{Alvarez-Gaume:1983ihn}. Leveraging the Green-Schwarz-Sagnotti mechanism~\cite{Green:1984bx,Green:1984sg,Sagnotti:1992qw}, local anomalies can be cancelled by the tree-level exchange of the (anti-)self-dual 2-forms in the gravity and tensor multiplets. This requires the anomaly polynomial to factorize as\footnote{For uniformity in notation we have introduced $b_0\defeq -a$.}
\begin{equation}
    \hat{I}_8^{\text{1-L}}(\calR,\calF) = \frac{1}{2}\Omega_{\alpha\beta}X_4^\alpha \wedge X_4^\beta \,, \qquad X_4^\alpha = -\frac{1}{2}b_0^\alpha\tr\calR^2 + 2\sum_i\frac{b_i^\alpha}{\lambda_i}\tr\calF_i^2 \,,
\end{equation}
where $i\in\{1,2,\ldots,\kappa\}$ runs over the simple groups in $G=\prod_{i=1}^\kappa G_i$, the anomaly vectors $b_I$ ($I=0,i$) live in $\R^{1,T}$ and the normalization constants $\lambda_i$ are given in table~\ref{tab:lambda-constants}. Matching terms, the irreducible terms impose the following constraints:
\begin{align}
    \tr\calR^4 : \phantom{{\sum n_R^iB_R^i}}{H-V} &= 273-29T \,, \label{eq:gravitational-constraint}\\
    \tr\calF_i^4 : \phantom{{H-V}}{\sum_R n_R^iB_R^i} &= B_\adj^i \,. \label{eq:B-constraint}
\end{align}
It will be useful to introduce both $\H^\ch$ which consists of only the \emph{charged} hypermultiplets and $\Delta\defeq H^\ch-V$, in terms of which equation~\eqref{eq:gravitational-constraint} amounts to $\Delta\leq 273-29T$. The remaining, reducible terms determine all inner products of the $b_I$ amongst themselves,
\begin{eqnalign}
\label{eq:bI.bJ-defn}
    b_0\cdot b_0 &= 8 \,, &
    b_0\cdot b_i &= \frac{1}{6}\Big(\sum_R n_R^iA_R^i - A_\adj^i\Big) \,,\\
    b_i\cdot b_i &= \frac{1}{3}\Big(\sum_R n_R^iC_R^i - C_\adj^i\Big) \,,\qquad &
    b_i\cdot b_j &= \sum_{R,S} n_{(R,S)}^{i,j}A_R^iA_S^j \,, \quad (i\neq j) \,,
\end{eqnalign}
where we have introduced the short-hand $u\cdot v\defeq\Omega_{\alpha\beta}u^\alpha v^\beta$. The positive numbers $n_R^i$ give the number of hypermultiplets transforming in each irreducible representation $R$ of $G_i$, and the corresponding indices $A_R^i$, $B_R^i$ and $C_R^i$ are defined by
\begin{equation}
    \lambda_i\tr_R\calF_i^2 = A_R^i\tr\calF_i^2 \,, \qquad
    \lambda_i^2\tr_R\calF_i^4 = B_R^i\tr\calF_i^4 + C_R^i(\tr\calF_i^2)^2 \,.
\end{equation}
With this normalization we have $A_\adj^i=2h_i^\lor$, where $h_i^\lor$ is the dual Coxeter number of $G_i$, and $A_R^i$, $B_R^i$, $C_R^i$ are nearly always integers: the only exceptions are for $A_1$, $A_2$, $B_3$ and $D_4$ which have $C_R^i\in\frac{1}{2}\Z$. Vectors $b_I\in\R^{1,T}$ with inner products determined by~\eqref{eq:bI.bJ-defn} exist exactly when the matrix of inner products $b_I\cdot b_J$ has at most one positive and at most $T$ negative eigenvalues. When this is the case, the vectors $b_I$ are unique up to $\O(1,T;\R)$ transformations.\footnote{We caution that for $T\geq 9$ the vectors $b_I$ are not always unique up to $\O(1,T;\R)$ transformations when the matrix $b_I\cdot b_J$ is negative semi-definite with rank $<T$.}

A careful analysis for the groups $\SU(2)$, $\SU(3)$ and $G_2$ reveals that these are subject to an additional condition modulo $12$, $6$ and $3$, respectively, which ensures $b_i\cdot b_i\in\Z$~\cite{Bershadsky:1997sb,Lee:2020ewl,Davighi:2020kok}. Given that all local anomalies are cancelled, the lattice $\Lambda=\bigoplus_I b_I\Z$ is integral~\cite{Kumar:2010ru}, i.e.\ all $b_I\cdot b_J$ in equation~\eqref{eq:bI.bJ-defn} are integers. In addition, it was shown in~\cite{Seiberg:2011dr} that the string charge lattice $\Gamma$ must be unimodular. Therefore it is not enough for $\Lambda$ to be an integral lattice: it must embed into a unimodular lattice of signature $(1,T)$ and the eigenvalue bounds on $b_I\cdot b_J$ are no longer sufficient to ensure anomalies may be cancelled when $T\geq 1$. There are two possibilities for $T=1$: either $\Gamma$ is the odd unimodular lattice $\Gamma_{1,1}$ with $\Omega=\diag(1,-1)$ and $b_0=(3,1)$ or $\Gamma$ is the even unimodular lattice $U$ with $\Omega=\begin{psmallmatrix}
    0 & 1\\ 1 & 0
\end{psmallmatrix}$ and $b_0=(2,2)$. See appendix~\ref{app:lattices} for further discussion, including the determination of $b_i\in\Gamma$ from $b_i\cdot b_i$ and $b_0\cdot b_i$ alone and the dismissal of cases where $\Gamma=U$ with $b_0=(1,4)$, which \emph{a priori} is a distinct possibility.

Finally, the moduli space is parametrized by a vector $j$, which we require satisfies both $j\cdot j>0$ and $j\cdot b_i>0$. The former ensures the positivity of the moduli space metric and the latter ensure that all of the gauge kinetic terms, which take the form $-j\cdot b_i\tr\calF_i^2$, have the correct sign. For $T=0$ one can always simply take $j=1$ since $b_i>0$. In fact, we will see in section~\ref{sec:classification} that for $T=1$ the existence of such a vector is also guaranteed: it follows from anomaly cancellation and unimodularity.

\medskip

In summary, the low-energy consistency conditions discussed above are all satisfied provided the constraints of equations~\eqref{eq:gravitational-constraint} and~\eqref{eq:B-constraint} are met and integer vectors $b_i\in\Gamma$ realizing the inner products of equation~\eqref{eq:bI.bJ-defn} can be found, with $\Gamma=\Z$, $b_0=3$ for $T=0$ and either $\Gamma=\Gamma_{1,1}$, $b_0=(3,1)$ or $\Gamma=U$, $b_0=(2,2)$ for $T=1$. We will refer to models which meet all of the consistency conditions except perhaps for~\eqref{eq:gravitational-constraint} as \emph{admissible}, and we refer to admissible models which additionally satisfy~\eqref{eq:gravitational-constraint} as \emph{anomaly-free}. That is, admissible models comprise the (supersymmetric) field theory landscape and anomaly-free models comprise the supergravity landscape. The strategy we adopt in the following sections is to enumerate all anomaly-free models and then impose additional quantum gravitational conditions to see how the supergravity landscape compares to the quantum gravity landscape of \emph{consistent} models.

\section{Multi-hypergraphs and simplicial complexes}
\label{sec:multihypergraphs-and-simplicial-complexes}

In~\cite{Hamada:2023zol} it was explained how considering anomaly-free models with hypermultiplets charged under at most two gauge factors naturally leads to a certain multi-graph structure. Vertices of the multi-graph encoded choices of simple gauge factor and corresponding hypermultiplets satisfying the constraint of equation~\eqref{eq:B-constraint}. Meanwhile edges described bi-charged hypermultiplets which when formed from the two incident vertices yield an admissible model with exactly two simple gauge factors. A standard ``branch-and-prune'' algorithm for generating cliques was then adapted to systematically construct all anomaly-free models.
In this section we generalize these ideas to account for hypermultiplets charged under any number of gauge factors. Following our noses, we are lead naturally to consider multi-hypergraphs\footnote{The ``hyper'' in multi-hypergraph refers to their having \emph{hyper}edges which join two or more vertices and is completely unrelated to the ``hyper'' in hypermultiplet. This coincidence of nomenclature should not cause any confusion in what follows.} and simplicial complexes in place of multi-graphs and cliques. These will have some additional structure appropriate for encoding the physics of anomaly-cancellation. At this point we keep the discussion general; these methods can be used for any simple groups and any $T$, although in later sections we will specialize to $T\leq 1$.

\smallskip

The multi-hypergraph $\G$ consists of a set of labelled vertices $\V$ and hyperedges $\E$. Each vertex $\v\in\V$ corresponds to a choice of simple group and charged hypermultiplets which both satisfy the constraint of equation~\eqref{eq:B-constraint} and admit at least one choice for lattice $\Gamma$ (and corresponding $b_0$) and integer vector $b\in\Gamma$ realizing the required norm $b\cdot b$ and inner-product $b_0\cdot b$. That is, to each vertex is associated a simple gauge group, $G(\v)$, charged hypermultiplets
\begin{equation}
    \H^\ch(\v) \defeq \bigoplus_R n_R(\v)\times R \,, \qquad n_R(\v)>0 \,,\; \dim R>1 \,,
\end{equation}
and set of possible anomaly vectors and corresponding lattice, $\{(b^{(a)},\Gamma^{(a)})\}_a$, realizing the inner products implied by $\H^\ch(\v)$ through equation~\eqref{eq:bI.bJ-defn}. Quantities such as $H^\ch(\v)$ and $\Delta(\v)$ are defined in the obvious way. A complete list of vertices for each simple group may be generated up to some maximum value of $\Delta(\v)$ using the methods described in appendix~C.1 of~\cite{Hamada:2023zol}.

Each $(r\geq2)$-edge $\e^{(r)}\in\E$ is incident to $r$ vertices,
\begin{equation}
    \iota(\e^{(r)}) = \big(\v_1(\e^{(r)}), \ldots, \v_r(\e^{(r)})\big) \in \underbrace{\V\times\cdots\times\V}_r \,.
\end{equation}
By far the most common are $2$-edges, which we will refer to simply as \emph{edges}. These encode bi-charged hypermultiplets
\begin{equation}
    \H^\ch(\e^{(2)}) = \bigoplus_{R,S}n_{R,S}(\e^{(2)})(R,S) \,, \qquad n_{R,S}(\e^{(2)})\geq 0 \,,\;\; \dim R>1 \,,\;\; \dim S>1
\end{equation}
for which the inner-product $b_i\cdot b_j$ determined via~\eqref{eq:bI.bJ-defn} is achieved by some choice of lattice and vectors $b_i^{(a)}$ for the two vertices of $\iota(\e^{(2)})$. $R$ and $S$ run over representations of $G(\v_1(\e^{(2)}))$ and $G(\v_2(\e^{(2)}))$, respectively. We require that
\begin{equation}
    n_R\big(\v_1(\e^{(2)})\big) \geq \sum_S n_{R,S}(\e^{(2)})\dim S \,,
\end{equation}
and similarly for $\v_2(\e^{(2)})$ so that merging the hypermultiplets of $\v_1(\e^{(2)})$ and $\v_2(\e^{(2)})$ to form the bi-charged hypermultiplets required by $\H^\ch(\e^{(2)})$ does not leave behind a negative multiplicity of hypermultiplets charged under only a single gauge factor. Edges for which $\H^\ch(\e^{(2)})=0$ imply that $b_i\cdot b_j=0$ and we refer to as \emph{trivial}.

Higher hyperedges with $r>2$ are introduced similarly, with
\begin{equation}
    \H^\ch(\e^{(r)}) = \bigoplus_{R_1,\ldots,R_r}n_{R_1,\ldots,R_r}(\e^{(r)})(R_1,\ldots,R_r) \,, \qquad n_{R_1,\ldots,R_r}(e^{(r)})\geq 0 \,,\;\; \dim R_j > 1 \,,
\end{equation}
but must be non-trivial, i.e.\ at least one $n_{R_1,\ldots,R_r}(\e^{(r)})$ must be non-zero. Again we require that merging hypermultiplets to form $r$-charged hypermultiplets does not leave behind a negative multiplicity anywhere. That is, for each $j\in\{1,2,\ldots,r\}$ there must exist an $(r-1)$-edge $\e_j^{(r-1)}$ satisfying
\begin{eqnalign}
\label{eq:graph-hierarchical}
    \iota(\e_j^{(r-1)}) &= \big( \v_1(\e^{(r)}),\ldots,\widehat{\v_j(\e^{(r)})},\ldots,\v_r(\e^{(r)}) \big) \,,\\
    n_{R_1,\ldots,\widehat{R_j},\ldots,R_r}(\e_j^{(r-1)}) &\geq \sum_{R_j}n_{R_1,\ldots,R_r}(\e^{(r)})\dim R_j \,,
\end{eqnalign}
where as usual hats indicate omission.

\begin{figure}[t]
    \centering
    \newcommand{\s}{1.9}
    \begin{tikzpicture}
        \node[scale=0.7] (A3) at (-\s,0) {};
        \node[scale=0.7] (G2) at (\s,0) {};
        \node[scale=0.7] (C3) at (0,1.732*\s) {};

        \node[below=3] at (A3) {$\v_1$};
        \node at ($(C3.center)+(-0.4,-0.15)$) {$\v_2$};
        \node[below left] at (G2) {$\v_3$};

        \draw[very thick] (C3) to (A3) to (G2) to (C3);

        \fill[white] ($0.5*(C3.center)+0.5*(A3.center)-0.23*(0.5,0.866)+0.1*(0.866,-0.5)$) -- ($0.5*(C3.center)+0.5*(A3.center)-0.23*(0.5,0.866)-0.1*(0.866,-0.5)$) -- ($0.5*(C3.center)+0.5*(A3.center)+0.27*(0.5,0.866)-0.1*(0.866,-0.5)$) -- ($0.5*(C3.center)+0.5*(A3.center)+0.27*(0.5,0.866)+0.1*(0.866,-0.5)$);
        \node at ($0.5*(C3.center)+0.5*(A3.center)$) {\scriptsize$\e_1^{(2)}$};
        
        \fill[white] ($0.5*(A3.center)+0.5*(G2.center)+(-0.25,-0.1)$) rectangle ($0.5*(A3.center)+0.5*(G2.center)+(0.23,0.1)$);
        \node at ($0.5*(A3.center)+0.5*(G2.center)$) {\scriptsize$\e_2^{(2)}$};
        
        \fill[white] ($0.5*(C3.center)+0.5*(G2.center)-0.2*(0.5,-0.866)+0.1*(0.866,0.5)$) -- ($0.5*(C3.center)+0.5*(G2.center)-0.2*(0.5,-0.866)-0.1*(0.866,0.5)$) -- ($0.5*(C3.center)+0.5*(G2.center)+0.18*(0.5,-0.866)-0.1*(0.866,0.5)$) -- ($0.5*(C3.center)+0.5*(G2.center)+0.18*(0.5,-0.866)+0.1*(0.866,0.5)$);
        \node at ($0.5*(C3.center)+0.5*(G2.center)$) {\scriptsize$\e_3^{(2)}$};

        \draw[very thick] (C3) to[out=-20,in=80] (G2);
        \node (e42) at ($0.5*(C3.center)+0.5*(G2.center)+0.75*(0.866,0.5)$) {};
        \fill[white] ($(e42.center)+0.22*(-0.5,0.866)+0.2*(0.866,0.5)$) -- ($(e42.center)+0.22*(-0.5,0.866)-0.2*(0.866,0.5)$) -- ($(e42.center)-0.17*(-0.5,0.866)-0.2*(0.866,0.5)$) -- ($(e42.center)-0.17*(-0.5,0.866)+0.2*(0.866,0.5)$);
        \node at (e42) {\scriptsize$\e_4^{(2)}$};

        \draw[very thick,rotate=-30] ($(G2.center)+(0.4,0)$) ellipse (0.4 and 0.3);
        \node at ($(G2.center)+0.8*(0.866,-0.5)+(0.25,-0.08)$) {\scriptsize$\e_6^{(2)}$};

        \draw[very thick] ($(C3.center)+(0,0.7)$) ellipse (0.56 and 0.7);
        \node at ($(C3.center)+(0.9,0.7)$) {\scriptsize$\e_5^{(2)}$};
        
        \fill[draw,thick,myTeal,fill opacity=0.4] ($(C3.center)+0.25*(-0.5,0.866)$) to[out=125,in=100,looseness=10] ($(C3.center)+(-0.005,0.27)$) to ($(C3.center)+(0.005,0.27)$) to[out=80,in=55,looseness=10] ($(C3.center)+0.25*(0.5,0.866)$) arc (-71:251:0.4 and 0.5) -- cycle;
        \node at ($(C3.center)+(0,0.83)$) {\scriptsize$\e_2^{(3)}$};

        \fill[draw,thick,myTeal,fill opacity=0.4] ($(C3.center) + (0,-0.25)$) to[out=-100,in=40] ($(A3.center) + 0.25*(0.866,0.5)$) to[out=20,in=160] ($(G2.center) + 0.25*(-0.866,0.5)$) to[out=140,in=-80] cycle;
        \node at (0,0.577*\s) {\scriptsize$\e_1^{(3)}$};

        \node[draw,thick,circle,fill=myYellow,scale=0.7] (C3) at (-\s,0) {};
        \node[draw,thick,circle,fill=myYellow,scale=0.7] (G2) at (\s,0) {};
        \node[draw,thick,circle,fill=myYellow,scale=0.7] (A3) at (0,1.732*\s) {};

        \node[left] at (11.25,6.5) {
        \begin{tabular}{*{8}{>$c<$}}
            \toprule
            \v_i & G(\v_i) & \H^\ch(\v_i) & \Delta(\v_i) & b_i\cdot b_i & b_0\cdot b_i & b_i\in\Gamma_{1,1} & b_i\in U \\ \midrule
            \v_1 & \SU(4) & 56\times\rep{4}\oplus\rep{15}\oplus\rep[\prime]{20} & 244 & 18 & 12 & \text{--} & (3,3) \\
            \v_2 & \Sp(3) & 28\times\rep{6}\oplus4\times\rep[\prime]{14}\oplus\rep{21} & 224 & 8 & 8 & (3,1) & (2,2) \\
            \v_3 & G_2 & 24\times\rep{7}\oplus\rep{14} & 168 & 8 & 8 & (3,1) & (2,2) \\
            \bottomrule
        \end{tabular}
        };

        \node[left] at (11.25,2) {
        \begin{tabular}{*{4}{>$c<$}}
            \toprule
            \e_i^{(r)} & \iota(\e_i^{(r)}) & \H^\ch(\e_i^{(r)}) \\ \midrule
            \e_1^{(2)} & (\v_1,\v_2) & 7(\rep{4},\rep{6})\oplus(\rep{4},\rep[\prime]{14}) \\
            \e_2^{(2)} & (\v_1,\v_3) & 6(\rep{4},\rep{7}) \\
            \e_3^{(2)} & (\v_2,\v_3) & 4(\rep{6},\rep{7}) \\
            \e_4^{(2)} & (\v_2,\v_3) & \tfrac{3}{2}(\rep{6},\rep{7})\oplus\tfrac{1}{2}(\rep[\prime]{14},\rep{7}) \\
            \e_5^{(2)} & (\v_2,\v_2) & \scalebox{0.7}{$3(\rep{6},\rep{6})\oplus\tfrac{1}{2}(\rep{6},\rep[\prime]{14})\oplus\tfrac{1}{2}(\rep[\prime]{14},\rep{6})$} \\
            \e_6^{(2)} & (\v_3,\v_3) & 2(\rep{7},\rep{7}) \\
            \e_1^{(3)} & (\v_1,\v_2,\v_3) & (\rep{4},\rep{6},\rep{7}) \\
            \e_2^{(3)} & (\v_2,\v_2,\v_2) & \tfrac{1}{2}(\rep{6},\rep{6},\rep{6}) \\
            \bottomrule
        \end{tabular}
        };
    \end{tikzpicture}
    \caption{A small multi-hypergraph within $\G$ induced by the three vertices $\v_1$, $\v_2$ and $\v_3$. Each $2$-edge gives, via equation~\eqref{eq:bI.bJ-defn}, a value for $b_i\cdot b_j$ which is compatible with the possible vectors associated to the two vertices it joins (e.g.\ the edge between $\v_1$ and $\v_2$ gives $b_i\cdot b_j=12$ and both edges between $\v_2$ and $\v_3$ give $b_i\cdot b_j=8$).}
    \label{fig:graph-example}
\end{figure}

\smallskip

The hierarchical structure of hyperedges which is required to avoid negative multiplicities is reminiscent of a simplicial complex, and indeed we can identify each admissible model with an \emph{embedding} of a simplicial complex into $\G$. A simplicial complex $\K$ consists of a set of simplices (i.e.\ points, lines, triangles, tetrahedra, etc.) with the restriction that for each $\sigma\in\K$ the faces of $\sigma$ are also in $\K$, and any non-trivial intersection of two simplices is a face of each. For our purposes we will impose an additional property, namely that for a simplicial complex $\K$ with $\kappa$ points there are exactly $\binom{\kappa}{2}$ lines, one connected each pair of points.
An embedding of a simplicial complex $\K$ into $\G$ maps the points (i.e.\ $0$-simplices) of $\K$ to vertices in $\G$ and $(r-1)$-simplices of $\K$ to $r$-edges of $\G$ for $r\geq 2$. This must be structure-preserving, namely must maintain face/incidence relations. In particular, for each $(r\geq2)$-simplex its image together with each of the images of its $(r-1)$-simplex faces must satisfy equation~\eqref{eq:graph-hierarchical}.

\medskip

Let us close this section with a small example for $T=1$. Figure~\ref{fig:graph-example} shows a multi-hypergraph induced by three carefully chosen vertices of $\G$. Consider the following three models with three simple factors, which we can view as ways to label a simplicial complex as just described:\footnote{$\SU(4)$ has several irreducible representations of dimension $20$, which we will denote by $\rep{20}=\ydiagram{2,1}$, $\rep[\prime]{20}=\ydiagram{2,2}$ and $\rep[\prime\prime]{20}=\ydiagram{3}$.}
\begin{center}
    \begin{tikzpicture}
        \begin{scope}[yshift=0]
            \node[draw,thick,circle,fill=myYellow,scale=0.7] (G2a) at (-1,0) {};
            \node[draw,thick,circle,fill=myYellow,scale=0.7] (G2b) at (1,0) {};
            \node[draw,thick,circle,fill=myYellow,scale=0.7] (C3) at (0,1.733) {};
            \draw[very thick] (G2a) to node[below] {\scriptsize$\e_6^{(2)}$} (G2b) to node[right] {\scriptsize$\e_4^{(2)}$} (C3) to node[left] {\scriptsize$\e_4^{(2)}$} (G2a);
            \node[above] at (C3) {$\v_2$};
            \node[left] at (G2a) {$\v_3$};
            \node[right] at (G2b) {$\v_3$};

            \node[right] at (2,0.87) {$
            \begin{aligned}
                G &= \Sp(3)\times G_2\times G_2\\
                \H^\ch &= 7(\rep{6},\triv,\triv)\oplus (-3)(\rep[\prime]{14},\triv,\triv)\oplus (\rep{21},\triv,\triv)\oplus (-6)(\triv,\rep{7},\triv)\\
                &\quad \oplus (\triv,\rep{14},\triv) \oplus (-6)(\triv,\triv,\rep{7})\oplus (\triv,\triv,\rep{14})\oplus \tfrac{3}{2}(\rep{6},\rep{7},\triv)\\
                &\quad \oplus \tfrac{1}{2}(\rep[\prime]{14},\rep{7},\triv) \oplus \tfrac{3}{2}(\rep{6},\triv,\rep{7})\oplus \tfrac{1}{2}(\rep[\prime]{14},\triv,\rep{7})\oplus 2(\triv,\rep{7},\rep{7})
            \end{aligned}$};
        \end{scope}
        
        \begin{scope}[yshift=-3cm]
            \node[draw,thick,circle,fill=myYellow,scale=0.7] (C3) at (-1,0) {};
            \node[draw,thick,circle,fill=myYellow,scale=0.7] (G2) at (1,0) {};
            \node[draw,thick,circle,fill=myYellow,scale=0.7] (A3) at (0,1.733) {};
            \draw[very thick] (C3) to node[below] {\scriptsize$\e_3^{(2)}$} (G2) to node[right] {\scriptsize$\e_2^{(2)}$} (A3) to node[left] {\scriptsize$\e_1^{(2)}$} (C3);
            \node[above] at (A3) {$\v_1$};
            \node[left] at (C3) {$\v_2$};
            \node[right] at (G2) {$\v_3$};

            \node[right] at (2,0.87) {$
            \begin{aligned}
                G &= \SU(4)\times\Sp(3)\times G_2\\
                \H^\ch &= (-42)(\rep{4},\triv,\triv)\oplus(\rep{15},\triv,\triv)\oplus(\rep[\prime]{20},\triv,\triv)\oplus (-28)(\triv,\rep{6},\triv)\\
                &\quad \oplus(\triv,\rep{21},\triv)\oplus(-24)(\triv,\triv,\rep{7})\oplus(\triv,\triv,\rep{14})\\
                &\quad \oplus7(\rep{4},\rep{6},\triv)\oplus(\rep{4},\rep[\prime]{14},\triv)\oplus6(\rep{4},\triv,\rep{7})\oplus4(\triv,\rep{6},\rep{7})
            \end{aligned}$};
        \end{scope}
        
        \begin{scope}[yshift=-6cm]
            \node[draw,thick,circle,fill=myYellow,scale=0.7] (C3) at (-1,0) {};
            \node[draw,thick,circle,fill=myYellow,scale=0.7] (G2) at (1,0) {};
            \node[draw,thick,circle,fill=myYellow,scale=0.7] (A3) at (0,1.733) {};
            \draw[very thick,fill=myTeal,fill opacity=0.4] ($(C3.center)$) to node[below, opacity=1] {\scriptsize$\e_3^{(2)}$} ($(G2.center)$) to node[right, opacity=1] {\scriptsize$\e_2^{(2)}$} ($(A3.center)$) to node[left, opacity=1] {\scriptsize$\e_1^{(2)}$} ($(C3.center)$);
            \node[draw,thick,circle,fill=myYellow,scale=0.7] (C3) at (-1,0) {};
            \node[draw,thick,circle,fill=myYellow,scale=0.7] (G2) at (1,0) {};
            \node[draw,thick,circle,fill=myYellow,scale=0.7] (A3) at (0,1.733) {};
            \node[above] at (A3) {$\v_1$};
            \node[left] at (C3) {$\v_2$};
            \node[right] at (G2) {$\v_3$};
            \node at (0,0.577) {\scriptsize$\e_1^{(3)}$};
            
            \node[right] at (2,0.87) {$
            \begin{aligned}
                G &= \SU(4)\times\Sp(3)\times G_2\\
                \H^\ch &= (\rep{15},\triv,\triv)\oplus(\rep[\prime]{20},\triv,\triv)\oplus(\triv,\rep{21},\triv)\oplus(\triv,\triv,\rep{14})\\
                &\quad \oplus(\rep{4},\rep[\prime]{14},\triv)\oplus(\rep{4},\rep{6},\rep{7})
            \end{aligned}$};
        \end{scope}
    \end{tikzpicture}
\end{center}
Notice that the first example includes repeated vertices and edges, meaning that its embedding in $\G$ is degenerate. We encourage the enthusiastic reader to work through the calculation of $\H^\ch$ for the last of these examples, going vertex-by-vertex and edge-by-edge: the process is reminiscent of the inclusion-exclusion principle. The first two would be admissible if not for their negative hypermultiplet multiplicities, but as it is they should be discarded as unphysical. The third, however, is admissible; in fact, it has $\Delta =244$ exactly and so is anomaly-free without having to add neutral hypermultiplets. This is one of the rare examples of an anomaly-free model with hypermultiplets charged under three or more gauge factors, in this case with a single tri-fundamental of $\SU(4)\times\Sp(3)\times G_2$.

\section{Classification}
\label{sec:classification}

In this section we describe the natural generalization of the ideas presented in~\cite{Hamada:2023zol} to systematically enumerate anomaly-free models. We will not dwell on those details which carry over essentially unchanged, instead focusing on the special properties that are present and additional improvements that can be made for $T\leq 1$.

\begin{table}[p]
    \centering
    \footnotesize
    \renewcommand{\arraystretch}{0.95}
    \begin{tabular}{ccclcc}
        \toprule
        $b_i\cdot b_i$ & $b_0\cdot b_i$ & $G(\v)$ & $\H^\ch(\v)$ & $\Delta(\v)$ & Notes\\
        \midrule
        $-12$ & $-10$ & $E_8$ & $\emptyset$ & $-248$ & $\mathbb{F}_{12}$ NHC \\\cmidrule{3-6}
        $-8$ & $-6$ & $E_7$ & $\emptyset$ & $-133$ & $\mathbb{F}_8$ NHC\\ \cmidrule{3-6}
        $-7$ & $-5$ & $E_7$ & $\tfrac{1}{2}\rep{56}$ & $-105$ & $\mathbb{F}_7$ NHC \\ \cmidrule{3-6}
        \multirow{2}{*}{$-6$} & \multirow{2}{*}{$-4$} & $E_6$ & $\emptyset$ & $-78$ & $\mathbb{F}_6$ NHC \\
        & & $E_7$ & $\rep{56}$ & $-77$ \\ \cmidrule{3-6}
        \multirow{3}{*}{$-5$} & \multirow{3}{*}{$-3$} & $E_6$ & $\rep{27}$ & $-51$ \\
        & & $E_7$ & $\tfrac{3}{2}\rep{56}$ & $-49$ \\
        & & $F_4$ & $\emptyset$ & $-52$ & $\mathbb{F}_5$ NHC \\ \cmidrule{3-6}
        \multirow{4}{*}{$-4$} & \multirow{4}{*}{$-2$} & $\SO(N)$ & $(N-8)\times\rep{N}$ & $\frac{(N-7)(N-8)}{2}-28$ & \scalebox{0.8}{$N\geq 8$, $\mathbb{F}_4$ NHC for $N=8$}\\
        & & $E_6$ & $2\times\rep{27}$ & $-24$ \\
        & & $E_7$ & $2\times\rep{56}$ & $-21$ \\
        & & $F_4$ & $\rep{26}$ & $-26$ \\ \cmidrule{3-6}
        \multirow{5}{*}{$-3$} & \multirow{5}{*}{$-1$} & $\SU(3)$ & $\emptyset$ & $-8$ & $\mathbb{F}_3$ NHC \\
        & & $\SO(N)$ & \tiny$(N-7)\times\rep{N}\oplus\big(2^{\lfloor\frac{10-N}{2}\rfloor}\big)\times\rep{2^{\lfloor\frac{N-1}{2}\rfloor}}$ & $\frac{(N-6)(N-7)}{2}-5$ & $7\leq N\leq 12$ \\
        & & $E_6$ & $3\times\rep{27}$ & $3$ \\
        & & $E_7$ & $\tfrac{5}{2}\rep{56}$ & $7$ \\
        & & $F_4$ & $2\times \rep{26}$ & $0$ \\
        & & $G_2$ & $\rep{7}$ & $-7$ \\ \cmidrule{3-6}
        \multirow{7}{*}{$-2$} & \multirow{7}{*}{$0$} & $\SU(N)$ & $(2N)\times\rep{N}$ & $N^2+1$ & $N\geq 2$\\
        & & $\SO(N)$ & \tiny$(N-6)\times\rep{N}\oplus\big(2\cdot 2^{\lfloor\frac{10-N}{2}\rfloor}\big)\times\rep{2^{\lfloor\frac{N-1}{2}\rfloor}}$ & $\frac{(N-5)(N-6)}{2}+17$ & $7\leq N\leq 13$\\
        & & $E_6$ & $4\times\rep{27}$ & $30$ \\
        & & $E_7$ & $3\times\rep{56}$ & $35$ \\
        & & $F_4$ & $3\times \rep{26}$ & $26$ \\
        & & $G_2$ & $4\times\rep{7}$ & $14$ \\ \cmidrule{3-6}
        \multirow{8}{*}{$-1$} & \multirow{8}{*}{$1$} & $\SU(N)$ & $(N+8)\times\rep{N}\oplus\rep{N(N-1)/2}$ & $\frac{(N+7)(N+8)}{2}-27$ & $N\geq 3$ \\
        & & $\SU(2)$ & $10\times\rep{2}$ & $17$ \\
        & & $\SU(6)$ & $15\times\rep{6}\oplus\tfrac{1}{2}\rep{20}$ & $65$ \\
        & & $\SO(N)$ & \tiny$(N-5)\times\rep{N}\oplus\big(3\cdot 2^{\lfloor\frac{10-N}{2}\rfloor}\big)\times\rep{2^{\lfloor\frac{N-1}{2}\rfloor}}$ & $\frac{(N-4)(N-5)}{2}+38$ & $7\leq N\leq 12$ \\
        & & $\Sp(N)$ & $(2N+8)\times\rep{2N}$ & $N(2N+15)$ & $N\geq 2$ \\
        & & $E_6$ & $5\times\rep{27}$ & $57$ \\
        & & $E_7$ & $\tfrac{7}{2}\rep{56}$ & $63$ \\
        & & $F_4$ & $4\times \rep{26}$ & $52$ \\
        & & $G_2$ & $7\times\rep{7}$ & $35$ \\ \cmidrule{3-6}
        $-1$ & $-1$ & $\SU(N)$ & $(N-8)\times\rep{N} \oplus \rep{N(N+1)/2}$ & $\frac{(N-7)(N-8)}{2}-27$ & $N\geq 8$ \\ \midrule
        \multirow{12}{*}{$0$} & \multirow{12}{*}{$2$} & $\SU(N)$ & $16\times\rep{N}\oplus 2\times\rep{N(N-1)/2}$ & $15N+1$ & $N\geq 4$\\
        & & $\SU(2)$ & $16\times\rep{2}$ & $29$ \\
        & & $\SU(3)$ & $18\times\rep{3}$ & $46$ \\
        & & $\SU(6)$ & $17\times\rep{6}\oplus\rep{15}\oplus\tfrac{1}{2}\rep{20}$ & $92$\\
        & & $\SU(6)$ & $18\times\rep{6}\oplus \rep{20}$ & $93$\\
        & & $\SO(N)$ & \tiny$(N-4)\times\rep{N}\oplus\big(4\cdot 2^{\lfloor\frac{10-N}{2}\rfloor}\big)\times\rep{2^{\lfloor\frac{N-1}{2}\rfloor}}$ & $\frac{(N-3)(N-4)}{2}+58$ & $7\leq N\leq 14$ \\
        & & $\Sp(N)$ & $16\times\rep{2N}\oplus\rep{(N-1)(2N+1)}$ & $30N-1$ & $N\geq 2$ \\
        & & $\Sp(3)$ & $\tfrac{35}{2}\rep{6}\oplus\tfrac{1}{2}\rep[\prime]{14}$ & $91$ \\
        & & $E_6$ & $6\times\rep{27}$ & $84$ \\
        & & $E_7$ & $4\times\rep{56}$ & $91$ \\
        & & $F_4$ & $5\times \rep{26}$ & $78$ \\
        & & $G_2$ & $10\times\rep{7}$ & $56$ \\
        \bottomrule
    \end{tabular}
    \caption{All type-\tS{S} ($b_i\cdot b_i<0$) and type-\tN{N} ($b_i\cdot b_i=0$) vertices for $T=1$.}
    \label{tab:type-S/N-vertices}
\end{table}

\begin{table}[t]
    \centering
    \begin{tabular}{c*{2}{>{$\;}r<{\;$}}*{2}{>{$\;}c<{\;$}}}
        \toprule
        Type & b_i\cdot b_i\!\! & b_0\cdot b_i\!\! & b_i\in\Gamma_{1,1} & b_i\in U \\
        \midrule
        \multirow{10}{*}{\tS{S}}
        & -12 & -10 & (-2,\phantom{+}4) & (1,-6)\;\;\text{or}\;\;(-6,1) \\
        &  -8 &  -6 & (-1,\phantom{+}3) & (1,-4)\;\;\text{or}\;\;(-4,1) \\
        &  -7 &  -5 & (-3,-4) & \text{--} \\
        &  -6 &  -4 & \text{--} & (1,-3)\;\;\text{or}\;\;(-3,1) \\
        &  -5 &  -3 & (-2,-3)  & \text{--} \\
        &  -4 &  -2 & (\phantom{+}0,\phantom{+}2) & (1,-2)\;\;\text{or}\;\;(-2,1) \\
        &  -3 &  -1 & (-1,-2) & \text{--} \\
        &  -2 &   0 & \text{--} & (1,-1)\;\;\text{or}\;\;(-1,1) \\
        &  -1 &   1 & (\phantom{+}0,-1) & \text{--} \\
        &  -1 &  -1 & (\phantom{+}0,\phantom{+}1) & \text{--} \\ \cmidrule{2-5}
        \tN{N}
        &   0 &   2 & (\phantom{+}1,\phantom{+}1) & (1,\phantom{+}0)\;\;\text{or}\;\;(\phantom{+}0,1) \\
        \bottomrule
    \end{tabular}
    \caption{All possible Gram matrix elements for type-\tS{S} and type-\tN{N} vertices when $T=1$ and the possible choices for the vector $b_i$.}
    \label{tab:type-S/N-bi-vectors}
\end{table}

It is helpful to categorize vertices based on $b_i$ being (\tS{S})pace-like, (\tN{N})ull or (\tT{T})ime-like:
\begin{equation}
\label{eq:SNT-definition}
    \text{Type-\tS{S}}: \quad b_i\cdot b_i < 0 \,, \qquad \text{Type-\tN{N}}: \quad b_i\cdot b_i = 0 \,, \qquad \text{Type-\tT{T}}: \quad b_i\cdot b_i > 0 \,.
\end{equation}
Of course for $T=0$ there are only type-\tT{T} vertices. For $T=1$, vertices of types \tS{S} and \tN{N} are quite limited: see table~\ref{tab:type-S/N-vertices} for a complete list. There are only ten possible $(b_i\cdot b_i,b_0\cdot b_i)$ pairs for type-\tS{S} vertices and all type-\tN{N} vertices have $(b_i\cdot b_i,b_0\cdot b_i)=(0,2)$. In table~\ref{tab:type-S/N-bi-vectors} we record all vectors which realize these norms and inner products for the two lattices.

In specializing to $T\leq 1$ the consistency conditions on the lattice $\Lambda=\bigoplus_Ib_I\Z$ become much more stringent and the multi-hypergraph $\G$ gains some additional structure which we can leverage to great effect. In particular, we show in section~\ref{sec:edges-and-structure} that there are no edges whatsoever between type-\tS{S} vertices and that trivial edges are especially rare and constraining. With this knowledge, we describe the general structure of anomaly-free models for $T\leq 1$ which then motivates the ``branch-and-prune'' algorithm outlined in section~\ref{sec:branch-and-prune} to systematically generate all anomaly-free models.

\subsection{Edge restrictions and the structure of anomaly-free models}
\label{sec:edges-and-structure}

From the structure of $\R^{1,1}$ it is clear that any edge between type-\tS{S} vertices must be non-trivial. In fact, using $b_i^{\tS{S}}\cdot b_j^{\tS{S}}>0$ ($i\neq j$) and $j\cdot b_i^{\tS{S}}>0$, we claim that there are no edges between type-\tS{S} vertices in $\G$ whatsoever. This can be established in several steps.
\begin{enumerate}
    \item Due to their small hypermultiplet multiplicities, no vertices with $b_i\cdot b_i<-4$ can form a non-trivial edge to any other type-\tS{S} vertices. The only candidates with $b_i\cdot b_i<-4$ for merging are $E_7$ with $\rep{56}$ and $\frac{3}{2}\rep{56}$ which could potentially merge to give $\frac{1}{2}(\rep{2},\rep{56})$ or $\frac{1}{2}(\rep{3},\rep{56})$ of $\SU(2)\times E_7$ or $\frac{1}{2}(\rep{3},\rep{56})$ of $\SU(3)\times E_7$, but no type-\tS{S} vertex for $\SU(2)$ or $\SU(3)$ has the necessary $n_{\rep{2}}\geq 28$ or $n_{\rep{3}}\geq 28$. Therefore we conclude that any potential non-trivial edge must be between type-\tS{S} vertices with $b_i\cdot b_i\geq -4$.
    
    \item Merging $n_R\times R$ and $n_S\times S$ into a bi-charged hyper $(R,S)$ for a non-trivial edge requires at least one of $n_R\geq H_R$ or $n_S\geq H_S$. From table~\ref{tab:type-S/N-vertices} we can read off that the only ``hypermultiplet-rich'' type-\tS{S} vertices are those with $b_i\cdot b_i=-2$ or $b_i\cdot b_i=-b_0\cdot b_i=-1$, so any non-trivial edge between type-\tS{S} vertices must involve at least one of these.
    
    \item From table~\ref{tab:type-S/N-bi-vectors}, we see that if $b_i\cdot b_i=-b_0\cdot b_i=-1$ then $b_i$ is anti-parallel to or has negative inner-product with all $b_j$ which have $-4\leq b_j\cdot b_j\leq -1$, thereby either violating $j\cdot b_i>0$ or $b_i\cdot b_j\geq 0$.
    
    \item From table~\ref{tab:type-S/N-bi-vectors} we see that if $b_i\cdot b_i=-2$ the only potential edges are to other vertices with $b_j\cdot b_j=-2$ or to vertices with $b_j\cdot b_j=-4$. The first case is quickly ruled out since either $b_i=b_j$, violating $b_i\cdot b_j\geq 0$, or $b_i=-b_j$, violating $j\cdot b_i>0$. The second case can also be ruled out since $b_i\cdot b_j\geq 0$ only allows for $b_i\cdot b_j=\Omega\big((1,-1),(-2,1)\big)=+3$ odd, but this is impossible to achieve since the $b_j\cdot b_j=-4$ vertices either have $\rep{N}$ of $\SO(N)$ or $\frac{1}{2}\rep{56}$ of $E_7$ available to merge and both $A_{\rep{N}}=2$ and $A_{\frac{1}{2}\rep{56}}=6$ are even.
\end{enumerate}
We conclude that there are no \tS{S}--\tS{S} edges in $\G$ and that admissible models can contain at most one type-\tS{S} vertex.\footnote{This is consistent with what is known in F-theory: for $T=1$ two non-Higgsable clusters cannot coexist~\cite{Morrison:2012np}.} It is also easy to see that trivial edges are only possible between type-\tT{T} and type-\tS{S} vertices or between two type-\tN{N} vertices. When two type-\tN{N} vertices are joined by a trivial edge in an admissible model, this forces $b_i=b_j$ and \emph{all} of their inner products with all of the $b_I$ must be identical. Similarly, when two type-\tT{T} vertices both have trivial edge to the type-\tS{S} vertex in an admissible model then they must have $b_i,b_j$ parallel and all of their inner products to all $b_I$ must be proportional. That is, a general admissible model therefore has a structure for $b_I\cdot b_J$ like the following example:
\begin{equation}
    b_I\cdot b_J = \scalebox{0.8}{$\left(\begin{array}{c|c|ccc|ccccc}
        8 & \# & + & + & + & 2 & 2 & 2 & 2 & 2 \\ \hline
        \# & - & 0 & 0 & + & + & + & + & + & + \\ \hline
        + & 0 & + & + & + & + & + & + & + & + \\
        + & 0 & + & + & + & + & + & + & + & + \\
        + & + & + & + & + & + & + & + & + & + \\ \hline
        2 & + & + & + & + & 0 & 0 & 0 & 1 & 1 \\
        2 & + & + & + & + & 0 & 0 & 0 & 1 & 1 \\
        2 & + & + & + & + & 0 & 0 & 0 & 1 & 1 \\
        2 & + & + & + & + & 1 & 1 & 1 & 0 & 0 \\
        2 & + & + & + & + & 1 & 1 & 1 & 0 & 0
    \end{array}\right)$} \,.
\end{equation}
From the off-diagonal zero entries alone we would know that $b_2\|b_3$, $b_5=b_6=b_7$ and $b_8=b_9$, so that the corresponding rows and columns must be proportional or equal.

If we write $\kappa_\tS{S}$, $\kappa_\tN{N}$ and $\kappa_\tT{T}$ for the number of vertices of each type in an anomaly-free model, it is not too difficult to show that
\begin{eqnalign}
\label{eq:kSNT-bounds-general}
    \kappa_\tN{N} \geq 1 \quad&\implies\quad \kappa_\tS{S}+\kappa_\tT{T} \leq 9 \,,\\
    \exists\; i,j\;:\;b_i^{\tN{N}}\cdot b_j^{\tN{N}} = 1 \quad&\implies\quad \kappa_\tS{S}+\kappa_\tT{T} + \lceil\tfrac{\kappa_\tN{N}}{2}\rceil \leq 9 \,,\\
    \kappa_\tS{S} = 1 \quad&\implies\quad \kappa_\tN{N} \leq 18 \,,\\
    \kappa_\tS{S}=\kappa_\tT{T} = b_i^{\tN{N}}\cdot b_j^{\tN{N}} = 0 \quad &\implies\quad \kappa_\tN{N} \leq 8 \,.
\end{eqnalign}
In fact, the first three lines hold even for admissible models. The first two bounds follow easily from the fact that type-\tN{N} vertices can form at most $9$ non-trivial edges since $n_R(\v^\tN{N})\leq 18$. Similarly, the third bound follows from the fact that $n_R(\v^\tS{S})\leq 36$ whenever $H_R\leq 18$. For the fourth bound, since all edges are trivial $\Delta$ is just given by the sum $\Delta=\sum\Delta(\v^{\tN{N}})\geq 29\kappa_\tN{N}$ and imposing $\Delta\leq 244$ limits $\kappa_\tN{N}\leq 8$.\footnote{$\kappa_\tS{S}=\kappa_\tT{T}=0$ and $\kappa_\tN{N}=8$ with $b_i\cdot b_j$ all zero is achieved by exactly one anomaly-free model: $G=\SU(2)^8$ with $\H=12(\triv^8)\oplus[16(\rep{2},\triv^7)\oplus(\text{7 others})]$.} In later sections when we forbid $\SU(2)$, $\SU(3)$ and $\Sp(2)$, all non-trivial representations which remain have $H_R\geq 3$ (set by $\tfrac{1}{2}\rep{6}$ of $\Sp(3)$) and by an identical argument we have
\begin{eqnalign}
\label{eq:kSNT-bounds}
    \kappa_\tN{N} \geq 1 \quad&\implies\quad \kappa_\tS{S}+\kappa_\tT{T} \leq 6 \,,\\
    \exists\; i,j\;:\;b_i^{\tN{N}}\cdot b_j^{\tN{N}} = 1 \quad&\implies\quad \kappa_\tS{S}+\kappa_\tT{T} + \lceil\tfrac{\kappa_\tN{N}}{2}\rceil \leq 6 \,,\\
    \kappa_\tS{S} = 1 \quad&\implies\quad \kappa_\tN{N} \leq 12 \,,\\
    \kappa_\tS{S}=\kappa_\tT{T} = b_i^{\tN{N}}\cdot b_j^{\tN{N}} = 0 \quad &\implies\quad \kappa_\tN{N} \leq 8 \,.
\end{eqnalign}
since now type-\tN{N} vertices can form at most $6$ non-trivial edges.

\smallskip

From the strong constraints on $\Lambda\subset\R^{1,1}$ we can also show that the existence of a vector $j\in\R^{1,1}$ satisfying both $j\cdot j=1$ and $j\cdot b_i>0$ for all $i$ is guaranteed. Since all type-\tT{T} vectors are future-directed and all type-\tN{N} vectors lie on the future light-cone, $j\cdot b_i^{\tT{T}}>0$ and $j\cdot b_i^{\tN{N}}>0$ are satisfied iff $j$ is a future-directed, time-like vector. If there is a type-\tS{S} vector present then the allowed $j$s are restricted but never eliminated entirely since $j\cdot b_i^{\tS{S}}$ is positive for some cone of future-directed, time-like vectors $j$. In anomaly-free models consisting of only a single type-\tS{S} vertex the moduli space has two components, one with $j$ future-directed and $j\cdot b_0>0$ and the other with $j$ past-directed and $j\cdot b_0<0$. In all other cases, the conditions $j\cdot b_i>0$ require $j$ to be future-directed and thus force $j\cdot b_0>0$ as well. Therefore we do not impose $j\cdot b_0>0$ as an independent condition, since it always follows from (or is allowed by) the other consistency conditions.

\subsection{Branch and prune}
\label{sec:branch-and-prune}

After having built the multi-hypergraph $\G$, admissible models can be constructed recursively using the ideas discussed in section~4.5 of~\cite{Hamada:2023zol}. Here we only point out some improvements that can be made for $T\leq 1$ in particular (leveraging the structure discussed in section~\ref{sec:edges-and-structure}). First of all, since trivial edges are so rare it is unnecessary to distinguish models which are \emph{irreducible}, as described in section~4.4 of~\cite{Hamada:2023zol}. Second, because there can be at most one type-\tS{S} vertex in an admissible model it is useful for this to the be \emph{first} vertex as admissible models of ever-larger size are built iteratively, if at all.

We first present some useful ideas which can be used to bound the values of $\Delta$ for admissible models. These will serve two purposes: (i) they can be used during the pruning step to abandon models for which no future branching will bring $\Delta$ down in accordance with the bound $\Delta\leq 273-29T$, and (ii) they can be used to limit the size of the multi-hypergraph $\G$ by providing bounds on $\Delta$ of vertices which can ever appear in an anomaly-free model. In contrast to~\cite{Becker:2023zyb} where universal bounds were derived in the more limited setting where the hypermultiplet representations are restricted to a small handful of options, here the bounds are determined empirically.

Given an admissible model, represented by a labelled simplicial complex $\K$, for which the charged hypermultiplets are $\H^\ch(\K)=\bigoplus_R n_R(\K)\times R$ where $n_R(\K)>0$ and $R$ runs over representations of $G(\K)$, we define
\begin{align}
    H^\ch(\K;\calR) &\defeq \sum_{R\;:\;(H_R,n_R(\K))\in\calR}\hspace{-20pt} n_R(\K)H_R \,, &
    \Delta(\K;z,\calR) &\defeq \Delta(\K) - z\,H^\ch(\K;\calR) \,,
\end{align}
where $\calR\subset\R^2$ is a region in the $(H,n)$-plane. Clearly $H^\ch(\K;\emptyset)=0$ and $\Delta(\K)=\Delta(\K;z,\emptyset)=\Delta(\K;0,\calR)$. These two quantities satisfy some obvious monotonicity properties:
\begin{eqnalign}
    \calR' &\subset \calR &\implies&& H^\ch(\K;\calR') &\leq H^\ch(\K;\calR) \,,\\
    z' &< z &\implies&& \Delta(\K;z',\calR) &\geq \Delta(\K;z,\calR) \,,\\
    z\geq 0 \,,\;\; \calR' &\subset \calR &\implies&& \Delta(\K;z,\calR') &\geq \Delta(\K;z,\calR) \,.
\end{eqnalign}
It will useful to define the following semi-infinite rectangular regions:
\begin{eqnalign}
    \calR[a,b] \defeq [0,a]\times[b,\infty) \,, \qquad \calR[\infty,b] \defeq [0,\infty)\times[b,\infty) \,.
\end{eqnalign}
Then we have $H^\ch(\K)=H^\ch(\K;\calR[\infty,0])$. The motivation for defining these regions in particular is that hypermultiplets $n_R\times R$ can potentially merge with $n_S\times S$ only when $(H_R,n_R)\in\calR[n_S,H_S]$, i.e.\ when $H_R\leq n_S$ and $n_R\geq H_S$.

We will write $\K\subseteq \K'$ when there exists some subset of vertices of $\K'$ such that the simplicial subcomplex induced by these vertices is $\K$ (and similarly $\K\subset\K'$ for proper simplicial subcomplexes). We then write $\K'\setminus\K$ for the ``complement'' simplicial subcomplex, i.e.\ the simplicial subcomplex induced by the other vertices. Physically, decomposing a simplicial complex $\K'$ representing an admissible model into $\K\subset\K'$ and $\K'\setminus\K$ corresponds to restricting the gauge group $G(\K')=\prod_{i\in I'}G_i$ to the subgroups $\prod_{i\in I\subset I'}G_i$ and $\prod_{i\notin I}G_i$ and restricting the hypermultiplets appropriately.

The utility of the quantities $H^\ch(\K;\calR)$ and $\Delta(\K;z,\calR)$ comes from the following inequality which relates a simplicial complex to its simplicial subcomplexes. For every $\K_1\subset\K$, if we write $\K_2=\K\setminus\K_1$ then
\begin{equation}
\label{eq:delta-ineq}
    \Delta(\K) \geq \Delta(\K_1;z,\calR_\ast^{(1)}) + \Delta(\K_2;1-z,\calR_\ast^{(2)}) \,, \quad \forall\;z\in[0,1] \,,
\end{equation}
where the pair $(\calR_\ast^{(1)},\calR_\ast^{(2)})$ is the fixed point of the following iteration:
\begin{eqnalign}
    \calR_0^{(1)} = \calR_0^{(2)} &\defeq \calR[\infty,0] \,,\\
    \calR_{\ell+1}^{(1)} &\defeq \bigcup_{R\;:\;(H_R,n_R(\K_2))\in\calR_\ell^{(2)}} \hspace{-20pt} \calR[n_R(\K_2),H_R] \,,\\
    \calR_{\ell+1}^{(2)} &\defeq \bigcup_{R\;:\;(H_R,n_R(\K_1))\in\calR_\ell^{(1)}} \hspace{-20pt} \calR[n_R(\K_1),H_R] \,.
\end{eqnalign}
Notice that $\calR_{\ell+1}^{(i)}\subseteq \calR_\ell^{(i)}$ and there are only a finite number of possible values for $\calR_\ell^{(i)}$, so $\calR_\ell^{(i)}$ are eventually constant and $\calR_\ast^{(i)}$ exist. We have the following inclusions which are often sufficient in practice:
\begin{eqnalign}
\label{eq:Rstar-bound}
    \calR_\ast^{(2)} &\subseteq \calR_1^{(2)}
    = \bigcup_R \calR[n_R(\K_1),H_R]
    \subseteq
    \calR[ n_\text{max}(\K_1), H_\text{min}^\ch(\K_1)]\\
    n_\text{max}(\K) &\defeq \sup_R n_R(\K) \qquad
    H_\text{min}^\ch(\K) \defeq \inf_{R\;:\;n_R(\K)>1} H_R
\end{eqnalign}
We will use this bound on $\calR_\ast^{(2)}$, which importantly only depends on $\K_1$, momentarily.

\smallskip

During the pruning step one needs to discard all admissible models with $\Delta>273-29T$ for which no amount of future branching will produce an anomaly-free model. Let $\K_1$ be the simplicial complex under consideration with $\Delta(\K_1)>273-29T$, and let $\K$ denote some (larger, yet-to-be-constructed) model such that $\K_1\subset\K$. Then, using equations~\eqref{eq:delta-ineq} and~\eqref{eq:Rstar-bound} along with $H_R\geq 3$, we have the following:
\begin{eqnalign}
\label{eq:pruning-condition}
    \Delta(\K) \geq \max_{z\in[0,1]} \Big[\Delta\big(\K_1;1-z,\calR[\infty,0]\big) + \hspace{-10pt} \inf_{\substack{\K_2^{\tN{N}\tT{T}}\\\text{admissible}}} \hspace{-10pt} \Delta\big(\K_2^{\tN{N}\tT{T}};z,\calR[n_\text{max}(\K_1),3]\big)\Big] \,.
\end{eqnalign}
Models for which the right-hand side above is larger than $273-29T$ should be pruned. The superscript in $\K_2^{\tN{N}\tT{T}} = \K\setminus\K_1$ emphasizes that $\K$ will not contain any additional type-\tS{S} vertices other than those in $\K_1$, based on the choice made above to have type-\tS{S} vertices included first, if at all. The admissible models which minimize $\Delta(\K^{\tN{N}\tT{T}};z,\calR[a,3]\big)$ for fixed $a$ turn out to be very simple and the infimum above is piece-wise linear in $z$, allowing for a straightforward maximization over $z$. For example, for $a=5$ we find
\begin{eqnalign}
    \inf_{\substack{\K^{\tN{N}\tT{T}}\\\text{admissible}}} \Delta\big(\K^{\tN{N}\tT{T}};z,\calR[5,3]\big) = \begin{cases}
        \min\{56,\,61-64z,\,91-105z\} & z\in[0,\tfrac{13}{15}] \,,\\
        -\infty & z\in(\tfrac{13}{15},1] \,,
    \end{cases}
\end{eqnalign}
In appendix~\ref{app:bounds} we discuss the determination of this function for all $a$ in more detail. In doing so we provide strong empirical evidence that the claimed bounds are correct and that the above argument is not subject to any logical inconsistencies; at face value we are using properties of the full ensemble of admissible models in order to prune models, meaning that some of the larger admissible models are never constructed.

\section{Results}
\label{sec:results}

\begin{figure}[p]
    \centering
    \includegraphics[width=\textwidth]{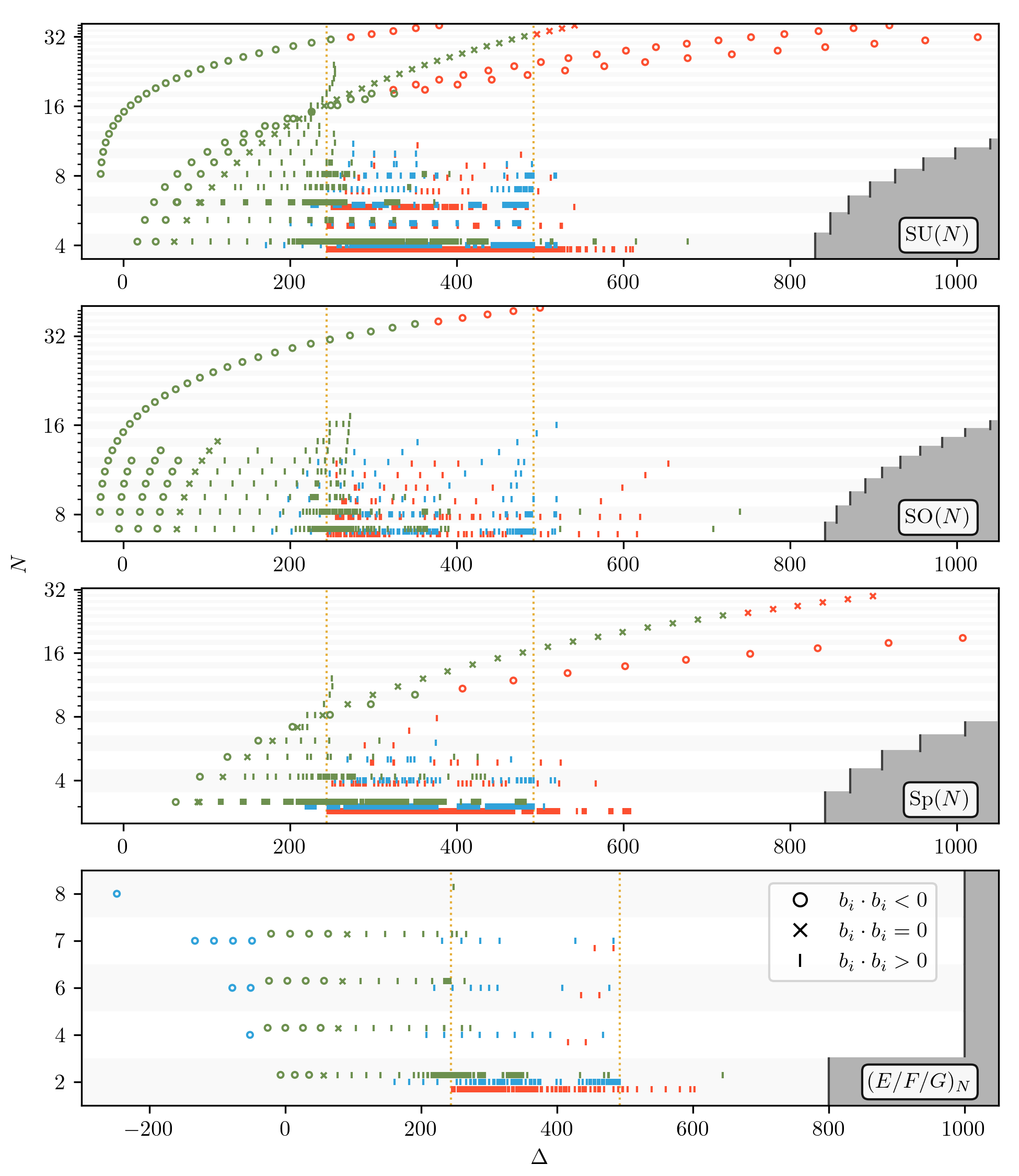}
    \caption{Vertices of $\G$ organized by $\Delta(\v)$ and $G(\v)$ for $T=1$. For each simple group, vertices are generated up to the grey regions at large $\Delta$ but only those which pass the initial pruning as described in appendix~\ref{app:graph-construction} are kept/shown. Colors indicate in what ways a vertex appears in anomaly-free models: red for vertices which do not appear in \emph{any} anomaly-free model, blue for vertices which appear in an anomaly-free model containing one of the eight type-\tS{S} vertices with $\Delta(\v)<-29$, and green for vertices which appear in one or more anomaly-free models but never along with a vertex which has $\Delta(\v)<-29$. The two vertical dotted lines indicate natural thresholds, one being $\Delta=244$ and the other $\Delta=244+248=492$ (related to the $E_8$ NHC).}
    \label{fig:vertex-usage}
\end{figure}

Our previous analysis for $T=0$ (see section~5.1 of~\cite{Hamada:2023zol}) covers cases where hypermultiplets are charged under at most two gauge factors. Applying the present, upgraded methods to constructing anomaly-free models for $T=0$ allows for the possibility of hypermultiplets charged under three or more gauge factors. We find that, given $\U(1)$, $\SU(2)$, $\SU(3)$ and $\Sp(2)$ factors are absent, there are no anomaly-free models with hypermultiplets charged under four or more factors and exactly $20$ with hypermultiplets charged under three factors: see appendix~\ref{app:3-charged-hypers} for the complete list. Nearly all of these models make use of the type-\tT{T} vertex which has $G(\v)=\SU(4)$ and $\H^\ch(\v)=64\times\rep{4}\oplus\rep[\prime]{20}$. In fact, in this context this vertex is better viewed as having gauge group $\SO(6)$ with $\rep{4}$ as the spinor representation. The reason is that many of the 20 models listed in appendix~\ref{app:3-charged-hypers} can be viewed as arising from Higgsing very simple, but anomalous, $\SO(N)$ models and making use of maximal subgroups of the form $\SO(N)\times\SO(M)\subset \SO(N+M)$ under which the spinor representation branches to either one or two bi-spinors. For example,
\begin{center}
    \begin{tikzpicture}
        \node[draw] (SO-20) at (0,0) {
        \begin{minipage}{2.5cm}
            \centering
            \footnotesize
            $\underline{\SO(20)}$\\
            \tiny
            $(-2)\times\triv\oplus \rep{209}\oplus \tfrac{1}{2}\rep{512}$
        \end{minipage}};
        \node[draw] (SO-13-7) at (5,1.5) {
        \begin{minipage}{2.75cm}
            \centering
            \footnotesize
            $\underline{\SO(13)\times\SO(7)}$\\
            \tiny
            $(-1)(\triv,\triv)\oplus(\rep{90},\triv)$\\
            $\oplus(\triv,\rep{27})\oplus \tfrac{1}{2}(\rep{64},\rep{8})$
        \end{minipage}};
        \node[draw] (SO-14-6) at (5,0) {
        \begin{minipage}{2.75cm}
            \centering
            \footnotesize
            $\underline{\SO(14)\times\SO(6)}$\\
            \tiny
            $(-1)(\triv,\triv)\oplus(\rep{104},\triv)$\\
            $\oplus(\triv,\rep[\prime]{20})\oplus (\rep{64},\rep{4})$
        \end{minipage}};
        \node[draw] (SO-7-7-6) at (10,1) {
        \begin{minipage}{3cm}
            \centering
            \footnotesize
            $\underline{\SO(7)^2\times\SO(6)}$\\
            \tiny
            $(\rep{27},\triv,\triv)\oplus (\triv,\rep{27},\triv)$\\
            $\oplus(\triv,\triv,\rep[\prime]{20})\oplus(\rep{8},\rep{8},\rep{4})$
        \end{minipage}};
        \node[draw] (SO-12-8) at (5,-1.5) {
        \begin{minipage}{3.5cm}
            \centering
            \footnotesize
            $\underline{\SO(12)\times\SO(8)}$\\
            \tiny
            $(-1)(\triv,\triv)\oplus(\rep{77},\triv)\oplus(\triv,\repss{35}{v})$\\
            $\oplus \tfrac{1}{2}(\rep{32},\repss{8}{s})\oplus \tfrac{1}{2}(\rep{32},\repss{8}{c})$
        \end{minipage}};
        \node[draw] (SO-8-6-6) at (10,-1) {
        \begin{minipage}{4cm}
            \centering
            \footnotesize
            $\underline{\SO(8)\times \SO(6)^2}$\\
            \tiny
            $(\repss{35}{v},\triv,\triv)\oplus (\triv,\rep[\prime]{20},\triv)\oplus (\triv,\triv,\rep[\prime]{20})$\\
            $\oplus(\repss{8}{s},\rep{4},\rep{4})\oplus(\repss{8}{c},\rep{4},\rep{4})$
        \end{minipage}};

        \draw[->] (SO-20) to (SO-13-7);
        \draw[->] (SO-20) to (SO-12-8);
        \draw[->] (SO-20) to (SO-14-6);
        \draw[->] (SO-13-7) to (SO-7-7-6);
        \draw[->] (SO-12-8) to (SO-8-6-6);
        \draw[->] (SO-14-6) to (SO-7-7-6);
        \draw[->] (SO-14-6) to (SO-8-6-6);
    \end{tikzpicture}
\end{center}
relates two of the models with tri-spinors to a common, anomalous $\SO(20)$ model with very simple hypermultiplet spectrum.

\smallskip

The remainder of this section we devote to exploring various aspects of the $T=1$ anomaly-free models. Implementing the techniques of~\cite{Hamada:2023zol} and ideas discussed in section~\ref{sec:classification} produces an exhaustive list of $608,\!355$ anomaly-free models containing no $\U(1)$, $\SU(2)$, $\SU(3)$ or $\Sp(2)$ gauge factors. Of these, $396,\!472$ require $\Gamma=\Gamma_{1,1}$, $184,\!347$ require $\Gamma=U$ and $27,\!536$ allow for $\Gamma$ either even or odd.
Figure~\ref{fig:vertex-usage} shows the distribution of vertices of the multi-hypergraph $\G$ according to whether they appear in an anomaly-free model. As might be expected, the number of vertices which participate in one or more anomaly-free models tapers quickly for $\Delta(\v)\gg 244$, although for the low-rank groups $\SU(4)$, $\SO(7)$, $\SO(8)$ and $G_2$ there are some outliers with $\Delta(\v)> 600$ which \emph{do} appear in anomaly-free models. We will see later when discussing some examples that these are in fact all related to one another.

\subsection{Distributions and gauge group correlations}
\label{sec:distributions-correlations}

\begin{figure}[p]
    \centering
    \includegraphics[width=\textwidth]{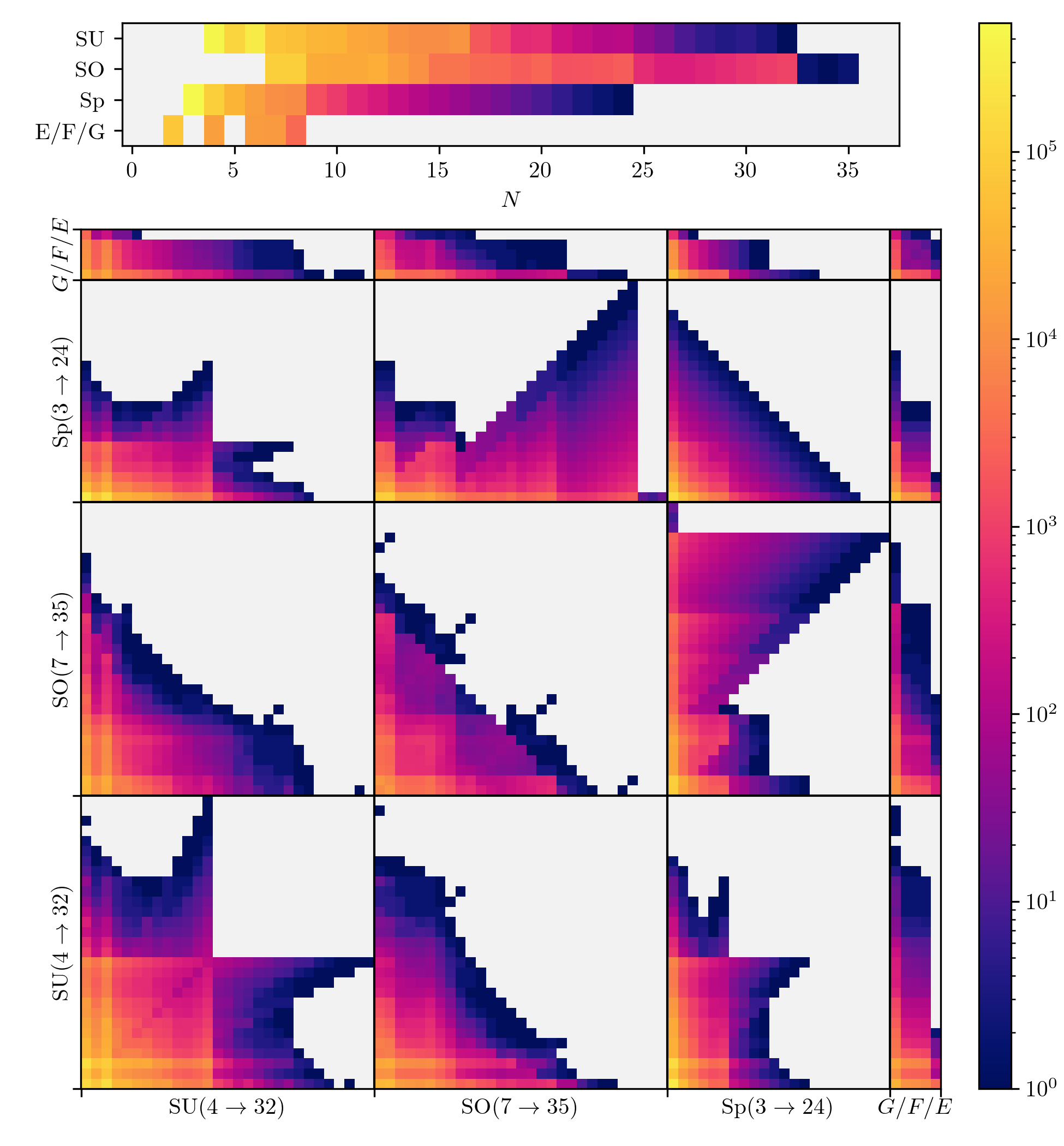}
    \caption{(Top) Frequency of individual gauge factors across anomaly-free models of any size. For the classical series the largest gauge factors which appear are $\SU(32)$, $\SO(35)$ and $\Sp(24)$ (see equations~\eqref{eq:example-su16-su32},~\eqref{eq:example-SO35-Sp3^9} and~\eqref{eq:example-SO32-Sp24}). (Bottom) Frequency of \emph{pairs} of gauge factors appearing together across anomaly-free models with at least two simple factors. For example, a model with gauge group $\SU(4)^2\times\SO(7)$ would count once towards the tally for $\SU(4)\times\SU(4)$ and twice towards the tally for $\SU(4)\times\SO(7)$.}
    \label{fig:gauge-factors-frequency}
\end{figure}

\begin{figure}[t]
    \centering
    \includegraphics[width=\textwidth]{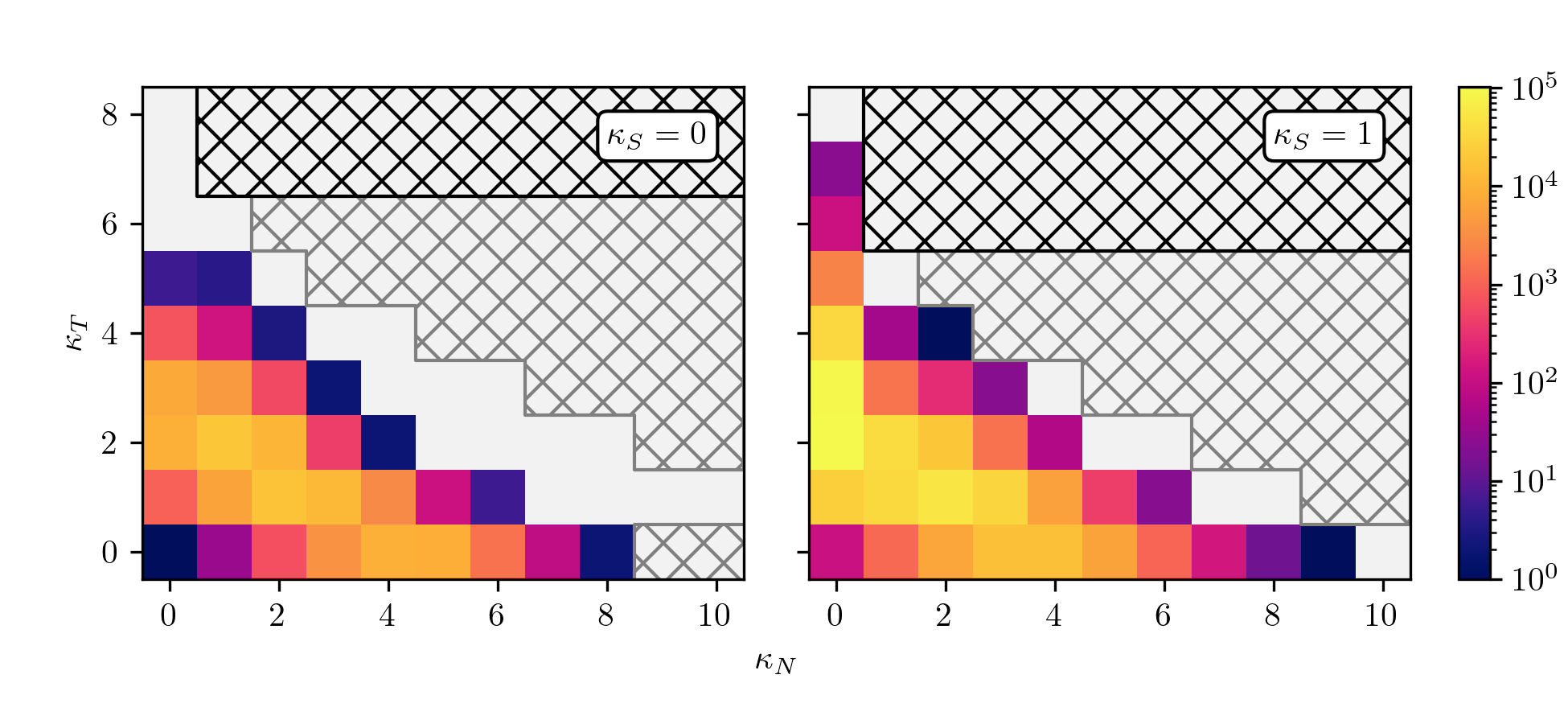}
    \caption{The frequency of anomaly-free models by $\kappa_\tS{S}$, $\kappa_\tN{N}$, $\kappa_\tT{T}$, the numbers of vertices of each type (cf.~equation~\eqref{eq:SNT-definition}). The hatched regions are ruled out by the bounds of equation~\eqref{eq:kSNT-bounds} (gray hatched regions are ruled out only with additional assumptions on $b_i^{\tN{N}}\cdot b_j^{\tN{N}}$, but nevertheless we see that they hold universally).}
    \label{fig:vertex-types-dist}
\end{figure}

Before turning to examples in the following section, let us discuss several gross features of the ensemble of $\approx \!600,\!000$ anomaly-free models for $T=1$. First of all, the frequency of models by total number of vertices, $\kappa$, is as follows:
\begin{center}
    \scalebox{0.91}{
    \begin{tabular}{c*{11}{>$c<$}}
        \toprule
        \#(models) & \multicolumn{11}{c}{$\kappa$}\\ \cmidrule{2-12}
         & 0 & 1 & 2 & 3 & 4 & 5 & 6 & 7 & 8 & 9 & 10 \\ \midrule
        w/ $\kappa_\tS{S}=0$ & 1 & 1,\!042 & 15,\!957 & \phantom{2}46,\!495 & \phantom{2}37,\!795 & \phantom{2}12,\!719 & \phantom{2}1,\!613 & \phantom{2,\!2}95 & \phantom{22}2 & \phantom{1}0 & 0\\
        w/ $\kappa_\tS{S}=1$ & 0 & \phantom{1,}118 & 26,\!933 & 144,\!987 & 202,\!122 & 100,\!861 & 15,\!687 & 1,\!718 & 195 & 14 & 1\\
        total & 1 & 1,\!160 & 42,\!890 & 191,\!482 & 239,\!917 & 113,\!580 & 17,\!300 & 1,\!813 & 197 & 14 & 1\\
        \bottomrule
    \end{tabular}
    }
\end{center}
The model with $\kappa=0$ vertices is simply the case where $V=0$ and $H=244$; we will see the model at the other extreme with ten simple factors below in equation~\eqref{eq:example-SO35-Sp3^9}. In figure~\ref{fig:vertex-types-dist} is presented a refinement of this data, the frequency of models by number of vertices of each type, $\kappa_\tS{S}$, $\kappa_\tN{N}$ and $\kappa_\tT{T}$. In the discussion around equation~\eqref{eq:kSNT-bounds} we saw how $\kappa_\tS{S}$, $\kappa_\tN{N}$ and $\kappa_\tT{T}$ can sometimes be bounded using very simple observations about the maximum number of edges that type-\tN{N} vertices can have. The bounds of~\eqref{eq:kSNT-bounds} are depicted as forbidden regions with hatching, and while the forbidden regions in gray were derived using additional assumptions on $b_i^{\tN{N}}\cdot b_j^{\tN{N}}$ we see that the ensemble adheres to the bounds universally regardless.

We can also easily extract information about the gauge groups which appear in the ensemble. In figure~\ref{fig:gauge-factors-frequency} shows the frequency of each gauge factor across all gauge groups of the ensemble, as well as the number of times that each \emph{pair} of gauge factors appears together. The sharp transitions between $\SU(16)/\SU(17)$ and $\Sp(8)/\Sp(9)$ are due to the dimension of the fundamental representation passing $16$, which abruptly disallows having edges to the $\SU(N)$ and $\Sp(N)$ type-\tN{N} vertices for which $n_F=16$ is fixed for any $N$.

\subsection{Higgsing and examples}
\label{sec:Higgsing-and-examples}

Most of the anomaly-free models are related via Higgsing. Checking that the generated list of anomaly-free models for $T=1$ without $\U(1)$, $\SU(2)$, $\SU(3)$ and $\Sp(2)$ factors is closed under Higgsing provides a nontrivial check both on our code and on the bounds for pruning discussed in appendix~\ref{app:bounds} for which we have strong empirical evidence but no mathematical proof. See appendix~\ref{app:Higgsing} for some details on how this check was automated. There are several well-known models which cannot be Higgsed:\footnote{There are a few more NHCs which appear for $T>1$, e.g.\ $G=G_2\times\SU(2)$ with $\H^\ch=\frac{1}{2}(\triv,\rep{2})\oplus \frac{1}{2}(\rep{7},\rep{2})$.}
\begin{center}
    \begin{tabular}{>$c<$|*{8}{>$c<$}}
        \toprule
        G & \SU(3)\footnotemark & \SO(8) & F_4 & E_6 & E_7 & E_7 & E_8 \\ \midrule
        \H^\ch & \text{--} & \text{--} & \text{--} & \text{--} & \text{--} & \tfrac{1}{2}\rep{56} & \text{--}\\
        \bottomrule
    \end{tabular}
\end{center}
\footnotetext{Note that because of our omission of $\SU(3)$ factors, in practice successive Higgsings terminate at $G_2+\rep{7}$ rather than $\SU(3)+\emptyset$.}
These so-called non-Higgsable clusters (NHCs) have a well-understood geometric description in F-theory.
All anomaly-free models can either be completely Higgsed to the model with $V=0$, $H=244$ or eventually filter down to one of the NHCs listed above. In the reverse direction, we find that $\mathcal{O}(10\%)$ of models do not come from the Higgsing of any other anomaly-free model. This figure should be taken with a grain of salt since it may change significantly if the omitted gauge factors are reinstated, but nevertheless these ``maximal'' models may be identified as being the most interesting.

\begin{figure}[p]
    \centering
    \includegraphics[width=\textwidth]{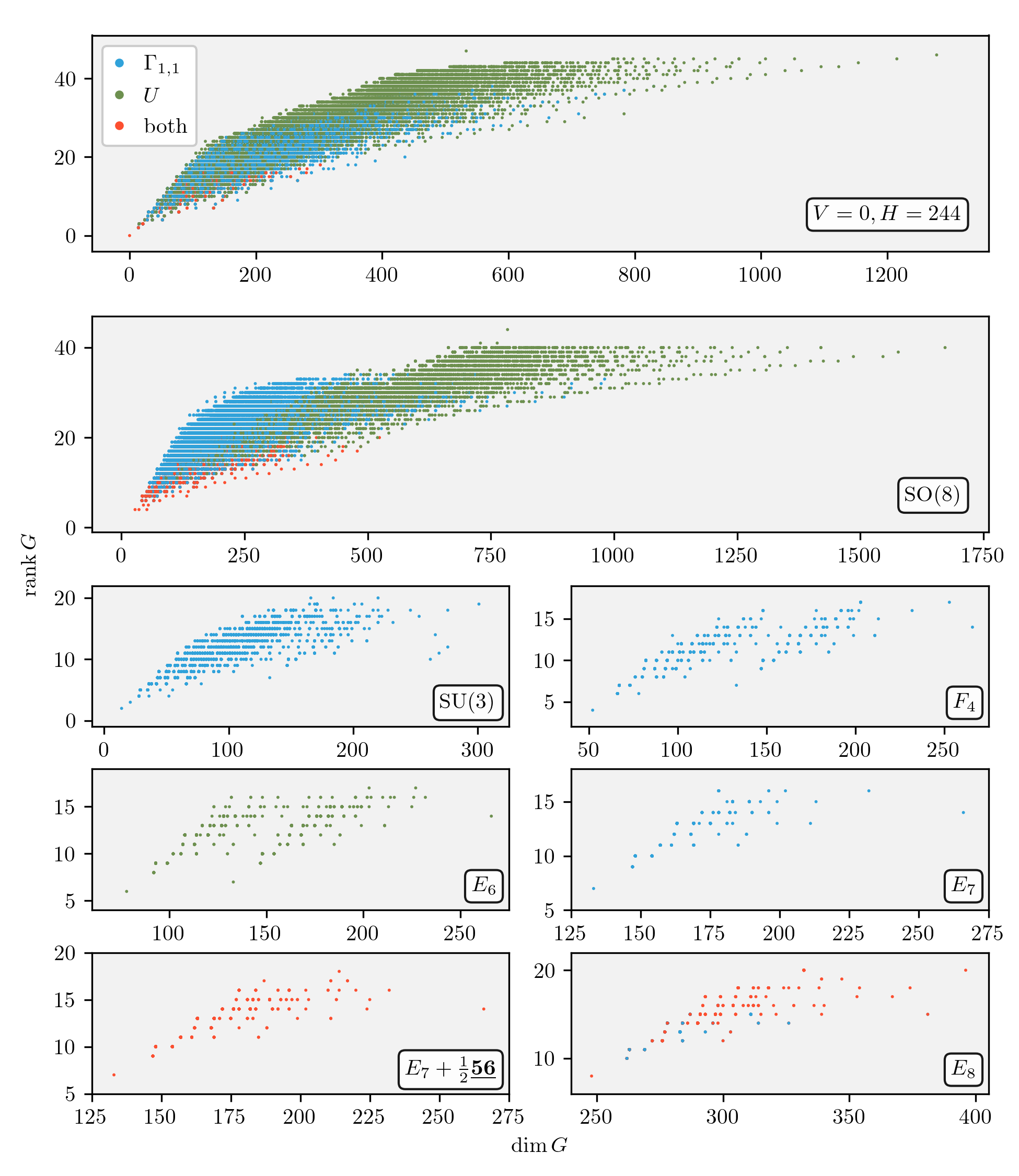}
    \caption{Anomaly-free models for $T=1$ by gauge group dimension and rank, grouped by the NHC to which they Higgs down. Colors indicate the lattice $\Gamma$: blue for odd, green for even and red if both odd and even are possible.}
    \label{fig:models-by-gauge-group}
\end{figure}

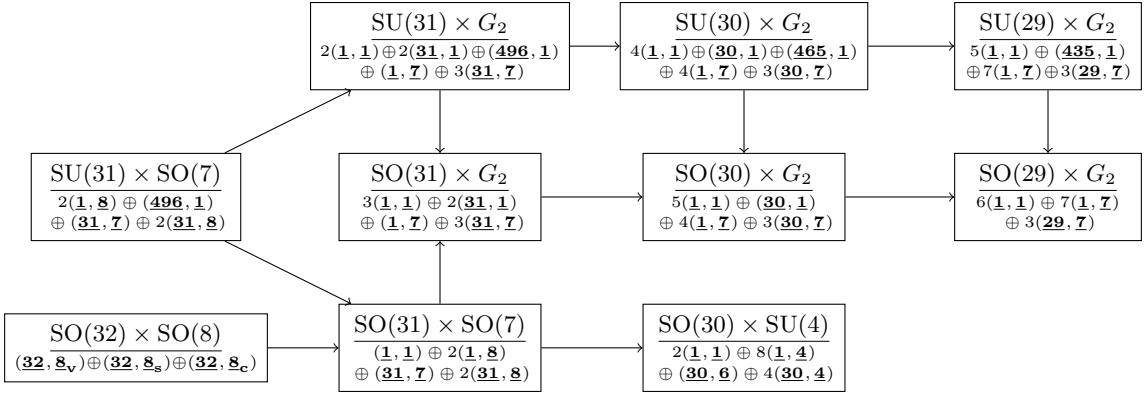
\begin{figure}[t]
    \centering
    \begin{tikzpicture}
        \node[draw] (SU31SO7) at (0,-2) {
        \begin{minipage}{2.5cm}
            \centering
            \footnotesize
            $\underline{\SU(31)\times\SO(7)}$\\
            \tiny
            $2(\triv,\rep{8})\oplus (\rep{496},\triv)$\\
            ${}\oplus (\rep{31},\rep{7})\oplus 2(\rep{31},\rep{8})$
        \end{minipage}};
        
        \node[draw] (SU31G2) at (4,0) {
        \begin{minipage}{3.15cm}
            \centering
            \footnotesize
            $\underline{\SU(31)\times G_2}$\\
            \tiny
            $2(\triv,\triv)\oplus 2(\rep{31},\triv)\oplus (\rep{496},\triv)$
            ${} \oplus (\triv,\rep{7}) \oplus 3(\rep{31},\rep{7})$
        \end{minipage}};
        
        \node[draw] (SU30G2) at (8,0) {
        \begin{minipage}{3cm}
            \centering
            \footnotesize
            $\underline{\SU(30)\times G_2}$\\
            \tiny
            $4(\triv,\triv)\oplus (\rep{30},\triv)\oplus (\rep{465},\triv)$
            ${}\oplus 4(\triv,\rep{7})\oplus 3(\rep{30},\rep{7})$
        \end{minipage}};
        
        \node[draw] (SU29G2) at (12,0) {
        \begin{minipage}{2.2cm}
            \centering
            \footnotesize
            $\underline{\SU(29)\times G_2}$\\
            \tiny
            $5(\triv,\triv)\oplus (\rep{435},\triv)$\\
            ${}\oplus 7(\triv,\rep{7})\oplus 3(\rep{29},\rep{7})$
        \end{minipage}};
        
        \node[draw] (SO32SO8) at (0,-4) {
        \begin{minipage}{3.2cm}
            \centering
            \footnotesize
            $\underline{\SO(32)\times\SO(8)}$\\
            \tiny
            $(\rep{32},\repss{8}{v})\oplus (\rep{32},\repss{8}{s})\oplus (\rep{32},\repss{8}{c})$
        \end{minipage}};
        
        \node[draw] (SO31SO7) at (4,-4) {
        \begin{minipage}{2.4cm}
            \centering
            \footnotesize
            $\underline{\SO(31)\times\SO(7)}$\\
            \tiny
            $(\triv,\triv)\oplus 2(\triv,\rep{8})$
            ${}\oplus (\rep{31},\rep{7})\oplus 2(\rep{31},\rep{8})$
        \end{minipage}};
        
        \node[draw] (SO30SU4) at (8,-4) {
        \begin{minipage}{2.4cm}
            \centering
            \footnotesize
            $\underline{\SO(30)\times\SU(4)}$\\
            \tiny
            $2(\triv,\triv)\oplus 8(\triv,\rep{4})$
            ${}\oplus (\rep{30},\rep{6})\oplus 4(\rep{30},\rep{4})$
        \end{minipage}};
        
        \node[draw] (SO31G2) at (4,-2) {
        \begin{minipage}{2.4cm}
            \centering
            \footnotesize
            $\underline{\SO(31)\times G_2}$\\
            \tiny
            $3(\triv,\triv)\oplus 2(\rep{31},\triv)$
            ${}\oplus (\triv,\rep{7})\oplus 3(\rep{31},\rep{7})$
        \end{minipage}};
        
        \node[draw] (SO30G2) at (8,-2) {
        \begin{minipage}{2.4cm}
            \centering
            \footnotesize
            $\underline{\SO(30)\times G_2}$\\
            \tiny
            $5(\triv,\triv)\oplus (\rep{30},\triv)$\\
            ${}\oplus 4(\triv,\rep{7})\oplus 3(\rep{30},\rep{7})$
        \end{minipage}};
        
        \node[draw] (SO29G2) at (12,-2) {
        \begin{minipage}{2.2cm}
            \centering
            \footnotesize
            $\underline{\SO(29)\times G_2}$\\
            \tiny
            $6(\triv,\triv)\oplus 7(\triv,\rep{7})$\\
            ${}\oplus 3(\rep{29},\rep{7})$
        \end{minipage}};

        \draw[->] (SU31SO7) to (SO31SO7);
        \draw[->] (SU31SO7) to (SU31G2);
        \draw[->] (SO32SO8) to (SO31SO7);
        \draw[->] (SO31SO7) to (SO31G2);
        \draw[->] (SU31G2)  to (SO31G2);
        \draw[->] (SU31G2)  to (SU30G2);
        \draw[->] (SO31G2)  to (SO30G2);
        \draw[->] (SU30G2)  to (SO30G2);
        \draw[->] (SO31SO7) to (SO30SU4);
        \draw[->] (SU30G2)  to (SU29G2);
        \draw[->] (SO30G2)  to (SO29G2);
        \draw[->] (SU29G2)  to (SO29G2);
    \end{tikzpicture}
    \caption{All anomaly-free models incorporating individual vertices with $\Delta(\v) > 600$ (see~\eqref{eq:large-Δ-vertices}). These are all related via Higgsing (discarding $\U(1)$ factors), as shown. Those with an $\SU(N\geq 29)$ factor have $b_I\cdot b_J=\begin{psmallmatrix}
        8 & -1 & 30\\
        -1 & -1 & 6\\
        30 & 6 & 28
    \end{psmallmatrix}$ and thus $\Gamma=\Gamma_{1,1}$ with $b_0=(3,1)$, $b_1=(0,1)$ and $b_2=(8,-6)$. As we will discuss later, these models are excluded by the brane string constraints. All others have $b_I\cdot b_J=\begin{psmallmatrix}
        8 & -2 & 30\\
        -2 & -4 & 12\\
        30 & 12 & 28
    \end{psmallmatrix}$ and either $\Gamma=\Gamma_{1,1}$ with $b_0=(3,1)$, $b_1=(0,2)$ and $b_2=(8,-6)$ or $\Gamma=U$ with $b_0=(2,2)$, $b_1=(1,-2)$ and $b_2=(1,14)$.}
    \label{fig:large-Δ-models}
\end{figure}

Figure~\ref{fig:models-by-gauge-group} shows the distribution of anomaly-free models by gauge group, separated by which model they Higgs down to. Let us now highlight some of the notable outliers in each class. Amongst models which can Higgs down to nothing, the model with gauge group of largest dimension
($\dim G=1278$) is
\begin{eqnalign}
\label{eq:example-su16-su32}
    G &= \SU(16)\times\SU(32) \,,\\
    \H &= 18(\triv,\triv) \oplus (\rep{16},\rep{32}) \oplus 2(\triv,\rep{496}) \,,\\
    b_I\cdot b_J &= \begin{psmallmatrix}
        8 &  0 & 2 \\
        0 & -2 & 1 \\
        2 &  1 & 0
    \end{psmallmatrix} \;\;\Longleftrightarrow\;\; \Gamma=U\,:\; b_0 = (2,2) \,,\; b_1 = (1,-1) \,,\; b_2 = (0,1) \,,
\end{eqnalign}
and the model with gauge group of largest rank ($\rank G=47$, the clear outlier in the first panel of figure~\ref{fig:models-by-gauge-group}) is\footnote{This model Higgses down to the $G=SU(18)\times SU(3)^{12}$ model found in~\cite{Kumar:2009ac} (see section~4.2 there).}
\begin{eqnalign}
    G &= \SU(18)\times\SU(6)^6 \,,\\
    \H &= 9(\triv;\triv^6) \oplus \big[(\rep{18};\rep{6},\triv^5) \oplus (\text{5 others}) \big] \oplus \big[(\triv;\rep{20},\triv^5) \oplus (\text{5 others}) \big] \,,\\
    b_I\cdot b_J &= \begin{psmallmatrix}
        8 &  0 & 2_6 \\
        0 & -2 & 1_6 \\
        2_6 &  1_6 & 0_{6\times 6}
    \end{psmallmatrix} \;\;\Longleftrightarrow\;\; \Gamma=U \,:\; b_0 = (2,2) \,,\; b_1 = (1,-1) \,,\; b_{2\to7} = (0,1) \,.
\end{eqnalign}
For models which Higgs down to the $\SU(3)$ NHC (via $G_2+\rep{7}$), the model with largest gauge group dimension ($\dim G = 301$) is
\begin{eqnalign}
    G &= E_7\times \SU(13) \,,\\
    \H &= 2(\triv,\triv) \oplus \tfrac{5}{2}(\rep{56},\triv) \oplus (\triv,\rep{13}) \oplus 5(\triv,\rep{78}) \,,\\
    b_I\cdot b_J &= \begin{psmallmatrix}
        8 & -1 & 5\\
        -1 & -3 & 0\\
        5 & 0 & 3
    \end{psmallmatrix} \;\;\Longleftrightarrow\;\; \Gamma=\Gamma_{1,1} \,:\; b_0=(3,1) \,,\; b_1=(-1,-2) \,,\; b_2=(2,1) \,.
\end{eqnalign}
For models which Higgs down to the $\SO(8)$ NHC, the model with largest gauge group dimension ($\dim G = 1672$) is
\begin{eqnalign}
\label{eq:example-SO32-Sp24}
    G &= \SO(32)\times\Sp(24) \,,\\
    \H &= 21(\triv,\triv) \oplus \tfrac{1}{2}(\rep{32},\rep{48}) \oplus (\triv,\rep{1127}) \,,\\
    b_I\cdot b_J &= \begin{psmallmatrix}
        8 & -2 & 2\\
        -2 & -4 & 1\\
        2 & 1 & 0
    \end{psmallmatrix} \;\;\Longleftrightarrow\;\; \Gamma=U \,:\; b_0 = (2,2) \,,\; b_1 = (1,-2) \,,\; b_2 = (0,1) \,.
\end{eqnalign}
and the model with largest gauge group rank ($\rank G = 44$, the clear outlier in the second panel of figure~\ref{fig:models-by-gauge-group}) is
\begin{eqnalign}
\label{eq:example-SO35-Sp3^9}
    G &= \SO(35)\times\Sp(3)^9 \,,\\
    \H &= 20(\triv;\triv^9) \oplus \big[\tfrac{1}{2}(\rep{35};\rep{6},\triv^8) \oplus (\text{8 others})\big] \oplus \big[\tfrac{1}{2}(\triv;\rep[\prime]{14},\triv^8) \oplus (\text{8 others})\big] \,,\\
    b_I\cdot b_J &= \begin{psmallmatrix}
        8 & -2 & 2_9\\
        -2 & -4 & 1_9\\
        2_9 & 1_9 & 0_{9\times 9}
    \end{psmallmatrix} \;\;\Longleftrightarrow\;\; \Gamma=U \,:\; b_0=(2,2) \,,\; b_1=(1,-2) \,,\; b_{2\to10}=(0,1) \,.
\end{eqnalign}
Note the unique roles played by the quaternionic irreducible representation $\rep[\prime]{14}$ of $\Sp(3)$ and the exceptional $\Sp(3)$ type-\tN{N} vertex: see table~\ref{tab:type-S/N-vertices}. Also notable is the spike of models out to $(\dim G, \rank G) = (981,33)$ which admit an odd lattice. In fact, these models incorporate the vertices with abnormally large $\Delta(\v)$ noted above, namely
\begin{eqnalign}
\label{eq:large-Δ-vertices}
    G(\v) &= \SO(8) \,, &\quad \H^\ch(\v) &= 32\times\repss{8}{v}\oplus 32\times\repss{8}{s}\oplus 32\times\repss{8}{c} \,, &\quad \Delta(\v) &= 740 \,,\\
    &= \SO(7) \,, & &= 31\times\rep{7}\oplus 64\times\rep{8} \,, & &= 708 \,,\\
    &= \SU(4) \,, & &= 30\times\rep{6}\oplus 128\times\rep{4} \,, & &= 677 \,,\\
    &= G_2 \,, & &= 94\times\rep{7} \,, & &= 644 \,.
\end{eqnalign}
In this context the above $\SU(4)$ vertex is better though of as having gauge group $\SO(6)$ with many fundamental and spinor hypermultiplets. It is also clear that the pattern continues to include a vertex with $\Sp(2)\sim\SO(5)$ and $29\times\rep{5}\oplus 128\times\rep{4}$, for which $\Delta(\v)=647$. See figure~\ref{fig:large-Δ-models} for a complete list of anomaly-free models in which the above four large-$\Delta$ vertices appear. We see that there are two anomaly-free models among this collection which Higgs down to the others, one with gauge group $\SU(31)\times\SO(7)$ and the other with $\SO(32)\times\SO(8)$.

Continuing, models of the following shape,
\begin{eqnalign}
    G &= E_7^2 \,,\\
    \H &= 62(\triv,\triv)\oplus \tfrac{8-k}{2}(\rep{56},\triv)\oplus \tfrac{k+8}{2}(\triv,\rep{56}) \,,\\
    b_I\cdot b_J &= \begin{psmallmatrix}
        8 & 2-k & 2+k\\
        2-k & -k & 0\\
        2+k & 0 & k
    \end{psmallmatrix} \,,
\end{eqnalign}
for which the lattices are
\begin{eqnalign}
    k &= 5\;: &\qquad \Gamma &= \Gamma_{1,1} \,, &\quad b_0 &= (3,1) \,, & b_1 &= (-2,-3) \,, & b_2 &= (3,2) \,,\\
    k &= 6\;: & \Gamma &= U\,, & b_0 &= (2,2)\,, & b_1 &= (1,-3)\,, & b_2 &= (1,3) \,,\\
    k &= 7\;: & \Gamma &= \Gamma_{1,1}\,, & b_0 &= (3,1)\,, & b_1 &= (-3,-4)\,, & b_2 &= (4,3) \,,\\
    k &= 8\;: & \Gamma &= \Gamma_{1,1}\,, & b_0 &= (3,1)\,, & b_1 &= (-1,3)\,, & b_2 &= (3,-1) \,,\\
    & &\text{or}\quad \Gamma &= U \,, & b_0 &= (2,2)\,, & b_1 &= (1,-4)\,, & b_2 &= (1,4)\,,
\end{eqnalign}
are those with largest gauge group dimension which Higgs down to the $F_4$, $E_6$, $E_7+\frac{1}{2}\rep{56}$ and $E_7$ NHCs for $k=5$, $k=6$, $k=7$ and $k=8$, respectively. In these sectors we find that the maximum gauge group ranks are 17 for the $F_4$ NHC (occuring $11$ times), 17 for the $E_6$ NHC (occuring twice), 16 for the $E_7+\frac{1}{2}\rep{56}$ NHC (occuring $11$ times), and 18 for the $E_7$ NHC (occuring once). Many of these maximal-rank gauge groups involve $\SU(4)$ or $\Sp(3)$ factors and so we expect these figures to change if $\SU(2)$, $\SU(3)$ and $\Sp(2)$ are reincorporated.

Finally, for models which Higgs down to the $E_8$ NHC the model with largest gauge group dimension ($\dim G = 396$) is
\begin{align}
\label{eq:e8-so8-so16}
    G &= E_8\times\SO(8)\times\SO(16) \,, \\
    \H &= (\triv,\repss{8}{v},\rep{16}) \oplus (\triv,\repss{8}{s},\rep{16}) \oplus (\triv,\repss{8}{c},\rep{16}) \oplus 2(\triv,\triv,\rep{128}) \,, \notag\\
    b_I\cdot b_J &= \begin{psmallmatrix}
        8 & -10 & 14_2\\
        -10 & -12 & 0_2\\
        14_2 & 0_2 & 12_{2\times 2}
    \end{psmallmatrix} \;\;\Longleftrightarrow\;\; \left\{\begin{aligned}
        \Gamma &= \Gamma_{1,1} \!\!\!&:\;\; b_0 &= (3,1) \,,\; b_1 = (-2,4) \,,\; b_{2,3} = (4,-2)\\
        \Gamma &= U &:\;\; b_0 &= (2,2) \,,\; b_1 = (1,-6) \,,\; b_{2,3} = (1,6) \,.
    \end{aligned}\right. \notag
\end{align}
In Appendix~\ref{app:orbifolds}, we show that this model has an asymmetric orbifold realization in string theory.
This also happens to have the largest rank ($\rank G = 20$), but there are others which match it such as
\begin{align}
    G &= E_8\times\SO(8)^3 \,, \\
    \H &= (\triv;\repss{8}{v},\repss{8}{v},\triv) \oplus (\triv;\repss{8}{s},\repss{8}{s},\triv) \oplus (\triv;\repss{8}{c},\repss{8}{c},\triv)\oplus (\triv;\triv,\repss{8}{v},\repss{8}{v}) \oplus (\triv;\triv,\repss{8}{s},\repss{8}{s}) \notag\\
    &\qquad \oplus (\triv;\triv,\repss{8}{c},\repss{8}{c}) \oplus (\triv;\repss{8}{v},\triv,\repss{8}{v}) \oplus (\triv;\repss{8}{s},\triv,\repss{8}{s}) \oplus (\triv;\repss{8}{c},\triv,\repss{8}{c}) \,, \notag\\
    b_I\cdot b_J &= \begin{psmallmatrix}
        8 & -10 & 14_3\\
        -10 & -12 & 0_3\\
        14_3 & 0_3 & 12_{3\times 3}
    \end{psmallmatrix} \;\;\Longleftrightarrow\;\; \left\{\begin{aligned}
        \Gamma &= \Gamma_{1,1} \!\!\!&:\;\; b_0 &= (3,1) \,,\; b_1 = (-2,4) \,,\; b_{2,3,4} = (4,-2)\\
        \Gamma &= U &:\;\; b_0 &= (2,2) \,,\; b_1 = (1,-6) \,,\; b_{2,3,4} = (1,6) \,.
    \end{aligned}\right. \notag
\end{align}
as well as 17 additional models like this using the same vertices but with various pairings of $\repss{8}{v}$, $\repss{8}{s}$ and $\repss{8}{c}$ into bi-charged hypermultiplets. Most can be reached by applying triality transformations ($\mathbf{v}\to\mathbf{s}\to\mathbf{c}\to\mathbf{v}$) repeatedly to individual $\SO(8)$ factors and from Higgsing $\SO(16)\to\SO(8)\times\SO(8)$ in the previous example.

\medskip

Also of note are anomaly-free models which have hypermultiplets charged under $E_8$. We find that there are exactly three such models:
\begin{eqnalign}
    G &= \SO(8)\times E_8 \,, &
    \H &= (\repss{8}{v},\triv)\oplus (\repss{8}{s},\triv)\oplus (\repss{8}{c},\triv)\oplus 2(\triv,\rep{248}) \,, \\
    G &= \SO(7)\times E_8 \,, &
    \H &= (\triv,\triv)\oplus 2(\rep{8},\triv)\oplus 2(\triv,\rep{248}) \,, \\
    G &= G_2\times E_8 \,, &
    \H &= 3(\triv,\triv)\oplus (\rep{7},\triv)\oplus 2(\triv,\rep{248}) \,, \\[5pt]
    b_I\cdot b_J &= \begin{psmallmatrix}
    8 & -1 & 10\\
    -1 & -3 & 0\\
    10 & 0 & 12
\end{psmallmatrix} \;\;\;\Longleftrightarrow& \Gamma&=\Gamma_{1,1} \,:\; b_0 = (3,1) \,,\; b_1 = (-1,-2) \,,\; b_2=(4,2) \,.
\end{eqnalign}
The largest of these, with gauge group $\SO(8)\times E_8$, is realized in string theory via an asymmetric orbifold (model~4 in~\cite{Baykara:2023plc}) and Higgses down to the others.

\medskip

Much like the case for $T=0$, we find that there are only a handful of models which make use of hypermultiplets charged under three gauge factors (see appendix~\ref{app:3-charged-hypers}) and \emph{no} instances of hypermultiplets charged under four or more gauge factors. That is, for $T=1$ if a hypermultiplet is charged under four or more non-abelian gauge factors then at least one of them must be $\SU(2)$, $\SU(3)$ or $\Sp(2)$. We can, however, concoct many examples of such hypermultiplets by Higgsing anomaly-free models with these small-rank factors absent and making use of the branching rules for special\footnote{In contrast to \emph{regular} maximal subgroups, which can be read off from the corresponding (extended) Dynkin diagram.} maximal subalgebras, e.g.\ $\SO(NM)\to\SO(N)\times\SO(M)$ under which $\rep{NM}\to(\rep{N},\rep{M})$ or $\SO(4NM)\to\Sp(N)\times\Sp(M)$ under which $\rep{4NM}\to(\rep{2N},\rep{2M})$. For example, starting from
\begin{align}
\label{eq:so8-so16^2}
    G &= \SO(8)\times \SO(16)^2 \,,\\
    \H &= (\triv;\rep{16},\rep{16})\oplus (\triv;\rep{128},\triv)\oplus (\triv;\triv,\rep{128}) \,, \notag\\
    b_I\cdot b_J &= \begin{psmallmatrix}
        8 & -2 & 6_2\\
        -2 & -4 & 0_2\\
        6_2 & 0_2 & 4_{2\times 2}
    \end{psmallmatrix} \;\;\Longleftrightarrow\;\; \left\{\begin{aligned}
        \Gamma &= \Gamma_{1,1} \!\!\!&:\; b_0 &= (3,1)\,,\; b_1 = (0,2) \,,\; b_{2,3} = (2,0)\,,\\
        \Gamma &= U &:\; b_0 &= (2,2)\,,\; b_1 = (1,-2) \,,\; b_{2,3} = (1,2)\,,
    \end{aligned}\right. \notag
\end{align}
(which is realized in string theory via an asymmetric orbifold: see appendix~\ref{app:orbifolds}) and breaking each $\SO(16)$ to its $\Sp(2)^2$ maximal subgroup results in a particularly simple anomaly-free model with a quad-fundamental of $\Sp(2)^4$:
\begin{align}
    G &= \SO(8)\times\Sp(2)^4 \,,\\
    \H &= (\triv;\rep{4},\rep{4},\rep{4},\rep{4})\oplus \big[(\triv;\rep{14},\triv,\triv,\triv)\oplus(\text{3 others})\big] \,, \notag\\
    b_I\cdot b_J &= \begin{psmallmatrix}
        8 & -2 & 12_4\\
        -2 & -4 & 0_4\\
        12_4 & 0_4 & 16_{4\times 4}
    \end{psmallmatrix} \;\;\Longleftrightarrow\;\; \left\{\begin{aligned}
        \Gamma &= \Gamma_{1,1} \!\!\!&:\; b_0 &= (3,1)\,,\; b_1 = (0,2) \,,\; b_{2\to5} = (4,0) \,,\\
        \Gamma &= U &:\; b_0 &= (2,2)\,,\; b_1 = (1,-2) \,,\; b_{2\to5} = (2,4) \,.
    \end{aligned}\right. \notag
\end{align}
Similarly, starting with
\begin{align}
    G &= \SO(8)\times\Sp(4)^2 \,,\\
    \H &= 4(\triv;\triv,\triv)\oplus 4(\triv;\rep{8},\rep{8})\oplus (\triv;\rep{42},\triv)\oplus (\triv;\triv,\rep{42}) \,, \notag\\
    b_I\cdot b_J &= \begin{psmallmatrix}
        8 & -2 & 6_2\\
        -2 & -4 & 0_2\\
        6_2 & 0_2 & 4_{2\times 2}
    \end{psmallmatrix} \;\;\Longleftrightarrow\;\; \left\{\begin{aligned}
        \Gamma &= \Gamma_{1,1} \!\!\!&:\; b_0 &= (3,1)\,,\; b_1 = (0,2) \,,\; b_{2,3} = (2,0) \,,\\
        \Gamma &= U &:\; b_0 &= (2,2)\,,\; b_1 = (1,-2) \,,\; b_{2,3} = (1,2) \,.
    \end{aligned}\right. \notag
\end{align}
and Higgsing each $\Sp(4)\to\SU(2)^3$, under which $\rep{8}\to(\rep{2},\rep{2},\rep{2})$, results in
\begin{align}
    G &= \SO(8)\times\SU(2)^6 \,,\\
    \H &= 4(\triv;\triv^6)\oplus 4(\triv;\rep{2}^6)\oplus \big[(\triv;\rep{5},\triv^5)\oplus(\text{5 others})\big] \,, \notag\\
    b_I\cdot b_J &= \begin{psmallmatrix}
        8 & -2 & 24_6\\
        -2 & -4 & 0_6\\
        24_6 & 0_6 & 64_{6\times 6}
    \end{psmallmatrix} \;\;\Longleftrightarrow\;\; \left\{\begin{aligned}
        \Gamma &= \Gamma_{1,1} \!\!\!&:\; b_0 &= (3,1)\,,\; b_1 = (0,2) \,,\; b_{2\to7} = (8,0) \,,\\
        \Gamma &= U &:\; b_0 &= (2,2)\,,\; b_1 = (1,-2) \,,\; b_{2\to7} = (4,8) \,.
    \end{aligned}\right. \notag
\end{align}
Examples very much like this have appeared previously in the literature (e.g.\ see section~(3.4) of~\cite{Kumar:2010am}).
As a final example of this strategy, consider
\begin{align}
    G &= \SO(8)\times\SU(4)^2\times G_2 \,, \notag\\
    \H &= (\triv;\triv,\rep[\prime]{20};\triv)\oplus (\triv;\triv,\triv;\rep{27})\oplus(\triv;\rep{4},\rep{4};\triv)\\
    &\qquad \oplus(\triv;\rep{6},\rep{6};\triv)\oplus (\triv;\triv,\rep{15};\rep{7})\oplus (\triv;\rep{4},\rep{4},\rep{7}) \,, \notag\\
    b_I\cdot b_J &= \begin{psmallmatrix}
         8 & -2 &  6 & 18 & 12 \\
        -2 & -4 &  0 &  0 &  0 \\
         6 &  0 &  4 & 12 &  8 \\
        18 &  0 & 12 & 36 & 24 \\
        12 &  0 &  8 & 24 & 16
    \end{psmallmatrix} \;\;\Longleftrightarrow\;\; \left\{\scalebox{0.85}{$\begin{aligned}
        \Gamma &= \Gamma_{1,1} \!\!\!&:\; b_0 &= (3,1)\,,\; b_1 = (0,2) \,,\; b_2=\tfrac{b_3}{3}=\tfrac{b_4}{2} = (2,0) \,,\\
        \Gamma &= U &:\; b_0 &= (2,2)\,,\; b_1 = (1,-2) \,,\; b_2=\tfrac{b_3}{3}=\tfrac{b_4}{2} = (1,2) \,.
    \end{aligned}$}\right. \notag
\end{align}
Higgsing $G_2\to \SU(2)^2$, under which $\rep{7}\to(\rep{2},\rep{2})\oplus(\rep{3},\triv)$, results in
\begin{align}
    G &= \SO(8)\times\SU(4)^2\times \SU(2)^2 \,, \notag\\
    \H &= (\triv;\triv^2;\triv^2)\oplus (\triv;\triv,\rep[\prime]{20};\triv^2) \oplus(\triv;\rep{4}^2;\triv^2)\oplus(\triv;\rep{6}^2;\triv^2)\\
    &\qquad \oplus (\triv;\triv^2;\rep{5},\triv)\oplus (\triv;\triv^2;\rep{2}^2)\oplus (\triv;\triv^2;\rep{3}^2) \notag\\
    &\qquad \oplus (\triv;\triv,\rep{15};\rep{3},\triv)\oplus (\triv;\triv,\rep{15};\rep{2}^2)\oplus (\triv;\rep{4}^2;\rep{3},\triv)\oplus (\triv;\rep{4}^2;\rep{2}^2) \,, \notag\\
    b_I\cdot b_J &= \scalebox{0.83}{$\begin{psmallmatrix}
         8 & -2 &  6 & 18 &  36 & 12 \\
        -2 & -4 &  0 &  0 &   0 &  0 \\
         6 &  0 &  4 & 12 &  24 &  8 \\
        18 &  0 & 12 & 36 &  72 & 24 \\
        36 &  0 & 24 & 72 & 144 & 48 \\
        12 &  0 &  8 & 24 &  48 & 16
    \end{psmallmatrix}$} \;\;\Longleftrightarrow\;\; \left\{\scalebox{0.83}{$\begin{aligned}
        \Gamma &= \Gamma_{1,1} \!\!\!&:\; b_0 &= (3,1)\,,\; b_1 = (0,2) \,,\; b_2=\tfrac{b_3}{3}=\tfrac{b_4}{6}=\tfrac{b_5}{2} = (2,0) \,,\\
        \Gamma &= U &:\; b_0 &= (2,2)\,,\; b_1 = (1,-2) \,,\; b_2=\tfrac{b_3}{3}=\tfrac{b_4}{6}=\tfrac{b_5}{2} = (1,2) \,.
    \end{aligned}$}\right. \notag
\end{align}
Additionally Higgsing the second $\SU(4)$ factor to $\SU(2)^2$, under which $\rep{4}\to(\rep{2},\rep{2})$, gives
\begin{align}
    G &= \SO(8)\times\SU(4)\times \SU(2)^4 \,, \notag\\
    \H &= 2(\triv;\triv;\triv^4)\oplus (\triv;\triv;\rep{5},\triv^3)\oplus (\triv;\triv;\triv,\rep{5},\triv^2)\oplus (\triv;\triv;\triv^2,\rep{5},\triv) \notag\\
    &\qquad \oplus(\triv;\rep{6};\rep{3},\triv^3)\oplus(\triv;\rep{6};\triv,\rep{3},\triv^2) \oplus (\triv;\triv;\triv^2,\rep{2}^2)\\
    &\qquad \oplus (\triv;\triv;\rep{3},\triv,\rep{3},\triv)\oplus (\triv;\triv;\triv,\rep{3}^2,\triv)\oplus (\triv;\triv;\triv^2,\rep{3}^2)\oplus (\triv;\triv;\rep{3}^3,\triv) \notag\\
    &\qquad \oplus (\triv;\triv;\rep{3},\triv,\rep{2}^2)\oplus (\triv;\triv;\triv,\rep{3},\rep{2}^2)\oplus (\triv;\triv;\rep{3}^2,\rep{2}^2) \notag\\
    &\qquad \oplus(\triv;\rep{4};\rep{2}^2,\triv^2) \oplus (\triv;\rep{4};\rep{2}^2,\rep{3},\triv)\oplus (\triv;\rep{4};\rep{2}^4) \,, \notag\\
    b_I\cdot b_J &= \scalebox{0.83}{$\begin{psmallmatrix}
         8 & -2 &  6 &  36_3 & 12 \\
        -2 & -4 &  0 &   0_3 &  0 \\
         6 &  0 &  4 &  24_3 &  8 \\
        36_3 &  0_3 & 24_3 & 144_{3\times3} & 48_3 \\
        12 &  0 &  8 &  48_3 & 16
    \end{psmallmatrix}$} \;\;\Longleftrightarrow\;\; \left\{\scalebox{0.83}{$\begin{aligned}
        \Gamma &= \Gamma_{1,1} \!\!\!&:\; b_0 &= (3,1)\,,\; b_1 = (0,2) \,,\; b_2=\tfrac{b_{3\to5}}{6}=\tfrac{b_6}{2} = (2,0) \,,\\
        \Gamma &= U &:\; b_0 &= (2,2)\,,\; b_1 = (1,-2) \,,\; b_2=\tfrac{b_{3\to5}}{6}=\tfrac{b_6}{2} = (1,2) \,.
    \end{aligned}$}\right. \notag
\end{align}
which is anomaly-free thanks to a very delicate mixture of hypermultiplets charged under up to five gauge factors.

\subsection{Bounding the quantum gravity landscape}
\label{sec:UV-conditions}

\begin{table}[p]
    \centering
    \small
    \newcommand{\mycol}{\multicolumn{1}{c|}{}}
    \newcommand{\myrow}[2]{\multicolumn{1}{c|}{\multirow{#1}{*}{#2}}}
    \begin{tabular}{crlrlc}
        \toprule
        $G$ & \multicolumn{1}{c}{$R$} & \multicolumn{1}{c}{$\lambda_i$} & \multicolumn{2}{c}{$\Delta_R$} & $\sum a_m^\vee\lambda_m$ \\ \midrule
        \\[-12pt]
        \myrow{4}{$\SU(N)$} & $\rep{N}$        &     $(1,0,\cdots,0)$ &  $\frac{N^2-1}{2N(k+N)}$ & $<1$    & $1$\\
        \mycol              & $\rep{N(N-1)/2}$ & $(0,1,0,\cdots,0)$ & $\frac{(N-2)(N+1)}{N(k+N)}$ & $<1$ & $1$\\
        \mycol              & $\rep{N(N+1)/2}$ & $(2,0,\cdots,0)$ &  $\frac{(N-1)(N+2)}{N(k+N)}$ & & $2$\\
        \mycol              & $\rep{N^2-1}$    & $(1,0,\cdots,0,1)$ & $\frac{N}{k+N}$ & $\leq 1$ &   $2$\\
        \\[-5pt]
        \myrow{3}{$\SO(2N)$}  & $\rep{2N}$    & $(1,0,\cdots,0)$ &  $\frac{2N-1}{2(k+2N-2)}$ & $<1$  &  $1$   \\
        \myrow{3}{$(N\geq4)$} 
        & $\rep{N(2N-1)}$ & $(0,1,0,\cdots,0)$ & $\frac{2N-2}{k+2N-2}$ & $\leq1$  & $2$ \\
        \mycol              & $\rep{(2N-1)(N+1)}$ & $(2,0,\cdots,0)$ & $\frac{2N}{k+2N-2}$ & & $2$ \\
        \mycol              & $\rep{2^{N-1}}$ & $(0,\cdots,\underline{0,1})$ & $\frac{N(2N-1)}{8(k+2N-2)}$ & &  $1$ \\
        \\[-5pt]
        \myrow{3}{$\SO(2N+1)$}  & $\rep{2N+1}$    &   $(1,0,\cdots,0)$   & $\frac{N}{k+2N-1}$ & $<1$ & $1$            \\
        \myrow{3}{$(N\geq3)$}              & $\rep{(2N+1)N}$ & $(0,1,0,\cdots,0)$ & $\frac{2N-1}{k+2N-1}$ & $\leq1$ &       $2$ \\
        \mycol           & $\rep{N(2N+3)}$ & $(2,0,\cdots,0)$ & $\frac{2N+1}{k+2N-1}$ & & $2$  \\
        \mycol              & $\rep{2^{N}}$ & $(0,\cdots,0,1)$ & $\frac{N(2N+1)}{8(k+2N-1)}$ & &  $1$ \\
        \\[-5pt]
        \myrow{3}{$\Sp(N)$} & $\rep{2N}$          &   $(1,0,\cdots,0)$    & $\frac{2N+1}{4(k+N+1)}$ & $<1$ &  $1$\\
        \mycol              & $\rep{(N-1)(2N+1)}$ & $(0,1,0,\cdots,0)$ & $\frac{N}{k+N+1}$ & $<1$ & $1$ \\
        \mycol              & $\rep{N(2N+1)}$     & $(2,0,\cdots,0)$ & $\frac{N+1}{k+N+1}$ & $\leq1$ &  $2$ \\
        \\[-5pt]
        \myrow{4}{$E_6$} & $\rep{27}$          &  $(1,0,\cdots,0)$  & $\frac{26}{3(k+12)}$ & $<1$ & $1$  \\
        \mycol           & $\rep{78}$          &  $(0,\cdots,0,1)$  & $\frac{12}{k+12}$ & $\leq1$ & $2$ \\
        \mycol           & $\rep{351}$         &  $(0,0,0,1,0,0)$  & $\frac{50}{3(k+12)}$ & & $2$ \\
        \mycol           & $\rep[\prime]{351}$ & $(0,0,0,0,2,0)$ & $\frac{56}{3(k+12)}$ & & $2$ \\
        \\[-5pt]
        \myrow{4}{$E_7$} & $\rep{56}$   &  $(0,\cdots,0,1,0)$  & $\frac{57}{4(k+18)}$ & $<1$ &   $1$ \\
        \mycol           & $\rep{133}$  &  $(1,0,\cdots,0)$  & $\frac{18}{k+18}$ & $\leq1$ & $2$ \\
        \mycol           & $\rep{912}$  & $(0,\cdots,0,1)$ & $\frac{105}{4(k+18)}$ & &  $2$ \\
        \mycol           & $\rep{1463}$ & $(0,\cdots,0,2,0)$ & $\frac{30}{k+18}$ & & $2$ \\
        \\[-5pt]
        \myrow{2}{$E_8$} & $\rep{248}$  & $(0,\cdots,0,1,0)$ & $\frac{30}{k+30}$ & $\leq1$ &  $2$\\
        \mycol           & $\rep{3875}$ & $(1,0,\cdots,0)$ & $\frac{48}{k+30}$ & & $2$\\
        \\[-5pt]
        \myrow{4}{$F_4$} & $\rep{26}$  & $(0,0,0,1)$ & $\frac{6}{k+9}$ & $<1$ & $1$  \\
        \mycol           & $\rep{52}$  & $(1,0,0,0)$ & $\frac{9}{k+9}$ & $\leq 1$ &  $2$ \\
        \mycol           & $\rep{273}$ & $(0,0,1,0)$ & $\frac{12}{k+9}$ & & $2$ \\
        \mycol           & $\rep{324}$ & $(0,0,0,2)$ & $\frac{13}{k+9}$ & & $2$ \\
        \\[-5pt]
        \myrow{4}{$G_2$} & $\rep{7}$  &  $(1,0)$  & $\frac{2}{k+4}$ & $<1$ & $1$  \\
        \mycol           & $\rep{14}$ &  $(0,1)$ & $\frac{4}{k+4}$ & $\leq 1$ & $2$ \\
        \mycol           & $\rep{27}$ & $(2,0)$ & $\frac{14}{3(k+4)}$ & & $2$ \\
        \mycol           & $\rep{64}$ & $(1,1)$ & $\frac{7}{k+4}$ & & $3$ \\
        \\[-12pt]
        \bottomrule
    \end{tabular}
    \caption{Some common, low-dimensional irreps and expressions for $\Delta_R$ and $\sum_{m=1}^r a_m^\vee \lambda_m$ in equations~\eqref{eq:Casimir_bounds} and~\eqref{eq:representation_bounds}. For some representations, the bound~\eqref{eq:representation_bounds} is satisfied independent of the value of the level $k$.}
    \label{tab:representation_bounds}
\end{table}

So far we have focused on the supergravity landscape, only having imposed those consistency conditions which can be understood from low-energy considerations alone. There are, however, additional requirements that arise from the UV which demonstrate that the string (or more generally, quantum gravity) landscape of consistent models is a strict subset of the anomaly-free models we have enumerated. In this section we explore the extent to which the following three criteria constrain the space of consistent models.
First, assuming completeness of the BPS spectrum, the string probe bound provided in~\cite{Kim:2019vuc,Kim:2019ths,Lee:2019skh,Tarazi:2021duw,Martucci:2022krl} states that
\begin{align}
\label{eq:cL_bound}
    \sum_{i=1}^\kappa \frac{k_i\dim G_i}{k_i+h^\vee_i}\leq c_L
\end{align}
must hold for all charges $Q\in\Gamma$ for which $c_L,c_R,k_\ell,k_i \geq 0$, where $h_i^\vee$ are dual-Coxeter numbers (see table~\ref{tab:lambda-constants}) and
\begin{eqnalign}
    c_L &= 3Q^2 + 9Q\cdot b_0 + 2 \,, & \quad k_\ell &= \tfrac{1}{2}(Q^2 - Q\cdot b_0 + 2) \,,\\
    c_R &= 3Q^2 + 3Q\cdot b_0 \,, & k_i &= Q\cdot b_i \,,
\end{eqnalign}
are central charges and levels for the current algebra hosted on the string.
Additionally,~\cite{Tarazi:2021duw} proposed two new criteria,
\begin{align}
    \Delta_R \defeq \frac{C_2(R)}{2(k_i+h_i^\vee)} &\leq 1, \label{eq:Casimir_bounds}
    \\
    \sum_{m=1}^r a_m^\vee \lambda_m &\leq k_i, \label{eq:representation_bounds}
\end{align}
where $C_2(R)$ is the quadratic Casimir of the representation $R$ of $G_i$, $a^\vee$ is the comark (e.g.\ see figure~14.1 in~\cite{DiFrancesco:1997nk}), and $\lambda$ is the Dynkin index. In table~\ref{tab:representation_bounds}, we summarize $\Delta_R$ and $\sum a_m^\vee \lambda_m$ for some common, low-dimensional irreducible representations. It is clear that~\eqref{eq:Casimir_bounds} is stronger than~\eqref{eq:representation_bounds} for low-dimensional representations. However, this is not the case for representations of larger dimension: for example the 3-index antisymmetric representation of $\SU(N)$ has
\begin{eqnalign}
    \Delta_R = \frac{3(N-3)(N+1)}{N(k_i+N)} \,, \qquad \sum a_m^\vee \lambda_m = 1 \,,
\end{eqnalign}
so that the bound~\eqref{eq:Casimir_bounds} reads $k_i \geq 2N - 6 - \frac{9}{N}$ and is stronger than~\eqref{eq:representation_bounds} for all $N\geq 5$.

\medskip

Let us first consider the ensemble of $T=0$ anomaly-free models. There, the brane probe bound of equation~\eqref{eq:cL_bound} is particularly straightforward to check since $\Gamma=\Z$ is one-dimensional. All $b_I$ are positive integers and it is easy to see that the brane probe bound reduces to
\begin{eqnalign}
\label{eq:cL_bound-T=0}
    \sum_{i=1}^\kappa \frac{Qb_i\dim G_i}{Qb_i + h_i^\vee} \leq 3Q^2 + 27Q + 2 \,, \qquad \forall\; Q\in\{0,1,2,\ldots\} \,.
\end{eqnalign}
An obvious consequence is that there is no constraint if $\sum_i\dim G_i\leq 32$. For simple gauge group ($\kappa=1$), a necessary and sufficient condition to satisfy~\eqref{eq:cL_bound-T=0} for all $Q\geq 0$ is simply
\begin{eqnalign}
\label{eq:k=1-T=0-brane-bound}
    \dim G_1 \leq 32\left(1 + \frac{h_1^\vee}{b_1}\right) \,.
\end{eqnalign}
It turns out that \emph{all} of the anomaly-free models with $\kappa=1$ satisfy this condition. Curiously, it is also easy to find examples of models which violate this condition but which are barely anomalous: for example, the admissible model with $G=\SU(14)$ and $\H^\ch=\rep{105}\oplus\rep{364}$ (``\texttt{vtx-A13-274-9-9}'' in the database~\cite{Loges:2023gh2}) violates the above inequality ($\dim G_1=195$, $h_1^\vee=14$, $b_1=3$) but has $H^\ch-V = 274 > 273$ too large to be anomaly-free by only one!
For $\kappa>1$ it suffices to check~\eqref{eq:cL_bound-T=0} out to only $Q=10$ since all anomaly-free models for $T=0$ have $\sum_i\dim G_i \leq 575$. Doing so, we find that \emph{no} anomaly-free models are inconsistent with the brane probe bound. Similarly, we find that equations~\eqref{eq:Casimir_bounds} and~\eqref{eq:representation_bounds} are met by all $T=0$ anomaly-free models.

\smallskip

Turning to the $T=1$ anomaly-free models, let us consider some necessary conditions which correspond to particular small choices of $Q$. Recall that we have either $\Gamma=\Gamma_{1,1}$ with $b_0=(3,1)$ or $\Gamma=U$ with $b_0=(2,2)$. Under the conditions $c_R\geq 0$ and $k_\ell\geq 0$, there are several small choices for $Q\in\Gamma$ which give relatively small $c_L>0$. We consider the effects of imposing the bounds for the following charges:
\begin{eqnalign}
    \Gamma &= \Gamma_{1,1} &:&\quad \begin{cases}
        Q = (0,-1) & c_L=8\,,\\
        Q = (1,1) & c_L=20\,,
    \end{cases}\\
    \Gamma &= U &:&\quad \hspace{11pt}Q = (\underline{1,0}) \hspace{19pt} c_L=20 \,.
\end{eqnalign}
Note that these small values of $c_L$ on the RHS of equation~\eqref{eq:cL_bound} are possible only if the corresponding charge gives $k_i\geq 0$ for all $i$. Larger charges give $c_L\geq 32$ and thus we should expect the above choices to be the most constraining.
\begin{itemize}
    \item $\Gamma=\Gamma_{1,1}$ and $Q=(0,-1)$:

    For odd lattices and this choice of charge, $k_i^{\tN{N}}=1$ is always positive but $k_i^{\tT{T}}$ and $k^{\tS{S}}$ can potentially have either sign.
    The number of the models excluded by the condition above is
    \begin{center}
    \scalebox{0.91}{
    \begin{tabular}{c*{11}{>$c<$}}
        \toprule
        \#(models excluded) & \multicolumn{11}{c}{$\kappa$}\\ \cmidrule{2-12}
         & 0 & 1 & 2 & 3 & 4 & 5 & 6 & 7 & 8 & 9 & 10 \\ \midrule
        w/ $\kappa_\tS{S}=0$ & 0 & 28 & 1,\!168 & 5,\!918 & 5,\!419 & 432 & 6 & 0 & 0 & 0 & 0
        \\
        w/ $\kappa_\tS{S}=1$ & 0 & \phantom{2}0  & \phantom{1,\!16}0 & \phantom{5,\!91}0 & \phantom{5,\!41}0 & \phantom{43}0 & 0 & 0 & 0 & 0 & 0
        \\
        total & 0 & 28 & 1,\!168 & 5,\!918 & 5,\!419 & 432 & 6 & 0 & 0 & 0 & 0
        \\
        \bottomrule
    \end{tabular}
    }
    \end{center}
    All models with a type-\tS{S} vertex ($\kappa_\tS{S}=1$) are free from this string probe bound because we require also the string probe have non-negative tension, $j\cdot Q\geq 0$.
    
    \item $\Gamma=\Gamma_{1,1}$ and $Q=(1,1)$:

    For odd lattices and this choice of charge, $k_i^{\tN{N}}$ is always zero and $k_i^{\tT{T}}$ is always positive since each $b_i^{\tT{T}}$ lies within the future light-cone. However, the sign of $k^{\tS{S}}$ is indefinite. We therefore have the following bound,
    \begin{eqnalign}
    \label{eq:odd-(1,1)-bound}
        \sum_{i\,:\,\tS{S},\tT{T}}\left.\frac{k_i\dim G_i}{k_i + h_i^\vee}\right|_{Q=(1,1)} \leq 20 \qquad\quad
        \begin{minipage}{6.5cm}
            for models with $\Gamma=\Gamma_{1,1}$ and without\\
            $b^{\tS{S}}=(-2,4),(-1,3),(0,2),(0,1)$,
        \end{minipage}
    \end{eqnalign}
    with the excluded type-\tS{S} vectors ensuring that all $k_i$ are non-negative. Even though this is a non-trivial condition to be satisfied, curiously we find that it excludes only \emph{one} anomaly-free model, namely,
    \begin{align}
        G &= \Sp(6)\times\Sp(4)\times\SO(10)^4 \,, \notag\\
        \H &= (\triv,\rep{42};\triv^4) \oplus \big[\tfrac{1}{2}(\rep{12},\triv;\rep{10},\triv^3)\oplus(\text{3 others})\big] \oplus \big[\tfrac{1}{2}(\triv,\rep{8};\rep{16},\triv^3)\oplus(\text{3 others})\big] \,, \notag\\
        \!\!b_I\cdot b_J &= \scalebox{0.95}{$\begin{psmallmatrix}
            8 & 1 & 6 & 2_4 \\
            1 & -1 & 0 & 1_4 \\
            6 & 0 & 4 & 2_4 \\
            2_4 & 1_4 & 2_4 & 0_{4\times 4}
        \end{psmallmatrix}$} \;\;\Longleftrightarrow\;\; \Gamma = \Gamma_{1,1} \,:\;\; \scalebox{0.95}{$\begin{aligned}
            b_0 &= (3,1) \,,\; b_1 = (0,-1) \,,\\[-3pt]
            b_2 &= (2,0) \,,\; b_{3\to6} = (1,1) \,,
        \end{aligned}$}
    \end{align}
    for which the LHS of~\eqref{eq:odd-(1,1)-bound} is $\frac{1\times 78}{1+7}+\frac{2\times 36}{2+5}+4\times\frac{0\times45}{0+8} = 20\!+\!\frac{1}{28}$.
    
    \item $\Gamma=U$ and $Q=(\underline{1,0})$:

    For even lattices and this choice of charge, $k_i^{\tN{N}}$ is either zero or one and $k_i^{\tT{T}}$ is positive. In addition, for any type-\tS{S} vector exactly one of $Q=(1,0)$ and $Q=(0,1)$ gives $k^{\tS{S}}>0$. Therefore we have the following bound,
    \begin{eqnalign}
    \label{eq:even-(1,0)-bound}
        \sum_i\left.\frac{k_i\dim G_i}{k_i + h_i^\vee}\right|_{Q=(\underline{1,0})} \leq 20 \qquad\quad
        \text{for models with }\Gamma=U \,,
    \end{eqnalign}
    with $Q$ chosen appropriately if a type-\tS{S} vertex is present. Again, despite this being a non-trivial condition we find that it excludes only \emph{one} anomaly-free model,
    \begin{eqnalign}
        G &= \SO(10)\times\SU(8)\times\SU(6) \,,\\
        \H &= (\triv,\rep{56},\triv)\oplus(\triv,\triv,\rep{35})\oplus(\rep{10},\rep{8},\triv)\oplus(\rep{16},\triv,\rep{6})\oplus(\triv,\rep{8},\rep{15}) \,,\\
        b_I\cdot b_J &= \begin{psmallmatrix}
            8 & 4_2 & 8\\
            4_2 & 2_{2\times 2} & 4_2\\
            8 & 4_2 & 8
        \end{psmallmatrix} \;\;\Longleftrightarrow\;\; \Gamma=U \,:\; b_0=(2,2) \,,\; b_{1,2}=(1,1) \,,\; b_3=(2,2) \,,
    \end{eqnalign}
    for which the LHS of~\eqref{eq:even-(1,0)-bound} is $\frac{1\times 45}{1+8} + \frac{1\times 63}{1+8} + \frac{2\times35}{2+6} = 20\!+\!\tfrac{3}{4}$.
\end{itemize}
We see that the models identified as inconsistent by the string probe bound nearly all have odd charge lattice. Some models for which both even and odd lattice are allowed by anomaly cancellation have the odd lattice forbidden by the above bounds. For example, the $E_8$ NHC violates the brane probe bound if one takes $\Gamma=\Gamma_{1,1}$, despite being anomaly-free.

\medskip

In contrast to what we find for the string probe bound, the conditions of equations~\eqref{eq:Casimir_bounds} and~\eqref{eq:representation_bounds} are quite strong. For example, equation~\eqref{eq:representation_bounds} alone excludes all models which contain a type-\tS{S} vertex with $b_i^2=b_0\cdot b_i=-1$; by choosing $Q=(Q_0,-1)$ with $Q_0>0$ sufficiently large we have $k^\tS{S}=1$ and all $k^{\tN{N}},k^{\tT{T}}>0$, but this is incompatible with there being a symmetric representation of $\SU(N)$.\footnote{See also section~4 of \cite{Tarazi:2021duw} for the bounds with general $T$. This also agrees with Claim 4.1 in \cite{Morrison:2023hqx}. We thank Hee-Cheol Kim for a comment.} This aligns nicely with section~6.3 of~\cite{Kumar:2010ru} where it was shown that these models have no F-theory realization. In this way we exclude $81,\!484$ models ($\approx\! 13.4\%$).

From equation~\eqref{eq:Casimir_bounds} it is possible to rule out individual vertices and thus kill all anomaly-free models which utilize them. For example, equation~\eqref{eq:Casimir_bounds} excludes the $b_i^2=-1$, $b_0\cdot b_i=1$ type-\tS{S} vertex with $\SU(6)$ gauge group and $15\times\rep{6}\oplus\tfrac{1}{2}\rep{20}$ charged hypermultiplets. The existence of the $\rep{20}$ representation requires $k^{\tS{S}}\geq 9/2$, but $Q=(1,1)$ gives $k^{\tS{S}}=1$. This excludes $8,\!476$ models ($\approx\! 1.4\%$).
Similarly, the type-\tN{N} vertex with $\SU(6)$ gauge group and $17\times\rep{6}\oplus\rep{15}\oplus\tfrac{1}{2}\rep{20}$ charged hypermultiplets can sometimes be forbidden: choosing $Q=(2,1)$ (odd lattice) or $Q=(1,1)$ (even lattice) leads to a violation of~\eqref{eq:Casimir_bounds} and is a valid choice for string charge in the absence of type-\tS{S} vertices with $b_i^2\leq-4$ or $b_i^2=b_0\cdot b_i=-1$.\footnote{It turns out that all the anomaly-free models with this type-\tN{N} vertex are free from type-\tS{S} vertices with $b_i^2\leq-4$ or $b_i^2=b_0\cdot b_i=-1$. Therefore, all the anomaly-free models with this type-\tN{N} vertex are excluded.} This condition excludes $70,\!707$ models ($\approx\! 11.6\%$).
All the other type-\tS{S} and type-\tN{N} vertices have both $\Delta_R\leq1$ and $\sum_{m=1}^r a_m^\vee \lambda_m\leq1$. Consequently, there are no further bounds from equations~\eqref{eq:Casimir_bounds} and~\eqref{eq:representation_bounds} for these vertices. By imposing these two bounds on the representation of type-\tT{T} vertices, we exclude the following numbers of models:
\begin{center}
    \scalebox{0.91}{
    \begin{tabular}{c*{11}{>$c<$}}
        \toprule
        \#(models excluded) & \multicolumn{11}{c}{$\kappa$}\\ \cmidrule{2-12}
         & 0 & 1 & 2 & 3 & 4 & 5 & 6 & 7 & 8 & 9 & 10 \\ \midrule
        w/ $\kappa_\tS{S}=0$ & 0 & 152 & 3,\!445 & 13,\!828 & 13,\!705 & \phantom{2}4,\!746 & \phantom{2,}593 & \phantom{2}25 & 0 & 0 & 0\\
        w/ $\kappa_\tS{S}=1$ & 0 & 24  & 5,\!412 & 35,\!008 & 58,\!268 & 33,\!223 & 4,\!976 & 416 & 26 & 0 & 0
        \\
        total & 0 & 176 & 8,\!857 & 48,\!836 & 71,\!973 & 37,\!969 & 5,\!569 & 441 & 26 & 0 & 0
        \\
        \bottomrule
    \end{tabular}
    }
\end{center}
All told, the number of models which survive the above bounds is as follows:
\begin{center}
    \scalebox{0.55}{
    \begin{tabular}{c*{11}{>$c<$}}
        \toprule
        \multirow{2}{*}{$\dfrac{\text{\#(survived)}}{\text{\#(models)}}$} & \multicolumn{11}{c}{$\kappa$}\\ \cmidrule{2-12}
         & 0 & 1 & 2 & 3 & 4 & 5 & 6 & 7 & 8 & 9 & 10 \\ \midrule
        $\kappa_\tS{S}=0$ & 1/1 & 890/1,\!042 & 12,\!512/15,\!957 & \phantom{2}32,\!667/\phantom{2}46,\!495 & \phantom{2}24,\!090/\phantom{2}37,\!795 & \phantom{2}7,\!973/\phantom{2}12,\!719 & \phantom{2}1,\!020/\phantom{2}1,\!613 & \phantom{2,\!2}70/\phantom{2,\!2}95 & \phantom{22}2/\phantom{22}2 & \phantom{1}0/\phantom{2}0 & 0/0\\
        $\kappa_\tS{S}=1$ & 0/0 & \phantom{2}94/\phantom{2,}118 & 21,\!521/26,\!933 & 109,\!979/144,\!987 & 143,\!854/202,\!122 & 67,\!638/100,\!861 & 10,\!712/15,\!687 & 1,\!302/1,\!718 & 169/195 & 14/14 & 1/1\\
        total & 1/1 & 984/1,\!160 & 34,\!033/42,\!890 & 142,\!646/191,\!482 & 167,\!944/239,\!917 & 75,\!611/113,\!580 & 11,\!732/17,\!300 & 1,\!372/1,\!813 & 171/197 & 14/14 & 1/1
        \\
        \bottomrule
    \end{tabular}
    }
\end{center}

\section{Discussion}
\label{sec:discussion}

In this work we have performed an exhaustive enumeration of anomaly-free models for 6D, $\mathcal{N}=(1,0)$ supergravity with $T\leq 1$, the only simplifying assumption being that the gauge group contains no $\U(1)$, $\SU(2)$, $\SU(3)$ or $\Sp(2)$ factors. In particular, we saw how a natural generalization of previous graph-theoretic techniques to multi-hypergraphs and simplicial complexes allows for general hypermultiplet representations. Implementing these ideas for $T=0$, we find exactly $20$ models with hypermultiplets charged under three or more gauge factors, most of which can be related via Higgsing. For $T=1$ there are $608,\!355$ anomaly-free models, with the number of simple gauge factors reaching as high as ten. We have made the data for this ensemble available at~\cite{Loges:2023gh2}, here highlighting some choices, extreme examples of models with large gauge groups (both dimension and rank), large values of $\Delta(\v)$ for individual vertices or notable hypermultiplet representations. As for $T=0$, very few models in this ensemble make use of hypermultiplets charged under three or more gauge factors, but we exemplify a general strategy which produces anomaly-free models with hypermultiplets charged under four or more gauge factors, necessarily including $\SU(2)$, $\SU(3)$ or $\Sp(2)$.

Our findings suggest that extending this exhaustive enumeration of anomaly-free models to include $\SU(2)$, $\SU(3)$ or $\Sp(2)$ or to $T\geq 2$ is currently infeasible. These three low-rank groups lead to a multi-hypergraph of unwieldy size due to their large number of low-dimension representations. Similarly, even for $T=2$ vertices are less constrained because the lattice $\Lambda=\Z^{1,2}$ is so much larger, and (extrapolating from figure~\ref{fig:vertex-usage}) we should expect to have to go out to $\Delta(\v)\gtrsim 900$ in order to capture all relevant vertices.

We also considered the effects of imposing three additional UV bounds, one bound which arises from the consistency of string probes and two proposed bounds which restrict the allowed hypermultiplet representations. Somewhat curiously, we find that anomaly cancellation is strong enough for $T=0$ so that \emph{all} models are consistent with the UV bounds we impose, suggesting that the field theory and quantum gravity landscapes may coincide for $T=0$. For $T=1$ they definitely do not coincide: all told we were able to rule out $\mathcal{O}(50\%)$ of anomaly-free models as inconsistent, with the string probe bound disproportionately killing models with odd charge lattice. Using the other two bounds, we are also able to provide an independent argument why a certain family of anomaly-free models with particular type-\tS{S} vertices, which had previously been shown to have no F-theory realization, is inconsistent with quantum gravity.

It is an interesting question to determine precisely how many/which of the models we have not been able to rule out as inconsistent actually have a realization in string theory. As mentioned above, some of the ``maximal'' examples of section~\ref{sec:Higgsing-and-examples} can be realized non-geometrically in the heterotic string using asymmetric orbifolds. It has also been shown that a large class of models have a geometric F-theory description~\cite{Kumar:2009ac}. Whether these two construction methods encompass all consistent models remains to be seen.

\acknowledgments
We thank Hee-Cheol Kim, Houri-Christina Tarazi and Cumrun Vafa for useful discussions and the Harvard Swampland Initiative, where some of this work was completed, for their hospitality.
The work of Y.H.\ and G.L.\ is supported in part by MEXT Leading Initiative for Excellent Young Researchers Grant Number JPMXS0320210099.
The work of Y.H. is also supported in part by JSPS KAKENHI Grant No.24H00976 and 24K07035.

\appendix
\section{Unimodular lattices of signature \texorpdfstring{$(1,1)$}{(1,1)}}
\label{app:lattices}

There are two unimodular lattices of signature $(1,1)$, both consisting of points $\Z^2$ but with different inner product $\Omega$: the odd lattice $\Gamma_{1,1}$ has $\Omega=\diag(1,-1)$ and the even lattice $U$ has $\Omega=\begin{psmallmatrix}
    0 & 1\\ 1 & 0
\end{psmallmatrix}$. For $\Gamma_{1,1}$, $b_0\cdot b_0=8$ uniquely determines $b_0=(3,1)$ up to discrete symmetries of the lattice. For $U$, there are two inequivalent choices: $b_0=(2,2)$ and $b_0=(1,4)$. However, the latter is superfluous for the following reason: since $b_0\cdot b_i\equiv b_i\cdot b_i\mod{2}$ for all $b_i$, we know that each vector takes the form $b_i=(u_i,2v_i)$, but the sub-lattice of vectors of this shape in $U$ is isomorphic to a different sub-lattice of index two for which $b_0=(2,2)$ through $\varphi\big((u,2v)\big)=(2u,v)$. Therefore if a choice for the $b_I\in U$ has $b_0=(1,4)$ there is always a different choice for which $b_0=(2,2)$ is a characteristic vector of the lattice. In~\cite{Monnier:2018nfs,Monnier:2018cfa} it was shown that $b_0=(1,4)$ can be ruled out from a bottom-up perspective by requiring that the model be well-defined on any spin manifold. By the above argument we see that this does not actually rule out any models since one can always pick $b_0=(2,2)$ instead.

For both of the two lattices we can determine each vector $b_i$ from $b_i\cdot b_i$ and $b_0\cdot b_i$ alone. First of all, it must be that $8b_i\cdot b_i - (b_0\cdot b_i)^2 = -d^2$ for some $d\in\Z$ since $b_0\Z \oplus b_i\Z$ is a sublattice of a unimodular lattice for each $i$. A straightforward calculation yields the following conditions and solutions:
\begin{itemize}
    \item $\Gamma_{1,1}$ with $b_0=(3,1)$: $b_i$ exists iff $b_0\cdot b_i\equiv\pm 3d\mod{8}$, in which case
    \begin{eqnalign}
        b_0\cdot b_i &\equiv +3d\mod{8} &\Longrightarrow\;\; b_i &= \big(\tfrac{3b_0\cdot b_i-d}{8},\tfrac{b_0\cdot b_i - 3d}{8}\big) \,,\\
        b_0\cdot b_i &\equiv -3d\mod{8} &\Longrightarrow\;\; b_i &= \big(\tfrac{3b_0\cdot b_i+d}{8},\tfrac{b_0\cdot b_i + 3d}{8}\big) \,.
    \end{eqnalign}
    There are cases where both conditions are met and $b_i$ is not uniquely determined. For example, $b_i\cdot b_i=24$, $b_0\cdot b_i=16$ gives either $b_i=(5,-1)$ or $b_i=(7,5)$.

    \item $U$ with $b_0=(2,2)$: $b_i$ exists iff $b_i\cdot b_i\equiv 0\mod{2}$ and $b_0\cdot b_i\equiv d\mod{4}$, in which case
    \begin{equation}
        b_i = \big(\tfrac{b_0\cdot b_i + d}{4},\tfrac{b_0\cdot b_i - d}{4}\big) \quad\text{or}\quad b_i = \big(\tfrac{b_0\cdot b_i - d}{4},\tfrac{b_0\cdot b_i + d}{4}\big) \,.
    \end{equation}
    These are clearly related by parity.
\end{itemize}

\section{Multi-hypergraph construction}
\label{app:graph-construction}

In our analysis one of our goals is be comprehensive when it comes to the allowed hypermultiplet representations. However, for small-rank simple groups this poses a computational problem because the number of irreducible representations with dimension less than $x$ grows very quickly with $x$, leading to a proliferation of vertices even with only moderate $\Delta(\v)$ and a multi-hypergraph $\G$ of unwieldy size. In practice, we find that $\G$ can be constructed for $T\leq 1$ for the groups $\SU(N\geq 5)$, $\SO(N\geq 7)$, $\Sp(N\geq 4)$, $E_N$, $F_4$ and $G_2$ without any special considerations, while $\SU(4)$ and $\Sp(3)$ come within reach with some additional ideas which we discuss below. $\SU(2)$, $\SU(3)$ and $\Sp(2)$ appear to be well out of reach without limiting the allowed hypermultiplet representations by hand. In addition to contributing a large number of vertices, these groups allow for many more non-trivial edges because of their low-dimension fundamental representations.

For $\SU(4)$ and $\Sp(3)$, generating all vertices up to $\Delta(\v)\lesssim 700$ for $T\leq 1$ is possible. We find that together these contribute $\V_{\tT{T}}=\mathcal{O}(1,\!000,\!000)$ vertices to $\G$, compared to $\mathcal{O}(50,\!000)$ for \emph{all} higher-rank groups combined. This would lead to $\E=\mathcal{O}(\V_{\tT{T}}^2)=\mathcal{O}(10^{12})$ hyperedges since $\G$ is not sparse, which is well out of our reach to manage. Instead, the approach we take is the following. After having generated vertices for $\G$ out to sufficiently large $\Delta(\v)$, we build all edges \emph{except} for those between type-\tT{T} vertices. Since the number of type-\tS{S} and type-\tN{N} vertices is quite small, the number of such edges is only $\mathcal{O}(N_G\V_{\tT{T}})$ where $N_G$ is the number of simple groups being considered. Then for each type-\tT{T} vertex $\v_{\tT{T}}$ we construct all $\mathcal{O}(N_G)$ models incorporating this vertex and a type-\tS{S} vertex. The vertex $\v_{\tT{T}}$ is then removed from $\G$ if \emph{all} of these should be pruned, i.e.\ have RHS of equation~\eqref{eq:delta-ineq} larger than $273$. Having ``pre-pruned'' the vertices of $\G$ in this way, we are left with $\V_{\tT{T}}=\mathcal{O}(20,\!000)$ vertices and completing the construction of $\G$ by building the $\E=\mathcal{O}(\V_{\tT{T}}^2)=\mathcal{O}(10^8)$ hyperedges is within reach. In figure~\ref{fig:vertex-usage} only vertices which survive this procedure are shown.

\medskip

We can understand the huge number of vertices for $\SU(2)$, $\SU(3)$ and $\Sp(2)$ even with reasonably small $\Delta(\v)$ in the following way. There are always collections of hypermultiplets which can be interchanged without changing any of the data relevant to anomaly-cancellation, i.e.\ the values of $b_i\cdot b_i$ and $b_0\cdot b_i$. These can be read off from the various tensor product decompositions for conjugate representations (recall that $R$ and $\overline{R}$ are indistinguishable when it comes to anomaly-cancellation). For example, the sums over all of $H_R$, $A_R$, $B_R$ and $C_R$ are the same for both sides of
\begin{eqnalign}
    \triv\oplus\rep{8} \quad\longleftrightarrow\quad \rep{3}\oplus\rep{6} \,,
\end{eqnalign}
for $\SU(3)$, which is clear from comparing $\rep{3}\otimes\repbar{3}=\triv\oplus\rep{8}$ and $\rep{3}\otimes\rep{3}=\repbar{3}\oplus\rep{6}$. These small-rank groups have many such replacement rules because they have many representations of small dimension. For $\SU(3)$ the list includes
\begin{eqnalign}
\label{eq:SU(3)-replacements}
    \triv\oplus\rep{8} \quad&\longleftrightarrow\quad \rep{3}\oplus\rep{6} \,, \\
    \rep{3}\oplus\rep{15} \quad&\longleftrightarrow\quad \rep{8}\oplus\rep{10} \,, \\
    \triv\oplus\rep{8}\oplus\rep{27} \quad&\longleftrightarrow\quad \rep{6}\oplus\rep{15}\oplus\rep[\prime]{15} \,, \\
    \rep{6}\oplus\rep{24} \quad&\longleftrightarrow\quad \rep{15}\oplus\rep[\prime]{15} \,, \\
    \rep{3}\oplus\rep{42} \quad&\longleftrightarrow\quad \rep{21}\oplus\rep{24} \,,
\end{eqnalign}
as well as any sums or differences of these (e.g.\ from the first two lines it is easy to see that we also have the replacement rule $\triv\oplus\rep{15}\longleftrightarrow\rep{6}\oplus\rep{10}$). In contrast, such replacements for larger groups are few and far between: for example the first two for $\SU(8)$ are
\begin{eqnalign}
    \triv\oplus\rep{63} \quad&\longleftrightarrow\quad \rep{28}\oplus\rep{36} \,,\\
    \rep{8}\oplus\rep{216} \quad&\longleftrightarrow\quad \rep{56}\oplus\rep{168} \,.
\end{eqnalign}
These replacements can be used repeatedly to transform one vertex into another as long as the multiplicity of charged hypermultiplets remains non-negative. The changes are more than just cosmetic, since they have a real effect on the possible edges a vertex can form. For example, there following four $\SU(3)$ vertices $\v_1$, $\v_2$, $\v_3$ and $\v_4$,
\begin{eqnalign}
    \H(\v_1) &= 58\times\rep{3} \oplus 3\times\rep{6} \,,\\
    \H(\v_2) &= 57\times\rep{3} \oplus 2\times\rep{6}\oplus \rep{8} \,,\\
    \H(\v_3) &= 56\times\rep{3} \oplus \rep{6}\oplus 2\times\rep{8} \,,\\
    \H(\v_4) &= 55\times\rep{3} \oplus 3\times\rep{8} \,,
\end{eqnalign}
all have $b_i\cdot b_i=15$ and $b_0\cdot b_i=11$ and are related via the replacements of~\eqref{eq:SU(3)-replacements}. However, $\v_2$ and $\v_3$ each only have one self-edge (with bi-charged hypers $15(\rep{3},\rep{3})$), whereas $\v_1$ and $\v_4$ each have three self-edges (with bi-charged hypers $15(\rep{3},\rep{3})$, $10(\rep{3},\rep{3})\oplus(\rep{3},\rep{6})$ and $5(\rep{3},\rep{3})\oplus(\rep{3},\rep{6})\oplus(\rep{6},\rep{3})$ for $\v_1$, and similarly for $\v_4$).

\section{Bounds and pruning condition}
\label{app:bounds}

In this appendix all statements only apply when $T\leq 1$ and $\U(1)$, $\SU(2)$, $\SU(3)$ and $\Sp(2)$ are omitted. Crucially, for the groups which remain all non-trivial representations have $H_R\geq 3$ (saturated only by $\tfrac{1}{2}\rep{6}$ of $\Sp(3)$).

\begin{table}[t]
    \centering
    \begin{tabular}{>$l<$>$l<$>$r<$>{$=}l<$}
        \toprule
        a     & g_a(z)                                       & z_a^\ast                        \\ \midrule
        0,1,2 & 56                                           & 1                               \\
        3     & \min\{56,\, 91-105z\}                        & \tfrac{13}{15} & 0.8666\ldots   \\
        4,5   & \min\{56,\, 61-64z,\, 91-105z\}              & \tfrac{13}{15} & 0.8666\ldots   \\
        6     & \min\{56,\, 61-64z,\, 83-98z,\, 93-108z\}    & \frac{31}{36}  & 0.86111\ldots  \\
        7     & 56-70z                                       & \tfrac{4}{5}   & 0.8            \\
        8,\ldots,15    & \min\{56-70z,\,68-96z\}             & \tfrac{17}{24} & 0.708333\ldots \\
        16,\ldots,25   & \min\{56-70z,\,94-160z\}            & \tfrac{47}{80} & 0.5875         \\
        26             & \min\{56-70z,\,78-130z,\,94-160z\}  & \tfrac{47}{80} & 0.5875         \\
        27             & \min\{56-70z,\,84-162z\}            & \tfrac{14}{27} & 0.518518\ldots \\
        28,\ldots,189  & \min\{56-70z,\,91-224z\}            & \tfrac{13}{32} & 0.40625        \\
        190,\ldots,209 & \min\{56-70z,\,91-224z,\,251-650z\} & \tfrac{13}{32} & 0.40625        \\
        210,\ldots,230 & \min\{56-70z,\,91-224z,\,253-693z\} & \tfrac{13}{32} & 0.40625        \\
        231,\ldots,252 & \min\{56-70z,\,91-224z,\,254-737z\} & \tfrac{13}{32} & 0.40625        \\
        253,\ldots,275 & \min\{56-70z,\,91-224z,\,254-782z\} & \tfrac{13}{32} & 0.40625        \\
        276,277,\ldots & \min\{56-70z,\,91-224z,\,253-828z\} & \tfrac{13}{32} & 0.40625        \\
        \bottomrule
    \end{tabular}
    \caption{Data for the bounds of equation~\eqref{eq:ga(z)-defn}.}
    \label{tab:delta-bounds}
\end{table}

\medskip

Using the notation introduced in section~\ref{sec:branch-and-prune}, we claim that, for each $a\in\{0,1,2,\ldots\}$,
\begin{eqnalign}
\label{eq:ga(z)-defn}
    \inf_{\substack{\K^{\tN{N}\tT{T}}\\\text{admissible}}} \Delta\big(\K^{\tN{N}\tT{T}};z,\calR[a,3]\big) = \begin{cases}
        g_a(z) & z\in[0,z_a^\ast] \,,\\
        -\infty & z\in(z_a^\ast,1] \,,
    \end{cases}
\end{eqnalign}
for the concave, piece-wise linear functions $g_a(z)$ and corresponding constants $z_a^\ast$ listed in table~\ref{tab:delta-bounds}. These bounds were determined empirically, as we describe further below. The models which set $g_a$ and $z_a^\ast$ turn out to be quite simple:
\begin{eqnalign}
    G &= G_2^\kappa \,, &\quad \H^\ch &= (10\times\rep{7})^\kappa \,, &\quad a &\geq 0 \,,\\
    &= \Sp(3)^\kappa \,, & &= (\tfrac{35}{2}\rep{6}\oplus \tfrac{1}{2}\rep[\prime]{14})^\kappa \,, & &\in\{3,4,5\} \,,\\
    &= \SU(4)^\kappa \,, & &= (16\times\rep{4}\oplus 2\times\rep{6})^\kappa \,, & &\in\{4,5,6\} \,,\\
    &= \SU(6)^\kappa \,, & &= (18\times\rep{6} \oplus \rep{20})^\kappa & &= 6 \,,\\
    &= \SU(4) \,, & &= 20\times\rep{4} \oplus 3\times\rep{6} & &= 6 \,,\\
    &= \SO(8)^\kappa \,, & &= (4\times\repss{8}{v} \oplus 4\times\repss{8}{c} \oplus 4\times\repss{8}{c})^\kappa & &\in\{8,\ldots,15\} \,,\\
    &= \SO(12)^\kappa \,, & &= (8\times\rep{12}\oplus 2\times\rep{32})^\kappa \,, & &\in\{16,\ldots,26\} \,,\\
    &= F_4^\kappa \,, & &= (5\times\rep{26})^\kappa \,, & &= 26 \,,\\
    &= E_6^\kappa \,, & &= (6\times\rep{27})^\kappa \,, & &= 27 \,,\\
    &= E_7^\kappa \,, & &= (4\times\rep{56})^\kappa \,, & &\geq 28 \,,\\
    &= \SU(20) \,, & &= 4\times\rep{20} \oplus 3\times\rep{190} & &\in\{190,\ldots,209\}\,, \\
    &= \SU(21) \,, & &= 3\times\rep{21} \oplus 3\times\rep{210} & &\in\{210,\ldots,230\}\,, \\
    &= \SU(22) \,, & &= 2\times\rep{22} \oplus 3\times\rep{231} & &\in\{231,\ldots,253\}\,, \\
    &= \SU(23) \,, & &= \rep{23} \oplus 3\times\rep{254} & &\in\{254,\ldots,275\}\,, \\
    &= \SU(24) \,, & &= 3\times\rep{276} & &\geq 276 \,.
\end{eqnalign}
Many of the above models consist of taking $\kappa$ copies of a type-\tN{N} vertex, and these are what set the values of $z_a^\ast$ since as soon $z$ is large enough so that $\Delta(\v^\tN{N};z,\calR[a,3])$ is negative, $\Delta(\K^{\tN{N}\tT{T}};z,\calR[a,3])$ can be made arbitrarily negative by taking $\kappa\to\infty$.

\smallskip

\begin{figure}[t]
    \centering
    \includegraphics[width=\textwidth]{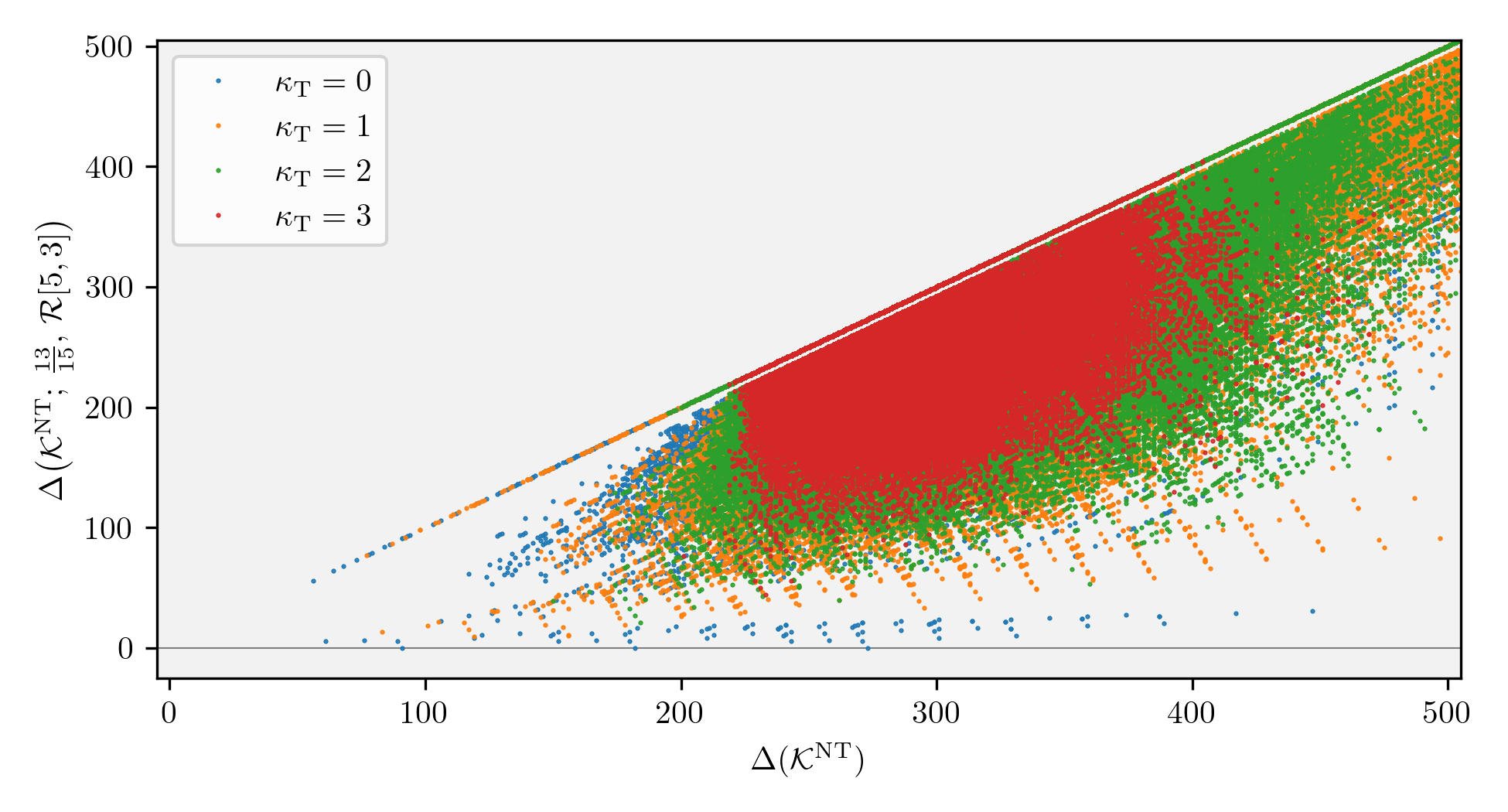}
    \caption{Values of $\Delta(\K^{\tN{N}\tT{T}};z,\calR[a,3])$ for $z=\frac{13}{15}$ and $a=5$ for the $\mathcal{O}(10^6)$ admissible models generated using vertices up to $\Delta(\v)\leq\max\{244,\dim G(\v)\}$ and \emph{without} pruning. From these data one may read off $g_5(z)$.}
    \label{fig:TN-model-bound}
\end{figure}

The functions $g_a(z)$ and constants $z_a^\ast$ were determined in the following way. First, a multi-hypergraph $\G$ was build using only type-\tN{N} and type-\tT{T} vertices with $\Delta(\v)\leq \max\{244,\dim G(\v)\}$, keeping $\G$ relatively small. Then, \emph{all} admissible models with up to three vertices were build using the usual branching procedure but with \emph{no} pruning, i.e.\ all admissible models were kept regardless of how large their value of $\Delta(\K)$.

From this subset of all admissible models one may empirically determine the data of table~\ref{tab:delta-bounds}. Although this does not comprise a minimization over all admissible models as required, the bounds are satisfied by a wider-and-wider margin as the number of type-\tT{T} vertices in the models increases. For example, figure~\ref{fig:TN-model-bound} shows the bound in effect for $a=5$: only models with one or fewer type-\tT{T} vertices come anywhere close to saturating the bound and it is easy to prove that the bounds hold for models with \emph{only} type-\tN{N} vertices. With enough case-work we suspect that one may be able to prove that the bounds hold unequivocally.

\section{Automated Higgsing}
\label{app:Higgsing}

As a non-trivial check of our results (both the code and the bounds discussed in appendix~\ref{app:bounds}), we have checked that the ensembles of anomaly-free models for both $T=0$ and $T=1$ are closed under Higgsing. By this we just mean the group-theoretic process of picking a maximal subgroup and computing the resulting hypermultiplet spectrum; we make no attempt at a detailed analysis of the scalar potential for any of the models. In this section we outline some of the details of this procedure.

Given a set of hypermultiplets $\H$ falling into representations of the gauge group $G$, the hypermultiplets $\H'$ which result upon breaking to a subgroup $G'\subset G$ are determined by
\begin{eqnalign}
    \H' = \varphi(\H) \oplus (-1)\varphi(\adj_G) \oplus \adj_{G'}
\end{eqnalign}
where $\varphi$ gives the branching rules. For example, $\Sp(7)$ has a maximal subgroup $\Sp(3)\times\Sp(4)$ for which the branching rules include
\begin{eqnalign}
    \varphi(\triv) &= (\triv,\triv) \,,\\
    \varphi(\rep{14}) &= (\rep{6},\triv)\oplus(\triv,\rep{8}) \,,\\
    \varphi(\rep{90}) &= (\triv,\triv)\oplus(\rep{14},\triv)\oplus(\triv,\rep{27})\oplus(\rep{6},\rep{8}) \,,\\
    \varphi(\rep{105}) &= (\rep{21},\triv)\oplus(\triv,\rep{36})\oplus(\rep{6},\rep{8}) \,,
\end{eqnalign}
and using these on the $T=1$ anomaly-free model with
\begin{equation}
    G = \Sp(7) \,, \qquad \H = 35\times\triv\oplus 16\times\rep{14}\oplus \rep{90} \,,
\end{equation}
results in the anomaly-free model with $G'=\Sp(3)\times\Sp(4)$ and hypermultiplets given by
\begin{eqnalign}
    \H' &= \big[35\varphi(\triv) \oplus 16\varphi(\rep{14}) \oplus \varphi(\rep{90})\big] \oplus (-1)\varphi(\rep{105}) \oplus \big[ (\rep{21},\triv)\oplus(\triv,\rep{36}) \big]\\
    &= 35(\triv,\triv)\oplus 16\big[(\rep{6},\triv)\oplus(\triv,\rep{8})\big] \oplus \big[(\triv,\triv)\oplus(\rep{14},\triv)\oplus(\triv,\rep{27})\oplus(\rep{6},\rep{8})\big]\\
    &\qquad \oplus(-1)\big[ (\rep{21},\triv)\oplus(\triv,\rep{36})\oplus(\rep{6},\rep{8}) \big] \oplus \big[ (\rep{21},\triv)\oplus(\triv,\rep{36}) \big] \\
    &= 36(\triv,\triv) \oplus 16(\rep{6},\triv) \oplus 16(\triv,\rep{8}) \oplus (\rep{14},\triv) \oplus (\triv,\rep{27}) \,.
\end{eqnalign}

Generically, anomaly-free models have gauge groups containing more than one simple factor and the procedure is more involved than simply checking maximal subgroups of each gauge factor independently. There are models for which more than one simple factor must break to a maximal subgroup simultaneously, corresponding, e.g., to giving a bi-fundamental hypermultiplet a VEV: see figure~\ref{fig:large-Δ-models} for many such examples. For a model with gauge group $G=\prod_{i=1}^\kappa G_i$ and hypermultiplets $\H$, introduce the following notation. First, let $L=(L_1,\ldots,L_k)$ denote a choice of subgroups for each $G_i$, where $L_i$ is either $G_i$ itself or one of its maximal subgroups, and let $|L|$ be the number of $L_i$ for which $L_i\neq G_i$ is a \emph{proper} subgroup. Write $G_L$ and $\H_L$ to denote the model which results upon Higgsing to the maximal subgroup of $G$ as specified by $L$ and write $G_L|$ and $\H_L|$ for the restriction to those gauge factors for which $L_i\neq G_i$ (i.e.\ ignoring gauge factors $G_i$ which were unaltered). Then we apply the following procedure:
\begin{enumerate}
    \item Let $\mathcal{L}_0=\{(G_1,\ldots,G_k)\}$. That is, $\mathcal{L}_0$ consists only of the trivial Higgsing which leaves the model unaltered.
    \item For each $\ell=1,2,\ldots,\kappa$ in turn\ldots
    \begin{enumerate}
        \item Construct the set of candidates $\mathcal{L}_\ell$ consisting of $L$ with $|L|=\ell$ which have the property that reverting any single $L_i\neq G_i$ in $L$ back to $G_i$ produces an element of $\mathcal{L}_{\ell-1}$.

        \item For each $L\in\mathcal{L}_\ell$, apply the Higgsing described by $L$ to the full model, removing by hand all $\U(1)$, $\SU(2)$, $\SU(3)$ and $\Sp(2)$ factors other than any $\SU(3)$ NHCs. If the resulting model is anomaly-free (i.e.\ all $n_R\geq 0$ in $\H_L$), check if it appears in the ensemble. Otherwise, if $\H_L|$ has $n_R<0$ for any non-trivial $R$, remove $L$ from $\mathcal{L}_\ell$.
    \end{enumerate}
\end{enumerate}
Following the above procedure finds all breaking patterns $G\to G'$ which are minimal, in the sense that if Higgsing $G$ to the maximal subgroup $G'\subset G$ produces an anomaly-free model then there is no intermediate subgroup $G''$ satisfying $G'\subset G''\subset G$ which also produces an anomaly-free model.

\section{Models with 3-charged hypermultiplets}
\label{app:3-charged-hypers}

Here we provide a complete list of anomaly-free models for $T\leq 1$ with hypermultiplets charged under three gauge factors when the groups $\U(1)$, $\SU(2)$, $\SU(3)$ and $\Sp(2)$ are forbidden. There are exactly 20 for both $T=0$ and $T=1$. The $T=0$ models are as follows:
\begin{center}
    \footnotesize
    \renewcommand{\arraystretch}{1.5}
    \begin{longtable}{*{3}{>$l<$}}
        \toprule
        \;G & \;\H & b_i\in\Z \\ \midrule
        \SU(4)^2\times\SO(7) & (\triv,\rep[\prime]{20},\triv)\oplus (\triv,\triv,\rep{35})\oplus (\rep{6},\rep{6},\triv)\oplus (\triv,\rep{15},\rep{7})\oplus (\rep{4},\rep{4},\rep{8}) & 2,6,4\\
        \SU(4)^2\times G_2 & \begin{aligned}
            &(\triv,\triv,\triv)\oplus (\triv,\rep[\prime]{20},\triv)\oplus (\triv,\triv,\rep{27})\oplus (\rep{4},\rep{4},\triv)\\[-5pt]
            &\quad{} \oplus (\rep{6},\rep{6},\triv)\oplus (\triv,\rep{15},\rep{7})\oplus (\rep{4},\rep{4},\rep{7})
        \end{aligned} & 2,6,4 \\
        \SU(4)^2\times\SO(8) & (\rep[\prime]{20},\triv,\triv)\oplus(\triv,\rep[\prime]{20},\triv)\oplus(\triv,\triv,\repss{35}{v})\oplus (\rep{4},\rep{4},\repss{8}{s})\oplus (\rep{4},\rep{4},\repss{8}{c}) & 4,4,4\\
        \SU(4)^2\times\SO(8) & (\rep[\prime]{20},\triv,\triv)\oplus(\triv,\rep[\prime]{20},\triv)\oplus(\triv,\triv,\repss{35}{s})\oplus (\rep{4},\rep{4},\repss{8}{v})\oplus(\rep{4},\rep{4},\repss{8}{c}) & 4,4,4\\
        \SU(4)\times\SO(7)^2 & (\rep[\prime]{20},\triv,\triv)\oplus(\triv,\rep{27},\triv)\oplus(\triv,\triv,\rep{27})\oplus(\rep{4},\rep{8},\rep{8}) & 4,4,4\\
        \SU(4)^2\times\SO(7) & (\triv,\triv,\triv)\oplus(\rep[\prime]{20},\triv,\triv)\oplus(\triv,\rep[\prime]{20},\triv)\oplus (\triv,\triv,\rep{27})\oplus 2(\rep{4},\rep{4},\rep{8}) & 4,4,4 \\
        \SU(4)^3 & 2(\triv,\triv,\triv)\oplus (\rep[\prime]{20},\triv,\triv)\oplus (\triv,\rep[\prime]{20},\triv)\oplus (\triv,\triv,\rep[\prime]{20})\oplus 4(\rep{4},\rep{4},\rep{4}) & 4,4,4 \\
        \SU(4)^2\times\SU(8) & (\rep[\prime]{20},\triv,\triv)\oplus(\triv,\rep[\prime]{20},\triv)\oplus (\triv,\triv,\rep{70})\oplus 2(\rep{4},\rep{4},\rep{8}) & 4,4,2 \\
        \SU(4)\times\SO(7)\times\Sp(4) & (\rep[\prime]{20},\triv,\triv)\oplus(\triv,\rep{27},\triv)\oplus(\triv,\triv,\rep{42})\oplus (\rep{4},\rep{8},\rep{8}) & 4,4,2\\
        \SU(4)^2\times\Sp(4) & (\triv,\triv,\triv)\oplus (\rep[\prime]{20},\triv,\triv)\oplus (\triv,\rep[\prime]{20},\triv)\oplus (\triv,\triv,\rep{42})\oplus 2(\rep{4},\rep{4},\rep{8}) & 4,4,2\\
        \SU(4)\times\Sp(4)^2 & (\rep[\prime]{20},\triv,\triv)\oplus (\triv,\rep{42},\triv)\oplus(\triv,\triv,\rep{42})\oplus(\rep{4},\rep{8},\rep{8}) & 4,2,2\\
        \SU(4)^2\times\Sp(3) & \begin{aligned}
            &2(\rep{4},\triv,\triv)\oplus 2(\rep{6},\triv,\triv)\oplus 2(\triv,\rep{15},\triv)\oplus 2(\triv,\rep[\prime\prime]{20},\triv) \oplus \tfrac{1}{2}(\triv,\triv,\rep{6})\\[-5pt]
            &\quad{}\oplus (\triv,\triv,\rep{14})\oplus \tfrac{1}{2}(\triv,\triv,\rep[\prime]{14})\oplus 3(\rep{4},\rep{4},\triv)\oplus (\rep{4},\rep{6},\triv)\oplus (\rep{6},\rep{4},\triv)\\[-5pt]
            &\quad{} \oplus 2(\triv,\rep{4},\rep{6})\oplus \tfrac{1}{2}(\triv,\rep{6},\rep{6})\oplus \tfrac{1}{2}(\rep{4},\rep{4},\rep{6})
        \end{aligned} & 2,5,1\\
        \SU(4)^2\times\Sp(3) & \begin{aligned}
            &(\triv,\rep{4},\triv)\oplus(\rep{6},\triv,\triv)\oplus (\triv,\rep{6},\triv)\oplus (\rep{15},\triv,\triv)\oplus (\triv,\rep{15},\triv)\\[-5pt]
            &\quad{} \oplus (\triv,\rep[\prime\prime]{20},\triv)\oplus \tfrac{1}{2}(\triv,\triv,\rep{6})\oplus (\triv,\triv,\rep{14})\oplus \tfrac{1}{2}(\triv,\triv,\rep[\prime]{14})\\[-5pt]
            &\quad{} \oplus 3(\rep{4},\rep{4},\triv)\oplus (\rep{4},\rep{6},\triv)\oplus 2(\rep{6},\rep{4},\triv) \oplus (\rep{4},\triv,\rep{6})\\[-5pt]
            &\quad{}\oplus (\triv,\rep{4},\rep{6})\oplus \tfrac{1}{2}(\triv,\rep{6},\rep{6}) \oplus\tfrac{1}{2}(\rep{4},\rep{4},\rep{6})
        \end{aligned} & 3,4,1\\
        \SU(4)^2\times\Sp(3) & \begin{aligned}
            &(\triv,\rep{4},\triv)\oplus2(\rep{6},\triv,\triv)\oplus (\rep{15},\triv,\triv)\oplus  (\triv,\rep{15},\triv)\oplus (\triv,\rep[\prime\prime]{20},\triv)\\[-5pt]
            &\quad{} \oplus \tfrac{1}{2}(\triv,\triv,\rep{6})\oplus (\triv,\triv,\rep{14}) \oplus\tfrac{1}{2}(\triv,\triv,\rep[\prime]{14})\oplus 3(\rep{4},\rep{4},\triv)\\[-5pt]
            &\quad{} \oplus 2(\rep{4},\rep{6},\triv)\oplus (\rep{6},\rep{4},\triv)\oplus 2(\triv,\rep{4},\rep{6})\oplus \tfrac{1}{2}(\rep{6},\triv,\rep{6})\oplus\tfrac{1}{2}(\rep{4},\rep{4},\rep{6})
        \end{aligned} & 3,4,1\\
        \SU(4)^2\times\SO(7)^2 & \begin{aligned}
            &(\rep[\prime]{20},\triv,\triv,\triv)\oplus(\triv,\rep[\prime]{20},\triv,\triv)\\[-5pt]
            &\quad{} \oplus(\triv,\triv,\rep{7},\rep{7})\oplus(\rep{4},\rep{4},\rep{8},\triv)\oplus(\rep{4},\rep{4},\triv,\rep{8})
        \end{aligned} & 4,4,2,2\\
        \SU(4)^4 & \begin{aligned}
            &(\triv,\triv,\triv,\triv)\oplus (\triv,\triv,\rep[\prime]{20},\triv)\oplus(\triv,\triv,\triv,\rep[\prime]{20})\\[-5pt]
            &\quad{} \oplus (\rep{6},\rep{6},\triv,\triv)\oplus 2(\rep{4},\triv,\rep{4},\rep{4})\oplus 2(\triv,\rep{4},\rep{4},\rep{4})
        \end{aligned} & 2,2,4,4\\
        \SU(4)^3\times\SO(7) & \begin{aligned}
            &(\triv,\triv,\rep[\prime]{20},\triv)\oplus(\triv,\triv,\triv,\rep{27})\oplus(\rep{6},\rep{6},\triv,\triv)\\[-5pt]
            &\quad{} \oplus(\rep{4},\triv,\rep{4},\rep{8})\oplus(\triv,\rep{4},\rep{4},\rep{8})
        \end{aligned} & 2,2,4,4\\
        \SU(4)^3\times\Sp(4) & \begin{aligned}
            &(\triv,\triv,\rep[\prime]{20},\triv)\oplus (\triv,\triv,\triv,\rep{42})\oplus (\rep{6},\rep{6},\triv,\triv)\\[-5pt]
            &\quad{} \oplus (\rep{4},\triv,\rep{4},\rep{8})\oplus(\triv,\rep{4},\rep{4},\rep{8})
        \end{aligned} & 2,2,4,2\\
        \SU(5)\times\SU(4)\times\Sp(3)^2 & \begin{aligned}
            &(\triv,\rep{10},\triv,\triv)\oplus (\rep{10},\rep{6},\triv,\triv)\oplus (\rep{5},\triv,\rep{14},\triv)\\[-5pt]
            &\quad{} \oplus (\rep{5},\triv,\triv,\rep{14})\oplus(\triv,\rep{4},\rep{6},\rep{6})
        \end{aligned} & 2,3,2,2\\
        \SU(4)^5 & \begin{aligned}
            &(\triv,\triv,\triv,\triv,\rep[\prime]{20})\oplus(\rep{6},\rep{6},\triv,\triv,\triv)\oplus(\triv,\triv,\rep{6},\rep{6},\triv)\oplus(\rep{4},\triv,\rep{4},\triv,\rep{4})\\[-5pt]
            &\quad{} \oplus(\rep{4},\triv,\triv,\rep{4},\rep{4})\oplus(\triv,\rep{4},\rep{4},\triv,\rep{4}) \oplus(\triv,\rep{4},\triv,\rep{4},\rep{4})
        \end{aligned} & 2,2,2,2,4\\
        \bottomrule
    \end{longtable}
\end{center}
The $T=1$ models are as follows:
\begin{center}
    \footnotesize
    \renewcommand{\arraystretch}{1.5}
    \begin{longtable}{*{3}{>$l<$}}
        \toprule
        \;G & \;\H & b_i\in\Gamma \\ \midrule
        \SU(4)^2\!\times\! G_2 & \begin{aligned}
            &(\rep[\prime]{20},\triv,\triv)\oplus(\triv,\rep[\prime]{20},\triv) \oplus (\rep{4},\rep{4},\triv)\oplus (\rep{6},\rep{10},\triv)\\[-5pt]
            &\quad{}\oplus (\rep{10},\rep{6},\triv) \oplus (\rep{4},\rep{4},\rep{7})
        \end{aligned} & U:(4,4)^2,(1,1) \\
        \SU(4)^3 & \begin{aligned}
            &(\rep{15},\triv,\triv)\oplus (\triv,\rep{15},\triv)\oplus (\triv,\triv,\rep{15})\oplus (\rep[\prime]{20},\triv,\triv)\\[-5pt]
            &\quad{}\oplus 2(\rep{4},\rep{6},\triv)\oplus 2(\rep{4},\triv,\rep{6})\oplus 2(\rep{4},\rep{4},\rep{4})
        \end{aligned} & U:(3,3),(2,2)^2 \\
        \SU(4)\!\times\!\Sp(3)\!\times\!G_2 & \begin{aligned}
            &(\rep{15},\triv,\triv)\oplus (\rep[\prime]{20},\triv,\triv)\oplus (\triv,\rep{21},\triv)\oplus (\triv,\triv,\rep{14})\\[-5pt]
            &\quad{}\oplus (\rep{4},\rep[\prime]{14},\triv)\oplus (\rep{4},\rep{6},\rep{7})
        \end{aligned} & U:(3,3),(2,2)^2 \\
        \SU(4)^2\!\times\!\Sp(3) & \begin{aligned}
            &(\triv,\triv,\triv)\oplus 4(\rep{15},\triv,\triv)\oplus 4(\triv,\rep{6},\triv)\oplus 3(\triv,\triv,\rep{14})\\[-5pt]
            &\quad{}\oplus 3(\rep{4},\rep{4},\triv)\oplus 2(\rep{6},\triv,\rep{6})\oplus \tfrac{1}{2}(\rep{4},\rep{4},\rep{6})
        \end{aligned} & U:(3,3),(1,1)^2 \\
        \SU(4)^2\!\times\!\Sp(3) & \begin{aligned}
            &(\rep{6},\triv,\triv)\oplus (\rep{10},\triv,\triv)\oplus 3(\rep{15},\triv,\triv)\oplus 4(\triv,\rep{6},\triv)\\[-5pt]
            &\quad{}\oplus 3(\triv,\triv,\rep{14})\oplus 3(\rep{4},\rep{4},\triv)\oplus 2(\rep{6},\triv,\rep{6})\\[-5pt]
            &\quad{}\oplus \tfrac{1}{2}(\rep{4},\rep{4},\rep{6})
        \end{aligned} & U: (3,3),(1,1)^2 \\
        \SO(8)\!\times\!\SU(4)^3 & \begin{aligned}
            &(\triv,\rep{15},\triv,\triv)\oplus(\triv,\triv,\rep{15},\triv)\oplus (\triv,\triv,\triv,\rep{15})\\[-5pt]
            &\quad{}\oplus (\repss{8}{v},\rep{6},\triv,\triv)\oplus(\repss{8}{s},\triv,\rep{6},\triv)\oplus (\repss{8}{c},\triv,\triv,\rep{6})\\[-5pt]
            &\quad{}\oplus 2(\triv,\rep{4},\rep{4},\rep{4})
        \end{aligned} & U:(1,1),(2,2)^3 \\
        \SU(4)^2\!\times\! G_2 & (\rep{4},\rep{4},\triv)\oplus (\rep{4},\rep{20},\triv)\oplus (\rep{20},\rep{4},\triv)\oplus(\rep{4},\rep{4},\rep{7}) & \Gamma_{1,1}:(5,3),(3,5),(1,1)\\
        \SU(4)^2\!\times\!\Sp(3) & \begin{aligned}
            &(\triv,\triv,\triv)\oplus 3(\rep{6},\triv,\triv)\oplus 5(\triv,\rep{10},\triv)\oplus \tfrac{1}{2}(\triv,\triv,\rep{6})\\[-5pt]
            &\quad{}\oplus (\triv,\triv,\rep{14})\oplus \tfrac{3}{2}(\triv,\triv,\rep[\prime]{14})\oplus 2(\rep{4},\rep{4},\triv)\\[-5pt]
            &\quad{}\oplus 3(\triv,\rep{6},\rep{6})\oplus \tfrac{1}{2}(\rep{4},\rep{4},\rep{6})
        \end{aligned} & \Gamma_{1,1}:(1,1),(5,2),(2,1) \\
        \SU(4)^2\!\times\!\Sp(3) & \begin{aligned}
            &(\triv,\triv,\triv)\oplus 2(\triv,\rep{10},\triv)\oplus\tfrac{1}{2}(\triv,\triv,\rep{6})\oplus (\triv,\triv,\rep{14})\\[-5pt]
            &\quad{}\oplus \tfrac{3}{2}(\triv,\triv,\rep[\prime]{14})\oplus (\rep{6},\triv,\rep{6})\oplus 2(\triv,\rep{6},\rep{6})\\[-5pt]
            &\quad{}\oplus 5(\rep{4},\rep{4},\triv)\oplus \tfrac{1}{2}(\rep{4},\rep{4},\rep{6})
        \end{aligned} & \Gamma_{1,1}:(2,0),(4,2),(2,1) \\
        \SU(4)^2\!\times\! \Sp(3) & \begin{aligned}
            &(\triv,\triv,\triv)\oplus (\rep{10},\triv,\triv)\oplus (\triv,\rep{10},\triv)\oplus \tfrac{1}{2}(\triv,\triv,\rep{6})\\[-5pt]
            &\quad{}\oplus (\triv,\triv,\rep{14})\oplus \tfrac{3}{2}(\triv,\triv,\rep[\prime]{14})\oplus \tfrac{3}{2}(\rep{6},\triv,\rep{6})\\[-5pt]
            &\quad{}\oplus \tfrac{3}{2}(\triv,\rep{6},\rep{6})\oplus 5(\rep{4},\rep{4},\triv)\oplus \tfrac{1}{2}(\rep{4},\rep{4},\rep{6})
        \end{aligned} & \Gamma_{1,1}:(3,1)^2,(2,1) \\
        \scalebox{0.9}{$\SO(7)\!\times\!\SU(4)^2\!\times\!\Sp(3)$} & \begin{aligned}
            &(\triv;\triv,\rep{10},\triv)\oplus (\triv;\triv,\triv,\rep{14})\oplus \tfrac{3}{2}(\triv;\triv,\triv,\rep[\prime]{14})\\[-5pt]
            &\quad{}\oplus \tfrac{1}{2}(\rep{7};\triv,\triv,\rep{6})\oplus (\rep{8};\rep{4},\triv,\triv)\oplus (\rep{8};\triv,\rep{4},\triv)\\[-5pt]
            &\quad{}\oplus 3(\triv,\rep{4},\rep{4},\triv)\oplus (\triv;\rep{6},\triv,\rep{6})\oplus \tfrac{3}{2}(\triv;\triv,\rep{6},\rep{6})\\[-5pt]
            &\quad{} \oplus \tfrac{1}{2}(\triv;\rep{4},\rep{4},\rep{6})
        \end{aligned} & \Gamma_{1,1}:\scalebox{0.9}{(1,1),(2,0),(3,1),(2,1)} \\
        G_2\!\times\!\SU(4)^3 & \begin{aligned}
            &(\rep{7};\triv,\triv,\triv)\oplus 2(\triv;\rep{10},\triv,\triv)\oplus 2(\triv;\triv,\rep{10},\triv)\\[-5pt]
            &\quad{}\oplus 2(\triv;\triv,\triv,\rep{10})\oplus (\triv;\rep{6},\rep{6},\triv)\oplus (\triv;\rep{6},\triv,\rep{6})\\[-5pt]
            &\quad{}\oplus (\triv;\triv,\rep{6},\rep{6})\oplus 2(\triv;\rep{4},\rep{4},\rep{4})
        \end{aligned} & \Gamma_{1,1}:(-1,-2),(4,2)^3 \\
        \SU(8)\!\times\!\SU(4)^3 & \begin{aligned}
            &(\rep{36};\triv,\triv,\triv)\oplus(\triv;\rep[\prime]{20},\triv,\triv)\oplus(\triv;\triv,\rep[\prime]{20},\triv)\\[-5pt]
            &\quad{}\oplus(\triv;\triv,\triv,\rep[\prime]{20})\oplus 4(\triv;\rep{4},\rep{4},\rep{4})
        \end{aligned} & \Gamma_{1,1}:(0,1),(4,0)^3 \\
        \SO(8)\!\times\!\SU(4)^3 & \begin{aligned}
            &(\triv;\triv,\triv,\triv)\oplus (\triv;\rep[\prime]{20},\triv,\triv)\oplus(\triv;\triv,\rep[\prime]{20},\triv)\\[-5pt]
            &\quad{}\oplus(\triv;\triv,\triv,\rep[\prime]{20})\oplus 4(\triv;\rep{4},\rep{4},\rep{4})
        \end{aligned} & \begin{aligned}
            \Gamma_{1,1} &: (0,2),(4,0)^3\\[-5pt]
            U &: (1,-2),(2,4)^3
        \end{aligned} \\
        \scalebox{0.9}{$\SO(8)\!\times\!\SU(4)^2\!\times\!\SO(7)$} & \begin{aligned}
            &(\triv;\rep[\prime]{20},\triv,\triv)\oplus (\triv;\triv,\rep[\prime]{20},\triv)\oplus (\triv;\triv,\triv,\rep{27})\\[-5pt]
            &\quad{}\oplus 2(\triv;\rep{4},\rep{4},\rep{8})
        \end{aligned} & \begin{aligned}
            \Gamma_{1,1} &: (0,2),(4,0)^3\\[-5pt]
            U &: (1,-2),(2,4)^3
        \end{aligned} \\
        \SO(9)\!\times\!\SU(4)^3 & \begin{aligned}
            &(\triv;\rep[\prime]{20},\triv,\triv)\oplus(\triv;\triv,\rep[\prime]{20},\triv)\oplus(\triv;\triv,\triv,\rep[\prime]{20})\\[-5pt]
            &\quad{}\oplus 4(\triv;\rep{4},\rep{4},\rep{4})
        \end{aligned} & \begin{aligned}
            \Gamma_{1,1} &: (0,2),(4,0)^3\\[-5pt]
            U &: (1,-2),(2,4)^3
        \end{aligned} \\
        \scalebox{0.9}{$\SO(8)\!\times\!\Sp(4)\!\times\!\SU(4)^2$} & \begin{aligned}
            &(\triv;\rep{42},\triv,\triv)\oplus (\triv;\triv,\rep[\prime]{20},\triv)\oplus (\triv;\triv,\triv,\rep[\prime]{20})\\[-5pt]
            &\quad{}\oplus 2(\triv;\rep{8},\rep{4},\rep{4})
        \end{aligned} & \begin{aligned}
            \Gamma_{1,1} &: (0,2),(2,0),(4,0)^2\\[-5pt]
            U &: (1,-2),(1,2),(2,4)^2
        \end{aligned} \\
        \SO(8)\!\times\!\SU(4)^4 & \begin{aligned}
            &(\triv;\triv,\triv,\rep[\prime]{20},\triv)\oplus(\triv;\triv,\triv,\triv,\rep[\prime]{20})\oplus (\triv;\rep{6},\rep{6},\triv,\triv)\\[-5pt]
            &\quad{}\oplus 2(\triv;\rep{4},\triv,\rep{4},\rep{4})\oplus 2(\triv;\triv,\rep{4},\rep{4},\rep{4})
        \end{aligned} & \begin{aligned}
            \Gamma_{1,1} &: (0,2),(2,0)^2,(4,0)^2\\[-5pt]
            U &: (1,-2),(1,2)^2,(2,4)^2
        \end{aligned} \\
        \SO(8)\!\times\!\SU(4)^2\!\times\!G_2 & \begin{aligned}
            &(\triv;\triv,\rep[\prime]{20},\triv)\oplus(\triv;\triv,\triv,\rep{27})\oplus (\triv;\rep{4},\rep{4},\triv)\\[-5pt]
            &\quad{}\oplus (\triv;\rep{6},\rep{6},\triv)\oplus (\triv;\triv,\rep{15},\rep{7})\oplus (\triv;\rep{4},\rep{4},\rep{7})
        \end{aligned} & \begin{aligned}
            \Gamma_{1,1} &: \scalebox{0.9}{$(0,2),(2,0),(6,0),(4,0)$}\\[-5pt]
            U &: \scalebox{0.9}{$(1,-2),(1,2),(3,6),(2,4)$}
        \end{aligned} \\
        \SO(14)\!\times\!\SU(4)^3 & \begin{aligned}
            &(\triv;\triv,\rep[\prime]{20},\triv)\oplus (\triv;\triv,\triv,\rep[\prime]{20}) \oplus (\rep{14};\rep{6},\triv,\triv)\\[-5pt]
            &\quad{}\oplus 4(\triv;\rep{4},\rep{4},\rep{4})
        \end{aligned} & \begin{aligned}
            \Gamma_{1,1} &: (0,2),(4,-2),(4,0)^2\\[-5pt]
            U &: (1,-2),(4,-2),(2,4)^2
        \end{aligned} \\
        \bottomrule
    \end{longtable}
\end{center}

\section{Orbifold Models}
\label{app:orbifolds}

In this appendix, we provide asymmetric orbifold models~\cite{Dixon:1985jw,Dixon:1986jc,Narain:1986qm,Narain:1990mw} realizing the example anomaly-free models of equations~\eqref{eq:e8-so8-so16} and~\eqref{eq:so8-so16^2}.
\begin{itemize}
    \item A model for \eqref{eq:e8-so8-so16}:

    Starting from the $E_8\times E_8$ heterotic theory and taking $\Gamma^{4,4}(D_4)$ lattice with
    \begin{align}
        \phi_L=(0,0)\,,
        \quad\phi_R=\tfrac{1}{2}(1,1)\,,
        \quad V_L=\tfrac{1}{2}\left(1^4,0^4;0^8\right)\,,
    \end{align}
    the invariant lattice and its dual are
    \begin{eqnalign}
        I &= \big\{(p_L,P_L;0)\;\big|\;p_L\in\Lambda_R(D_4)\,,\,P_L\in\Lambda_R(E_8\times E_8)\big\} \,,\\
        I^* &= \big\{(p_L^*,P_L;0)\;\big|\;p_L^*\in\Lambda_W(D_4)\,,\,P_L\in\Lambda_R(E_8\times E_8)\big\} \,.
    \end{eqnalign}
    The spectrum of the model is
    \begin{eqnalign}
        G &= \mathrm{Spin}(8)\times \mathrm{Spin}(16)
        \times E_8 \,,\\
        \H &= (\triv,\rep{128},\triv)\oplus (\repss{8}{v},\rep{16},\triv)
        \oplus (\repss{8}{s},\rep{16},\triv)\oplus (\repss{8}{c},\rep{16},\triv).
    \end{eqnalign}
    
    \item A model for \eqref{eq:so8-so16^2}:

    Starting from the $E_8\times E_8$ heterotic theory and taking $\Gamma^{4,4}(D_4)$ lattice with 
    \begin{align}
        \phi_L=(0,0)\,,
        \quad\phi_R=\tfrac{1}{2}(1,1)\,,
        \quad V_L=\tfrac{1}{2}\left(1^4,0^4;1^4,0^4\right)\,,
    \end{align}
    The lattices $I$ and $I^*$ are the same as above. The spectrum of the model is
    \begin{eqnalign}
        G &= \mathrm{Spin}(8)\times \frac{\mathrm{Spin}(16)^2}{\mathbb{Z}_2} \,,\\
        \H &= (\triv,\rep{128},\triv)\oplus (\triv,\triv,\rep{128})\oplus (\triv,\rep{16},\rep{16}) \,,
    \end{eqnalign}
    where $\mathbb{Z}_2$ acts diagonally on $\mathrm{Spin}(16)^2$ and acts non-trivially on the vector representation. Therefore, we get $H=512$.
    After Higgsing, this model is dual to F-theory on elliptic $\mathbb{F}_4$.

\end{itemize}

\bibliographystyle{JHEP}
\bibliography{refs.bib}

\end{document}